\documentclass[ijoc,nonblindrev]{informs4}
\usepackage{eqndefns-left}

\OneAndAHalfSpacedXI 

\usepackage{algorithm,enumitem}
\usepackage{algpseudocode}
\usepackage{bbold,bm} 
\usepackage{comment}
\usepackage{subcaption}
\usepackage{booktabs}
\usepackage{tikz}
\newcommand*\circled[1]{\tikz[baseline=(char.base)]{
 \node[shape=circle,draw,inner sep=1pt] (char) {#1};}} 
\usepackage{natbib}
 \bibpunct[, ]{(}{)}{,}{a}{}{,}%
 \def\bibfont{\small}%
\newcommand{\R}{\mathbb{R}}
\newcommand{\Z}{\mathbf{Z}} 
\newcommand{\rmd}{\mathrm{d}}
\newcommand{\cO}{\mathcal{O}}

\newcommand{\E}{\mathbb{E}}
\newcommand{\pr}{\mathbb{P}}
\newcommand{\var}{\mathbb{V}}
\newcommand{\cov}{\mathbb{C}}

\newcommand{\bt}{\boldsymbol{\theta}}   
\newcommand{\hatt}{\widehat{\bt}}   
\newcommand{\tildet}{\widetilde{\bt}}   
\newcommand{\ttrue}{\bt^c}
\newcommand{\tboot}{\tildet} 
\newcommand{\tsim}{\bt} 
\newcommand{\gpr}[1]{G({#1}|\tboot,\hatt)}
\newcommand{\bu}{\mathbf{u}}

\newcommand{\bensuggest}[1]{\textcolor{red}{#1}}

\EquationsNumberedThrough 

\TheoremsNumberedThrough 
\ECRepeatTheorems %

\MANUSCRIPTNO{}

\begin{document}


\RUNAUTHOR{He, Feng and Song}

\RUNTITLE{Efficient input uncertainty quantification for ratio estimator}

\TITLE{Efficient input uncertainty quantification for ratio estimator}

\ARTICLEAUTHORS{%
	\AUTHOR{Linyun He}
	\AFF{Georgia Institute of Technology, \EMAIL{lhe85@gatech.edu}}
	\AUTHOR{Ben Feng}
	\AFF{University of Waterloo, \EMAIL{ben.feng@uwaterloo.ca}}
	\AUTHOR{Eunhye Song}
	\AFF{Georgia Institute of Technology, \EMAIL{eunhye.song@isye.gatech.edu}}
} 

\ABSTRACT{%
We study the construction of a confidence interval (CI) for a simulation output performance measure that accounts for input uncertainty when the input models are estimated from finite data.
In particular, we focus on performance measures that can be expressed as a ratio of two dependent simulation outputs' means.
We adopt the parametric bootstrap method to mimic input data sampling and construct the percentile bootstrap CI after estimating the ratio at each bootstrap sample.
The standard estimator, which takes the ratio of two sample averages, tends to exhibit large finite-sample bias and variance, leading to overcoverage of the percentile bootstrap CI.
To address this, we propose two new ratio estimators that replace the sample averages with pooled mean estimators via the $k$-nearest neighbor ($k$NN) regression: the $k$NN estimator and the $k$LR estimator.
The $k$NN estimator performs well in low dimensions but its theoretical performance guarantee degrades as the dimension increases.
The $k$LR estimator combines the likelihood ratio (LR) method with the $k$NN regression, leveraging the strengths of both while mitigating their weaknesses; the LR method removes dependence on dimension, while the variance inflation introduced by the LR is controlled by $k$NN.
Based on asymptotic analyses and finite-sample heuristics, we propose an experiment design that maximizes the efficiency of the proposed estimators and demonstrate their empirical performances using three examples including one in the enterprise risk management application.
}



\HISTORY{}

\maketitle

\section{Introduction}\label{sec:intro}
In stochastic simulation, input models refer to the probability distributions that generate random inputs fed into the simulation logic.
The input models are often estimated from finite real-world observations, and therefore subject to estimation error.
The variability in the simulation output induced by the estimation error is termed \textit{input uncertainty} (IU) in simulation literature.
IU must be quantified to make correct inference on the system performance.
Over the past few decades, considerable attention has been directed towards IU quantification (IUQ) in simulation analysis; see~\cite{Barton2022Survey} for a recent survey.
Moreover,~\cite{corlu2020stochastic} discuss the impact of IUQ in application areas such as pharmaceutical supply chain, energy system, and high-tech manufacturing.

A popular measure of IUQ is the confidence (frequentist) or credible (Bayesian) interval that covers the performance measure under the unknown true input distributions.
The methods that construct such intervals often adopt a \textit{nested simulation} experiment design, which samples multiple sets of estimated input models from their approximate sampling (frequentist) or posterior (Bayesian) distributions to cover the space of possible input models, run simulations at each set of input models to estimate the performance measure, and aggregate the estimates to construct a confidence/credible interval.
These methods implicitly assume that the simulation output is \textit{unbiased} with respect to the performance measure.
Therefore, given a set of input models, the sample mean of the simulation outputs is an unbiased estimator of the performance measure.

In this paper, we design a simulation experiment for IUQ when the unbiasedness assumption breaks.
In particular, we consider the case when the performance measure is \emph{a ratio of expected values of two dependent simulation outputs}.
Below, we provide two examples that commonly arise in simulation studies:

\begin{itemize}
    \item Conditional expectation of a simulation output:
        A conditional expectation can be expressed as the ratio of the expectation of a simulation output multiplied with an indicator function of an event to the expectation of the indicator, i.e., the probability of the event.
    \item Regenerative simulation estimator of a steady-state performance measure:
        When the simulator is employed with the regenerative structure, the renewal theory  ensures that the ``long run average cost'' can be written as the ratio of the accumulated cost and the length of each regenerative cycle~\citep{hendersonglynn2001}.
\end{itemize}
Both quantities are frequently studied in simulation applications to obtain engineering/managerial insights, thus bear significant relevance.

The standard ratio estimator replaces the expected values in both the numerator and denominator with the sample averages of the respective simulation outputs generated from multiple independent runs.
Even though the sample average estimator for both numerator and denominator are unbiased, the ratio estimator is not.
Depending on the sample size and variability of the simulation sample paths, the bias can be quite large even for a reasonably large number of simulation runs.
When the standard estimator is combined with the nested simulation experiment for IUQ, the required simulation budget increases by the factor of the number of sampled input model sets, which can be prohibitively large if each simulation run is expensive.

To improve simulation efficiency, we propose new ratio estimators and corresponding experiment designs that incorporate the $k$-nearest neighbor ($k$NN) regression and the likelihood ratio (LR) method.
Taking the frequentist view, we assume that the input models have known distribution families with unknown parameters estimated from data.
To approximate the sampling distribution of the estimated parameter, we adopt the parametric bootstrap method and sample from a \emph{bootstrap parameter set}, compute the ratio estimator at each parameter in the set, then construct the percentile bootstrap confidence interval (CI) aggregating all ratio estimators.
Unlike the standard ratio estimator, our experiment design does not run simulations at the bootstrap parameter set.
Instead, it generates a separate \emph{simulation parameter set} at which simulations are carried out, then utilizes the collected simulation outputs to estimate the ratios at the bootstrap parameters.

The first proposal is the $k$NN ratio estimator; at each bootstrap parameter vector, its $k$NNs among the simulation parameter set are first determined.
Then, the numerator and denominator of the ratio are estimated separately by pooling the simulation outputs from the $k$NNs.
These pooled $k$NN estimators are biased, but have reduced variance.
Controlling the sizes of the bootstrap and simulation parameter sets, and $k$ in the $k$NN regression, we show that the bootstrap CI constructed from the $k$NN ratio estimators is asymptotically consistent as the input data size increases even if a finite number of simulation runs are made at each simulation parameter vector.
This is a stark difference from the standard ratio estimator, which requires the number of simulation runs to increase to infinity for the percentile CI to be consistent; see Proposition~\ref{prop:standard.est.convergence}.

However, the computational efficiency of the $k$NN ratio estimator degrades as the parameter dimension increases.
The second estimator, $k$LR ratio estimator, addresses this issue by combining $k$NN with the LR method.
For each simulation run, we collect the inputs generated within the run.
Incorporating the sample likelihood ratio of these inputs within the $k$NN regression, we obtain unbiased estimators of the numerator and denominator from which the $k$LR estimator is constructed.
We show that the estimation error of the $k$LR does not depend on the parameter dimension, while the consistency of the percentile CI is still achieved with a finite number of simulation runs at each simulation parameter vector. Furthermore, we show that the $k$LR method's total simulation budget is required to grow faster than the input data size, $m$, for consistency, while the standard estimator's is required to grow faster than $m^{3/2}$. 

\subsection{Related Literature and our contributions}
We review two threads of the literature most relevant to this work, IUQ and nested simulation experiment design, to clearly identify our contribution to the literature.
For the IUQ, we focus our attention to frequentist's approaches.
For a more comprehensive review, see~\cite{Barton2022Survey}.

The objective of IUQ is to provide statistical inference on the true performance measure of the simulated system when the simulation is subject to both stochastic and input uncertainties.
Several CI-based IUQ procedures have been proposed in the literature.
\cite{cheng1994selecting} and~\cite{cheng1997sensitivity} estimate the unknown parameter vector of the input models by the maximum likelihood estimator (MLE), then approximate the simulation output mean performance measure as a function of the MLE by the delta method.
Combining this linearized model with the asymptotic normality of the MLE, they construct a normality-based CI for the mean.
\cite{linsongnelson2015} and~\cite{songnelson2015} study improving the computational efficiency of the method, and generalizing the parametric model to moment estimates, respectively.
These delta-method-based CIs tend to have poor coverage until the input data size is quite large so that the local linearization of the mean function is sufficient to characterize IU.

\cite{barton_resampling_2001} first introduce bootstrap CIs to IUQ, which are free of the linear approximation and more robust to non-normality in the finite-sample regime.
However, bootstrap CIs tend to show overcoverage since the simulation error is convoluted with IU.
Addressing the overcoverage by increasing the number of simulation runs at each bootstrap sample is computationally demanding.
Several approaches have been proposed to improve the simulation efficiency of the bootstrap CI such as sectioning~\citep{glynn2017}, subsampling~\citep{LamQian2022}, and bootstrap shrinkage~\citep{songlambarton2024}.

\cite{bartonnelsonxie2014} and~\cite{xienelsonbarton2014} adopt metamodeling to address both linearization and simulation efficiency issues via fitting a Gaussian process (GP) model to the mean performance measure by simulating only at the parameter vectors selected by an experiment design.
However, the size of the experiment design needs to grow in the parameter dimension to obtain a good fit, and the computational overhead in computing the GP predictive means may outweigh the simulation cost for a large experiment design~\citep{songlambarton2024}.

As mentioned earlier, IUQ can be considered an application of nested simulation experiment.
In general, nested simulation consists of outer scenarios (e.g., bootstrap samples) and inner replications at each outer scenario (e.g., simulation runs).
Since~\cite{gordy2010nested} analyze the efficiency of the standard nested simulation experiment design where the performance measure at each outer scenario is estimated from the inner replications made only at the outer scenario, several designs that pool the inner replications to improve the computational efficiency have been proposed.
\cite{liu2010stochastic},~\cite{broadie2015risk}, and~\cite{hong2017kernel} propose pooling via GP, linear regression, and kernel smoothing, respectively.
Pooling via the LR method has been also explored in the context of IUQ~\citep{zhou2018online,feng2019efficientIUQ,fengsong2024}.

All aforementioned works consider (unbiased) simulation output mean as a performance measure at each outer scenario.
In contrast, this paper studies the ratio estimator of two different expected simulation outputs whose estimator is biased in general. For a large class of discrete-event simulation models, the bias can be also translated to the initialization bias, which has not been studied in the IUQ literature so far.
Furthermore, to the best of our knowledge, this is the first work that analyzes the combination of the $k$NN and LR methods in nested simulation, which we prove to be asymptotically more efficient than the standard method.

The rest of the paper is organized as follows.
We formally define the IUQ problem in Section~\ref{sec:problem statement} and propose the new estimators in Section~\ref{sec:kNN_estimators}.
Section~\ref{sec:analysis} presents theoretical analyses of the proposed ratio estimators and the CIs constructed from them. Section~\ref{sec:para selection} provides a practical guidance on how to select the parameters for the estimation procedures.
We study the empirical performances in Section~\ref{sec:numerical}.
Section~\ref{sec:conclusion} concludes. All proofs of the theoretical statements are presented in the Online Supplement of this paper.

\noindent \textbf{Notation.} For positive sequences $\{a_n\}$ and $\{b_n\}$, we write $a_n = \cO(b_n)$ if there exist constant $c_0>0$ and $n_0>0$ such that $a_n\leq c_0 b_n$ for all $n\geq n_0$;
and write $a_n = o(b_n)$ or $b_n = \omega(a_n)$ if $\lim_{n\to \infty} a_n/b_n = 0$.
Moreover, for a sequence of random variables $\{X_n\}$, we write $X_n = \cO_{\pr}(a_n)$ if for every $\varepsilon>0$, there exists $c_1>0$ and $n_1>0$ such that $\pr(\lvert X_n/a_n\rvert <c_1)>1-\varepsilon$ for all $n>n_1$.
We define $[n]\triangleq\{1,2,\ldots,n\}$.

\section{Problem statement}\label{sec:problem statement}
Suppose that the stochasticity of a target system to simulate is characterized by $L\geq 1$ independent input-generating distributions with known families.
Let $p_l(\cdot|\vartheta_l^c)$ be the probability distribution function of the $l$th input parameterized by vector $\vartheta_l^c$.
Thus, $\ttrue = (\vartheta_1^c,\ldots,\vartheta_L^c)\in\R^d$ and $p(\cdot|\ttrue)=\prod_{l=1}^L p_l(\cdot|\vartheta_l^c)$, respectively, represent the parameter vector and the joint distribution function of all inputs in the system.
Without loss of generality, we assume $L=1$ in the remainder of the paper for ease of exposition.
However, our discussions and results can easily be extended to the case when $L>1$.

We focus on the performance measure, $\eta$, which can be expressed as the ratio,
\begin{equation}\label{def:eta}
	\eta(\tsim) = \frac{\E[Y|\tsim]}{\E[A|\tsim]},
\end{equation}
where $(Y, A)$ is a pair of dependent simulation outputs from the same replication and $\E[\cdot|\tsim]$ is the expectation  with respect to the inputs generated from the models parameterized by arbitrary $\tsim$.
Throughout the paper, we assume $\E[Y|\tsim]<\infty$
and $\E[A|\tsim]\neq 0$, ensuring that $|\eta(\tsim)| <\infty$ for all feasible values of $\tsim$.

Two examples of $\eta(\tsim)$ are a conditional expectation and a steady-state performance measure of a regenerative simulation model.
For the former,
consider random variable $X$ and some event $\mathcal{B}$.
Then, the conditional expectation $\E[X|\mathcal{B}]= \E[Y]/\E[A]$, where $Y \triangleq X\mathbf{1}\{\mathcal{B}\}$ and $A \triangleq \mathbf{1}\{\mathcal{B}\}$.

For regenerative simulation, consider a real-valued stochastic process, $\Xi(\tsim, t), t \geq 0$, generated from a simulator run with input parameter $\tsim$, where $t$ is the simulation time. Suppose $\Xi(\tsim, t)$ has state space $\mathbb{S}$.
We say $s_0\in\mathbb{S}$ to be a \emph{regenerative state}, if each time $\Xi(\tsim,\cdot)$ revisits $s_0$, $\Xi(\tsim,\cdot)$ becomes independent from its historical sample path.
The \emph{regenerative cycle} is defined as the time between two consecutive visits to $s_0$.
Taking a Markovian queue as an example, $s_0$ represents the empty state (no remaining jobs).
Once the system reaches $s_0$, new arrivals and services occur until the system becomes empty again, completing a regenerative cycle.
As a result, $\Xi(\tsim,t)$ within the regenerative cycles are independent from each other.

Under some regularity conditions, a generic discrete-event simulation (DES) model can be represented in the regenerative simulation framework~\citep{hendersonglynn2001}.
An important performance measure of interest in DES is a steady-state mean of real-valued reward function $w:\mathbb{S}\to \R$ defined as
$
	\eta(\tsim) = \lim_{t\to\infty} \frac{1}{t}\int_0^t w(\Xi(\tsim,s)) \rmd s,
$
where $\tsim$ is the input parameter adopted to run the simulator.
In practice, due to the limited simulation budget, one can adopt $\overline{w}(\tsim,T) = \frac{1}{T}\int_0^T w(\Xi(\tsim,s))\rmd s$ for some finite $T$ as a finite-time approximation of $\eta(\tsim)$.
Unfortunately, the estimator $\overline{w}(\tsim,t)$ typically has the so-called initial bias as the steady-state distribution of the system state is rarely known and therefore, $s_0$ is chosen arbitrarily.

The regenerative simulation framework provides an alternative way to define $\eta(\tsim)$.
Let $Y_j$ and $A_j$ be the cumulative reward within and the length of the $j$th regenerative cycle, respectively.
Then, the pair, $(Y_j, A_j)$, is i.i.d.\ for $j=1,2,\ldots$, thanks to the independence of regenerative cycles.
Under regularity conditions, the renewal reward theorem~\citep{ross1995stochastic} stipulates that the long-run reward rate, $\eta$, is equivalent to~\eqref{def:eta}. 
The regenerative simulation framework translates the initial bias problem of the steady-state simulation to the bias problem in the ratio estimator~\citep{glynn2006simulation}.

Our goal is to provide a CI that covers $\eta(\ttrue)$ with nominal probability $1-\alpha$.
In many applications, $\ttrue$ is unknown and must be estimated from data collected from the target system.
Let $\{X_i\}_{i\in [m]}\stackrel{\text{i.i.d.}}{\sim}p(\cdot|\ttrue)$ be the size-$m$ input data.
Given the data, the maximum likelihood estimator (MLE), $\hatt$, is known to have desirable properties such as strong consistency and statistical efficiency~\citep{Vaart_1998}.
Often in practice, $\hatt$ is adopted as a point estimator of $\ttrue$, and simulations are run at $\hatt$ to compute a $1-\alpha$ CI for $\eta(\hatt)$ accounting for the simulation error in estimating $\eta(\hatt)$.
However, the resulting CI does not provide $1-\alpha$ coverage for $\eta(\ttrue)$ in general since it ignores the IU in $\eta(\hatt)$ caused by the estimation error in $\hatt$.
When $m$ is small, the CI may severely undercover~\citep{Barton2022Survey}.

To account for IU, the sampling distribution of $\eta(\hatt)$ must be approximated from the size-$m$ input data at hand.
We adopt \textit{parametric bootstrap}~\citep{russellchengbook}:
In each bootstrap iteration $i$, we draw a size-$m$ sample from $p(\cdot|\hatt)$ and computes the estimator (e.g., MLE), $\tboot_i$, of $\hatt$ from the sample.
Given $\hatt$, we denote the sampling distribution of $\tboot_i$ by $\widetilde{f}(\cdot|\hatt)$ and its support by $\widetilde\Theta$.
Note that $\widetilde{f}(\cdot|\hatt)$ is deterministic given $\hatt$.
Furthermore, $\hatt$ and $\tboot$ have the same sampling distribution family with different parameters, i.e., $\hatt\sim\widetilde{f}(\cdot|\ttrue)$, as the same MLE is adopted to compute $\hatt$ and $\tboot_i$.

Algorithm~\ref{algo:parametric_bootstrapping} shows a generic parametric bootstrap procedure to construct a CI that incorporates IU, where $r$ stands for the simulation size at each $\tboot_i$.
Let $\widehat{\eta}$ be an estimator of $\eta$ and $q_{\alpha,\widetilde{n}}(\{\widehat{\eta}(\tboot_i)\})$ be the empirical $\alpha$-quantile of the size-$\widetilde{n}$ sample $\{\widehat{\eta}(\tboot_i)\}_{i\in {[\widetilde{n}]}}$ for $0<\alpha<1$.
From the simulation runs made by Algorithm~\ref{algo:parametric_bootstrapping}, one can produce the basic bootstrap CI,
\begin{equation}
	\label{eq:basic.bootstrap.CI}
	\left[2\widehat{\eta}(\hatt) - q_{1-\alpha/2,\widetilde{n}}(\{\widehat{\eta}(\tboot_i)\}), 2\widehat{\eta}(\hatt)-q_{\alpha/2,\widetilde{n}}(\{\widehat{\eta}(\tboot_i)\})\right],
\end{equation}
which is derived from the bootstrap principle that the distribution of $\eta(\tboot_i)-\eta(\hatt)$ approximates that of $\eta(\hatt)-\eta(\ttrue)$ well when  $m$ is large.
Instead,
Algorithm~\ref{algo:parametric_bootstrapping} returns the percentile bootstrap CI,
\begin{equation}
	\label{eq:percentile.bootstrap.CI}
	\left[ q_{\alpha/2,\widetilde{n}}(\{\widehat{\eta}(\tboot_i)\}), q_{1-\alpha/2,\widetilde{n}}(\{\widehat{\eta}(\tboot_i)\}) \right],
\end{equation}
which can be justified if there exists a monotonic transformation that makes the distributions of $\eta(\tboot_i)-\eta(\hatt)$ and $\eta(\hatt)-\eta(\ttrue)$ symmetric around $0$ for large $m$~\citep{davison_hinkley_1997}; see Section~\ref{subsec:CIs} for further discussions.
We adopt~\eqref{eq:percentile.bootstrap.CI} in this paper as it has been reported in the literature to show better empirical coverage than~\eqref{eq:basic.bootstrap.CI} when $\eta$ is asymmetric~\citep{efron1994introduction,songlambarton2024}.

\begin{algorithm}[tb]
	\caption{A generic parametric bootstrap IUQ procedure}
	\begin{algorithmic} \footnotesize
		\State \textbf{Inputs:} size-$m$ input data $\{X_i\}_{i\in [m]}$, input distribution $p(\cdot|\tsim)$, bootstrap size~$\widetilde{n}$.
		\State Calculate the MLE, $\hatt$ of $\ttrue$ from the input data.
		\For{$i=1$ to $\widetilde{n}$}
		\State{\textbf{Bootstrapping}: Generate bootstrap sample $\{X_j^i\}_{j\in [m]}\stackrel{\text{i.i.d.}}{\sim} p(\cdot|\hatt)$ then calculate the MLE $\tboot_i$ of $\hatt$ from the bootstrap sample.}
		\State{\textbf{Simulation}: For $j\in [r]$, simulate with inputs generated from $p(\cdot|\tboot_i)$ to obtain $Y_j(\tboot_i)$ and $A_j(\tboot_i)$.}
		\State{\textbf{Compute Ratio Estimator}: Compute $\widehat{\eta}(\tboot_i)$ from $\{Y_j(\tboot_i), A_j(\tboot_i)\}_{j\in[r]}$}.
		\EndFor
		\State Return~\eqref{eq:percentile.bootstrap.CI} as a CI for $\eta(\ttrue)$
	\end{algorithmic}
	\label{algo:parametric_bootstrapping}
\end{algorithm}

Although Algorithm~\ref{algo:parametric_bootstrapping} can be run with any estimator $\widehat{\eta}$, the coverage error of~\eqref{eq:percentile.bootstrap.CI} is affected by the estimation error in $\widehat{\eta}$.
The efficiency of $\widehat{\eta}$ is assessed by how fast $r$ and $\widetilde{n}$ must grow in $m$ to achieve the convergence.

A natural choice for $\widehat\eta$ is the standard ratio estimator that replaces the expectations in~\eqref{def:eta} with the respective sample means computed from $r$ i.i.d.\ simulation runs:
\begin{equation}\label{def: std estimator}
	\widehat{\eta}_{std}(\tsim) = \frac{\bar{Y} (\tsim)}{\bar{A}(\tsim) }
	= \frac{\frac{1}{r}\sum_{j=1}^r Y_j(\tsim) }{\frac{1}{r}\sum_{j=1}^r A_j(\tsim) }.
\end{equation}
Here, a simulation run may refer to a replication or an observation within a regenerative cycle.
Although $\bar{Y}(\tsim)\triangleq\frac{1}{r} \sum_{j=1}^r Y_j(\tsim)$ and $\bar{A} (\tsim)\triangleq\frac{1}{r} \sum_{j=1}^r A_j(\tsim)$ are unbiased estimators of $\E[Y|\tsim]$ and $\E[A|\tsim]$, respectively, $\widehat{\eta}_{std}(\tsim)$ is biased in general.
While $\widehat{\eta}_{std}(\tsim)$ is a consistent estimator for $\eta(\tsim)$ as $r\to\infty$, the bias can be large for finite $r$.
In Section~\ref{sec:analysis}, we make assumptions on the smoothness of $\widetilde{f}$, $Y$, $A$ and the estimation error, $\widehat{\eta}_{std}(\tboot) - \eta(\tboot)$, under which the following proposition can be shown.
\begin{proposition}\label{prop:standard.est.convergence}
	Suppose Assumptions~\ref{assump: ecdf} and~\ref{assump:mono-trans} hold.
	Then, by choosing $r=\omega(m^{1/2})$ and $\widetilde{n}=\omega(m)$, we have
	$\pr\left(
		\eta(\ttrue) \in
			[
				q_{\alpha/2,\widetilde{n}}(\{\widehat{\eta}_{std}(\tboot_i)\}|\hatt), q_{1-\alpha/2,\widetilde{n}}(\{\widehat{\eta}_{std}(\tboot_i)\}|\hatt)
			]
		\right)
		= 1-\alpha+o(1)$ for $0< \alpha<1$.
\end{proposition}
The proposition states that selecting $\widetilde n \sim r^2$ produces a CI with the correct coverage.
Indeed, this result matches the sample-size requirements in the classical nested simulation literature to minimize the MSE of the nested statistic when the performance measure at each outer scenario is an expectation, not a ratio~\citep{gordy2010nested}.
On the other hand, the sample size requirements in relation to $m$ are unique in the IUQ problem; even if infinite $\tilde{n}$ and $r$ are used, there still remains error due to the finite input data. Since $m$ cannot be controlled by the simulation experiment design, we represent other parameters ($\tilde{n}$ and $r$) in $m$.

Although in Proposition~\ref{prop:standard.est.convergence} above and in the remainder of the paper we study the asymptotic regime when $m\to\infty$, we clarify that the sample size $m$ is fixed in our problem. The asymptotics nonetheless provide insights on how well the proposed IUQ procedures work when $m$ is sufficiently large and a guidance on how to choose the experiment design parameters.

\section{Proposed ratio estimators and experiment design}\label{sec:kNN_estimators}
In this section, we propose two ratio estimators that utilize the $k$NN and the LR methods, and the associated experiment design to improve estimation accuracy and computational efficiency of the standard estimator.

Our experiment design 
generates two sets of input parameters: \emph{bootstrap parameter set} and \emph{simulation parameter set}.
The former, $\{\tboot_i\}_{i\in [\widetilde{n}]}$, is generated via parametric bootstrap as  in Algorithm~\ref{algo:parametric_bootstrapping}.
However, we do \textit{not} run simulations at the bootstrap parameters.
Instead, we run $r$ i.i.d.\ simulation runs at each parameter in the size-$n$ simulation parameter set, $\{\tsim_j\}_{j\in[n]}$. Additionally, the two sample sizes, $n$ and $\widetilde{n}$, can differ.

Let $f(\cdot|\hatt)$ represent the sampling distribution of $\tsim$ and $\Theta$ be its support, respectively.
We allow $f(\cdot|\hatt)\neq\widetilde{f}(\cdot|\hatt)$. Moreover, $f(\cdot|\hatt)$ may depend on $\{\tboot_i\}_{i\in [\widetilde{n}]}$ if so desired by the user; our analysis accommodates this case, but we adopt the notation, $f(\cdot|\hatt)$, throughout the paper for simplicity.
The flexibility in choosing $f$ helps us obtain $\widehat{\eta}$ with smaller asymptotic MSE, as analyzed in Section~\ref{sec:analysis}.

\begin{figure}[tbp]
	\centering
	\includegraphics[width=0.9\textwidth]{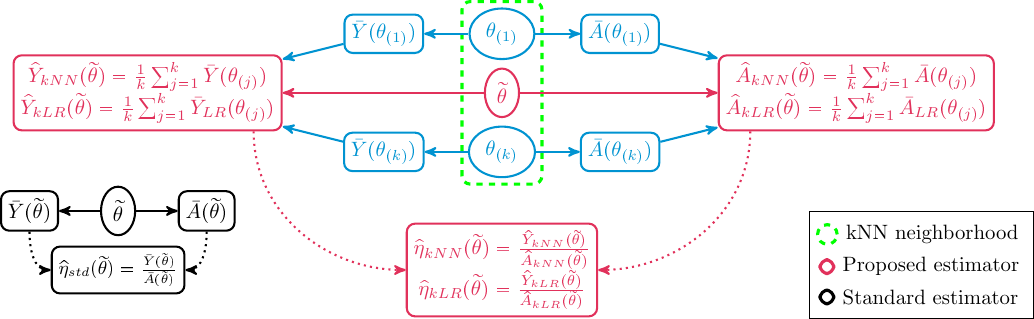}
	\caption{Schematic illustration of $\widehat{\eta}_{std}(\tboot)$ and the two proposed estimators, $\widehat{\eta}_{kNN}(\tboot)$ and $\widehat{\eta}_{kLR}(\tboot)$.}
	\label{fig:illustration}
\end{figure}

We propose two novel $k$NN-based ratio estimators, $\widehat{\eta}_{kNN}(\tboot)$ and $\widehat{\eta}_{kLR}(\tboot)$, utilizing the experiment design.
Both estimate $\eta(\tboot_i)$ at each bootstrap parameter $\tboot_i$ by pooling all $(Y,A)$ observations generated from $nr$ simulation runs made at the simulation parameters as illustrated in Figure~\ref{fig:illustration}.

Let $\tboot \in \widetilde{\Theta}$ be an arbitrary bootstrap parameter. We define $\widehat{\eta}_{kNN}(\tboot)$ as
\begin{equation}\label{def:kNN}
	\widehat{\eta}_{kNN}(\tboot) \triangleq \frac{\widehat{Y}_{kNN}(\tboot)}{\widehat{A}_{kNN}(\tboot)}
	= \frac{\frac{1}{k}\sum_{i=1}^{k}\bar{Y} (\tsim_{(i)})} {\frac{1}{k}\sum_{i=1}^{k}\bar{A}(\tsim_{(i)})},
\end{equation}
where $\tsim_{(i)}$ denotes the $i$th nearest neighbor (NN) of $\tboot$ in Euclidean distance among the simulation parameters, $\{\tsim_j\}_{j \in[n]}$.
In essence, we employ two separate $k$NN regression models for $\E[Y|\cdot]$ and $\E[A|\cdot]$, then $\widehat{\eta}_{kNN}(\tboot)$ is the ratio of the two models evaluated at $\tboot$.
Note that $\bar{Y}(\tsim_{(i)})$ and $\bar{A}(\tsim_{(i)})$ are correlated as they are computed from the same simulation sample paths.
This correlation is considered in our analysis.

Compared to $\widehat{\eta}_{std}(\tboot)$, the numerator and denominator of $\widehat{\eta}_{kNN}(\tboot)$ have smaller variances at the expense of biases since $\widehat{Y}_{kNN}(\tboot)$ and $\widehat{A}_{kNN}(\tboot)$ are no longer unbiased estimators of $\E[Y|\tboot]$ and $\E[A|\tboot]$, respectively.
The second method we propose eliminates the biases by combining $k$NN regression with the LR method.

Consider an arbitrary simulation parameter $\tsim\in\Theta$.
Let $\Z_j(\tsim) = \{Z_{j,1}(\tsim), \ldots, Z_{j,S(\tsim)}(\tsim)\}$ be the set of simulation inputs generated from $p(\cdot|\tsim)$ within the $j$th simulation run, where $S(\tsim)$ denotes the total number of inputs.
For example, in regenerative simulation, $S(\tsim)$ represents the number of inputs generated within a regeneration cycle, which is random.
Clearly, both $A_j(\tsim)$ and $Y_j(\tsim)$ are functions of $\Z_j(\tsim)$.
Assumption~\ref{assump:absolute.continuity} below lets us construct unbiased estimators of $\E[A|\tboot]$ and $\E[Y|\tboot]$ from $A_j(\tsim), Y_j(\tsim)$ and $\Z_j(\tsim)$.
\begin{assumption}\label{assump:absolute.continuity}
	For any $\tsim\in\Theta$ and $\tboot\in\widetilde{\Theta}$, $p(\cdot|\tboot)$ is absolutely continuous with respect to $p(\cdot|\tsim)$.
\end{assumption}
If $S(\tsim)$ is constant, then the following change of measure is valid under Assumption~\ref{assump:absolute.continuity}:
\begin{equation}\label{eq:LRdef}
	\E[ Y_j\left(\tsim\right) W_j(\tsim; \tboot)|\tsim]
	=\int y\left(\Z_j\right) \frac{\prod_{\ell=1}^{S\left(\tsim\right)}p(z_{j,\ell}|\tboot) }{\prod\nolimits_{\ell=1}^{S\left(\tsim\right)}p(z_{j,\ell}|\tsim) } \prod\nolimits_{\ell=1}^{S\left(\tsim\right)}p(z_{j,\ell}|\tsim)\rmd \mathbf{z}_j
	=\E[Y| \tboot].
\end{equation}
Note that $W_j(\tsim; \tboot) \triangleq \frac{\prod_{\ell=1}^{S\left(\tsim\right)}p(Z_{j,\ell}(\tsim)|\tboot) }{\prod\nolimits_{\ell=1}^{S\left(\tsim\right)}p(Z_{j,\ell}(\tsim)|\tsim) }$ is the LR between the input models parameterized by $\tboot$ and $\tsim$ evaluated at $\Z_j(\tsim)$.
When $S(\tsim)$ is random, the conclusion of~\eqref{eq:LRdef} is still valid but requires a slightly more complicated proof; see~\cite{feng2019efficientIUQ}.
Thus,~\eqref{eq:LRdef} suggests that $\bar{Y}_{LR} (\tsim) \triangleq \frac{1}{r}\sum_{j=1}^r Y_j(\tsim)W_j(\tsim; \tboot)$ is an unbiased estimator of $\E[Y| \tboot]$; $\bar{A}_{LR} (\tsim)$ is defined similarly. 
We propose the $k$LR ratio estimator, $\widehat{\eta}_{kLR}(\tboot)$, by combining $k$NN regression with the LR method:
\begin{equation}\label{def:kLR}
	\widehat{\eta}_{kLR}(\tboot) \triangleq \frac{\widehat{Y}_{kLR}(\tboot)}{\widehat{A}_{kLR}(\tboot)}
	\triangleq \frac{ \frac{1}{k}\sum_{i=1}^{k} \bar{Y}_{LR} \left(\tsim_{(i)}\right)} { \frac{1}{k}\sum_{i=1}^{k}\bar{A}_{LR} \left(\tsim_{(i)}\right)}
\end{equation}
Notice that $\widehat{Y}_{kLR}(\tboot)$ is an average of $k$ unbiased LR estimators of $\E[Y|\tboot]$, thus is itself an unbiased estimator of $\E[Y|\tboot]$.
So, unlike $\widehat{\eta}_{kNN}(\tboot)$, the numerator and denominator of $\widehat{\eta}_{kLR}(\tboot)$ are both unbiased estimators.

While the LR method eliminates the bias caused by the $k$NN regression, it may inflate variance:
$\bar{Y}_{LR} (\tsim)$
may have a large or even infinite variance when $\tsim$ and $\tboot$ significantly differ. 
In Section~\ref{subsec:kLR}, we show that the variance inflation can be mitigated by carefully choosing the value of $k$. 
Consequently, $\widehat{\eta}_{kLR}(\tboot)$ offsets the respective drawbacks of the $k$NN and LR methods by combining both.
\begin{algorithm}[tb]
	\caption{$k$NN and $k$LR IUQ procedure}
	\begin{algorithmic} \footnotesize
		\State \textbf{Inputs:} Inputs for Algorithm~\ref{algo:parametric_bootstrapping}, simulation parameter sampling distribution $f(\cdot|\hatt)$, simulation parameter set size $n$.
		\State Calculate the MLE, $\hatt$, of $\ttrue$ from the input data.
		\For{$i=1$ to $\widetilde{n}$}
		\State{\textbf{Bootstrapping}: Generate bootstrap sample $\{X_j^i\}_{j\in [m]}\stackrel{\text{i.i.d.}}{\sim} p(\cdot|\hatt)$ then calculate the MLE $\tboot_i$ of $\hatt$ from the bootstrap sample.}
		\EndFor
		\State{\textbf{Sampling}: Generate simulation parameter sample $\{\tsim_i\}_{i\in [n]} \stackrel{\text{i.i.d.}}{\sim} f(\cdot|\hatt)$}.
		\For{$i=1$ to $n$}
		\State{\textbf{Simulation}: For $j\in [r]$, simulate with inputs generated from $p(\cdot|\tsim_i)$ to obtain $Y_j(\tsim_i)$ and $A_j(\tsim_i)$.}
		\EndFor
		\For{$i=1$ to $\widetilde{n}$}
		\State{\textbf{Ratio Estimators}: Compute $\widehat{\eta}(\tboot_i)$ ($\widehat{\eta}_{kNN}(\tboot_i)$ or $\widehat{\eta}_{kLR}(\tboot_i)$) by pooling $\{Y_j(\tsim_i), A_j(\tsim_i)\}_{i\in[n], j\in[r]}$}.
		\EndFor
		\State Return~\eqref{eq:percentile.bootstrap.CI} as a CI for $\eta(\ttrue)$
	\end{algorithmic}
	\label{algo: qt-CI for kNN & kLR}
\end{algorithm}

We close this section by presenting Algorithm~\ref{algo: qt-CI for kNN & kLR} that summarizes the experiment design, simulation and computation of the ratio estimators and the bootstrap CI construction for the $k$NN and $k$LR ratio estimators.

\section{Theoretical performance of the ratio estimators}\label{sec:analysis}
In this section, we examine the asymptotic properties of the $k$NN and $k$LR estimators and the corresponding CIs constructed from them along with the sample-size requirements for $n$, $\widetilde{n}$, and $r$ with respect to $m$.

We begin with two assumptions that persist throughout our analyses.
\begin{assumption}\label{assump:theta-df}
	Given $\hatt$, the following statements hold:
	(i) $\widetilde{\Theta} \subset \Theta$;
	(ii) $f(\tsim|\hatt)$ is continuous and bounded in $\tsim\in \Theta$;
	(iii) $\pr(\lVert \tsim \rVert > t|\hatt) = \cO(t^{-\gamma})$ for some $\gamma>0$ as $t\to\infty$.
\end{assumption}
Assumption~\ref{assump:theta-df}(i) guarantees that for any $\tboot$, $\min_{j\in [n]}\lVert \tboot - \tsim_j \rVert \to 0$ almost surely as $n \to \infty$, which forms the fundamental theoretical basis of the $k$NN regression;~\ref{assump:theta-df}(ii) makes $f$ locally Lipschitz continuous; and~\ref{assump:theta-df}(iii) regulates the tail behavior of $\bt$ distribution and
enables us accommodate the case when $\Theta$ is unbounded.
For bounded $\Theta$,~\ref{assump:theta-df}(iii) holds trivially.

Assumption~\ref{assump:smoothness} stipulates differentiability of the moments of $Y$ and $A$ with respect to $\tsim$.
\begin{assumption}\label{assump:smoothness}
	Given $\hatt$, for any $\tsim\in \Theta$ and $t=1$ or $2$, $\E[Y^t|\tsim]f(\tsim|\hatt)$ and $\E[A^t|\tsim]f(\tsim|\hatt)$ are bounded and thrice differentiable in $\tsim$.
\end{assumption}
Assumption~\ref{assump:theta-df}(ii) and Assumption~\ref{assump:smoothness} together imply that the first two conditional moments of $Y$ and $A$ are bounded in the neighborhood of $\tsim$.
The following two-dimensional Taylor series expansion on $x_2/x_1$ is repeatedly referred to in our analyses: For $x_1 \neq 0$ and $x_1+\rmd x_1 \neq 0$,
\begin{equation}\label{eq:Taylor}
	\frac{x_2+\rmd x_2}{x_1+\rmd x_1} - \frac{x_2}{x_1}= - \frac{x_2 \rmd x_1}{x_1^2} + \frac{\rmd x_2}{x_1} - \frac{\rmd x_1 \rmd x_2}{x_1^2} + \frac{x_2}{x_1^3}(\rmd x_1)^2 + o((\rmd x_1)^2).
\end{equation}
For generic ratio estimator $\widehat{\eta}(\tboot) = {\widehat{Y}}/{\widehat{A}}$, define $d(Y) = \widehat{Y} - \E[Y|\tboot]$ and $d(A) = \widehat{A} - \E[A|\tboot]$.
From~\eqref{eq:Taylor},
\begin{equation}
	\widehat{\eta}(\tboot) - \eta(\tboot)
	= - \frac{\E[Y|\tboot] d(A) }{(\E[A|\tboot])^2} + \frac{d(Y)}{\E[A|\tboot]} - \frac{d(A) d(Y)}{(\E[A|\tboot])^2} + \frac{\E[Y|\tboot]}{(\E[A|\tboot])^3}(d(A))^2 + o((d(A))^2).\label{eq:bias-decomp}
\end{equation}
Hence, to derive the MSE for different choices of $\widehat\eta$, we can focus on the moments of $d(Y)$ and $d(A)$.

In Section~\ref{subsec:kNN}, we analyze the MSE of $\widehat{\eta}_{kNN}$, and establish a central limit theorem (CLT) for $\widehat{\eta}_{kNN}(\tboot)$ for any fixed $\tboot$.
Section~\ref{subsec:kLR} shows the bias and MSE of $\widehat{\eta}_{kLR}(\tboot)$ under a special case where the input distribution $p$ belongs in an exponential family.
Section~\ref{subsec:CIs} examines the asymptotic coverage probabilities of the  percentile bootstrap CIs constructed from the two ratio estimators' empirical quantiles.

\subsection{Analysis on the $k$NN ratio estimator}\label{subsec:kNN}
To analyze $\widehat{\eta}_{kNN}(\tboot)$, we utilize some fundamental results of the $k$NN regression summarized below.

Let $R_{n, k+1}\triangleq\lVert \tboot - \tsim_{(k+1)}\rVert$ be the $(k+1)$th NN's distance from $\tboot$ and $V_d$ be the volume of the unit ball in $\R^d$.
Then, $\widehat{f}_{n,k}(\tboot|\hatt) \triangleq {k}/({n V_d R_{n, k+1}^d})$ is the $k$NN density estimator of $f(\tboot|\hatt)$~\citep{loftsgaarden1965nonparametric}.
Although  $\widehat{f}_{n,k}(\tboot|\hatt)$ does not appear in the definition of the $k$NN ratio estimator~\eqref{def:kNN}, its asymptotic properties are utilized in our analyses.
For example, Assumption~\ref{assump:n and k} describes a relationship between $n$ and $k$ that ensures $\widehat{f}_{n,k}(\tboot|\hatt)$ to converge and is useful in our analyses.
\begin{assumption}\label{assump:n and k}
    As $n\to\infty$, we set (i) $k\to\infty$ and $\frac{k}{n}\to 0$;
    and (ii) $\frac{\log(n)}{k} \to 0$.
\end{assumption}
Under Assumption~\ref{assump:theta-df}(ii) and~\ref{assump:n and k}(i), $\widehat{f}_{n,k}(\tboot|\hatt) \to{f(\tboot|\hatt)}$ in probability~\citep{loftsgaarden1965nonparametric}.
If Assumption~\ref{assump:n and k}(ii) also holds and $f(\cdot|\hatt)$ is uniformly continuous on $\R^d$, then $\widehat{f}_{n,k}(\tboot|\hatt)$ is strongly uniformly consistent to $f(\tboot|\hatt)$~\citep{devroye1977strong}.

To analyze the properties of $\widehat{Y}_{kNN}(\tboot)$, we consider the following scaled estimator:
\begin{equation}\label{eq:Y_kNN-scaled}
	\widehat{Y}_{kNN}^s (\tboot)\triangleq \frac{\widehat{f}_{n,k}(\tboot|\hatt)}{f(\tboot|\hatt)} \widehat{Y}_{kNN} (\tboot)
	= \frac{ \sum_{i=1}^k \bar{Y} (\tsim_{(i)})}{n V_d R_{n,k+1}^d f(\tboot|\hatt)},
\end{equation}
Similarly, we define $\widehat{A}_{kNN}^s (\tboot)\triangleq\frac{\widehat{f}_{n,k}(\tboot|\hatt)}{f(\tboot|\hatt)} \widehat{A}_{kNN} (\tboot)$.
While the scaling does not change the $k$NN ratio estimator, i.e., $\frac{\widehat{Y}_{kNN}^s (\tboot)}{\widehat{A}_{kNN}^s (\tboot)}=\frac{\widehat{Y}_{kNN} (\tboot)}{\widehat{A}_{kNN}(\tboot)} = \widehat{\eta}_{kNN}(\tboot)$, it is helpful in the subsequent analysis.
Specifically, we show that $\widehat{Y}_{kNN}^s$ has smaller asymptotic bias than $\widehat{Y}_{kNN}$ and derive a tighter bound for the MSE of $\widehat{\eta}_{kNN}$.

Let $\psi(\boldsymbol{x}):\R^d \to \R$ be an arbitrary twice differentiable function and $\Delta_{\boldsymbol{x}}$ be the Laplace operator with respect to $\boldsymbol{x}$, i.e.,
$
	\Delta_{\boldsymbol{x}} \psi(\boldsymbol{x})\triangleq \sum_{i=1}^d \frac{\partial^2}{\partial x_i^2} \psi(\boldsymbol{x})
$.
Lemma~\ref{lem: knn bias and var} presents convergence results for the biases and variances of $\widehat{Y}_{kNN}^s$ and $\widehat{A}_{kNN}^s$.
\begin{lemma}\label{lem: knn bias and var}
	Suppose Assumptions~\ref{assump:theta-df},~\ref{assump:smoothness} and~\ref{assump:n and k} hold.
	Then, conditional on $\hatt$, for any $\tboot \in \widetilde{\Theta}$,
	\small
	\begin{align}\label{eq:lemma1eq1}
		 & \E [\widehat{Y}_{kNN}^s (\tboot)|\tboot, \hatt] - \E[Y|\tboot ]
		= \frac{\Delta_{\bt}(\E [Y|\tboot] f (\tboot|\hatt) ) }{2(d+2) V_d^{2/d} ( f(\tboot|\hatt) )^{1+2/d}} \left(\tfrac{k}{n}\right)^{{2}/{d}} + o\left(\left(\tfrac{k}{n}\right)^{{2}/{d}} \right), \mbox{ and } \\
		 & \var[\widehat{Y}_{kNN}^s(\tboot)|\tboot, \hatt]
		\leq \frac{\var[Y|\tboot]}{rk} +\frac{2(\E[Y|\tboot])^2 }{k} + 
		\frac{( \Delta_{\bt} (\E[Y|\tboot]f (\tboot|\hatt) ) )^2}{2(d+2)^2 V_d^{4/d}(f (\tboot|\hatt) )^{2+4/d}} \left(\tfrac{k}{n}\right)^{{4}/{d}} + o\left(\left(\tfrac{k}{n}\right)^{{4}/{d}}\right) + o\left(\tfrac{1}{k}\right). \label{eq:lemma1eq2}
	\end{align} \normalsize
	Similar statements can be made for
	$
		\E[\widehat{A}_{kNN}^s (\tboot)|\tboot, \hatt]- \E[A|\tboot]
	$
	and
	$\var[\widehat{A}_{kNN}^s (\tboot)|\tboot, \hatt]$, respectively.
\end{lemma}

The bias~\eqref{eq:lemma1eq1} is derived in Proposition 3 in~\cite{mack1981local} and here we rewrite it in our notation; the variance~\eqref{eq:lemma1eq2} is new.
As expected, the bias of $\widehat{Y}_{kNN}^s (\tboot)$ decreases in $n$ and increases in $k$; the more candidate $\bt$s we can pool from, the smaller the bias is; and the more $\bt$s we pool from, the larger the bias is.
The variance~\eqref{eq:lemma1eq2} has a more complex relationship with $k$---as $k$ increases, we pool more neighbors, which has a downward effect on the variance (reflected in the first two terms), but also an upward effect (the third term) from pooling a more diverse set of $\tsim$s. Both bias and variance degrade as the dimension, $d$, of the parameter space increases, which is a well-known drawback of the $k$NN regression.

Notice that $f(\tboot|\hatt)$ is in the denominator of the dominant term in~\eqref{eq:lemma1eq1}.
Thus, all else equal, the smaller $f(\tboot|\hatt)$ is, the more biased $\widehat{Y}^s_{kNN}$ is.
Intuitively, small $f(\tboot|\hatt)$ implies that the chance of sampling $\tsim\sim f(\cdot|\hatt)$ near $\tboot$ is low and the $k$NNs of $\tboot$ among the simulation parameter set is likely far away from $\tboot$.
A similar observation can be made for the third term of~\eqref{eq:lemma1eq2}.
Our experiment design allows flexibility in choosing $f\neq \tilde{f}$ to reduce finite-sample bias and variance as discussed in Section~\ref{sec:para selection}.

Proposition~\ref{prop:kNN-MSE} derives the bias and the MSE of $\widehat{\eta}_{kNN}(\tboot)$ from~\eqref{eq:bias-decomp}.
\begin{proposition}\label{prop:kNN-MSE}
	Suppose Assumptions~\ref{assump:theta-df},~\ref{assump:smoothness} and~\ref{assump:n and k} hold.
	Then, conditional on $\hatt$, for any $\tboot \in \widetilde{\Theta}$,
	$ \E[\widehat{\eta}_{kNN}(\tboot)|\tboot,\hatt] -\eta(\tboot)
		=\cO((\tfrac{k}{n})^{2/d}) + \cO(\tfrac{1}{k})$ and
	$\E[(\widehat{\eta}_{kNN}(\tboot) - \eta(\tboot))^2|\tboot,\hatt]
		= \cO((\tfrac{k}{n})^{4/d}) +\cO(\tfrac{1}{k})$.
\end{proposition}
As expected, the limitation of the $k$NN method due to its dependence on dimension carries over to the $k$NN ratio estimator's error. A similar observation can be made for Theorem~\ref{thm:CLT} that establishes a CLT for $\widehat{\eta}_{kNN}(\tboot)$, where $\overset{\mathcal{D}}{\to}$ denotes convergence in distribution.

\begin{theorem}\label{thm:CLT}
	Suppose Assumptions~\ref{assump:theta-df},~\ref{assump:smoothness} and~\ref{assump:n and k} hold.
	Additionally, for any $\tboot \in \widetilde{\Theta}$, suppose that $\E[\lvert \bar{Y}(\tboot) - \eta(\tboot) \bar{A}(\tboot) \rvert^3|\tboot,\hatt ]<\infty$
	and $\var[ \bar{Y}(\tboot) - \eta(\tboot) \bar{A}(\tboot)|\tboot,\hatt ]>0$.
	Let $k=o\left(n^{\frac{2}{2+d}}\right)$ and $r$ be a constant,
	then, conditional on both $\hatt$ and $\tboot$,
	$\sqrt{kr}\left(\widehat{\eta}_{kNN}(\tboot) - \eta(\tboot)\right) \overset{\mathcal{D}}{\to} \mathcal{N}\left( 0,\frac{ \var[ Y(\tboot) - \eta(\tboot) {A(\tboot)}|\tboot, \hatt ]}{V_d (\E[A|\tboot])^2} \right)$.
\end{theorem}
The scaling factor, $\sqrt{kr}$, in the CLT reflects that $\widehat{\eta}_{kNN}(\tboot)$ utilizes $kr$ simulation runs by pooling the $k$NNs of $\tboot$.
Indeed, $\var[ Y(\tboot) - \eta(\tboot) {A(\tboot)}|\tboot, \hatt ]/(\E[A|\tboot])^2$ is the asymptotic variance of the standard ratio estimator for a regenerative process scaled by $\sqrt{r}$ (cf.~Theorem~3 in~\citeauthor{hendersonglynn2001},~\citeyear{hendersonglynn2001}).
However, since none of the $kr$ runs are made at $\tboot$, there is a penalty for pooling in $\mathbb{R}^d$.
The volume of the unit ball in  $\mathbb{R}^d$, $V_d$, in the denominator of the variance term reflects the ``curse of dimensionality'' of the $k$NN regression: $V_d\to 0$ as $d\to \infty$. Moreover,  $k$ grows more slowly in $n$ for larger $d$.

\subsection{Analysis on the $k$LR ratio estimator}\label{subsec:kLR}
In this section, we study the statistical properties of the $k$LR estimator.
We focus on the case when the input model is a member of the exponential family, namely, the joint likelihood of $\Z_j(\tsim)$ can be written in canonical form $\prod_{\ell=1}^{S(\tsim)}p(Z_{j,\ell}|\tsim) = p_b(\Z_j)\exp(\tsim^\top U(\Z_j) - L(\tsim))$, where $p_b, U$ and $L$, are the base measure, sufficient statistic, and log-partition function, respectively.
We make the following assumptions to analyze the estimation error of the $k$LR estimator; $\text{Int}(\Theta)$ denotes the interior of $\Theta$.
\begin{assumption}\label{assump:exp_family}
	Let $\prod_{\ell=1}^{S(\tsim)}p(Z_{j,\ell}|\tsim)$ belong to an exponential family and is in the canonical form.
	There exist positive constants $c_1,c_2$ such that for any $\tboot\in \text{Int}(\Theta)$, (i) $\exp(L(\tboot)) < c_1 <\infty$; (ii) one can find neighborhood $N(\tboot)$ such that for all $\tsim\in N(\tboot)$, $\E[U(\Z)|\tsim]<c_2<\infty$; $\E[Y^2|\tsim]$ and $\E[A^2|\tsim]$ are bounded and thrice differentiable in $\tsim$.
\end{assumption}
Assumption~\ref{assump:exp_family}(i) assures that the joint likelihood is well defined for all $\tboot \in \widetilde{\Theta}$. In Assumption~\ref{assump:exp_family}(ii), by assuming the existence of $N({\tboot})$, we guarantee that for sufficiently large $n$ and $k$, all $k$NNs of $\tboot$ are found in $N({\tboot})$. Then, the moment conditions on $U(\Z), Y^2$, and $A^2$ guarantee that the $k$LR estimator constructed from the $k$NNs within $N(\tboot)$ has a bounded second moment.  

In the following, we analyze the convergence rates of the variances of $\widehat{Y}_{kLR}(\tboot)$ and $\widehat{A}_{kLR}(\tboot)$ given $\tboot$. Recall that both are unbiased estimators of $\E[Y|\tboot]$ and $\E[A|\tboot]$, respectively.

\begin{lemma}\label{lem:Y-kLR-MSE}
	Suppose Assumption~\ref{assump:theta-df},~\ref{assump:n and k} and~\ref{assump:exp_family} hold. Then,
	$
		\E[ ( \widehat{Y}_{kLR}(\tboot) - \E[Y|\tboot] )^2| \tboot,\hatt]
		= \frac{1}{rk}\var[Y|\tboot] + o(\frac{1}{rk}).
	$
	A similar statement can be made for $\widehat{A}_{kLR}(\tboot)$.
\end{lemma}
Plugging in the variance expressions in Lemma~\ref{lem:Y-kLR-MSE} into~\eqref{eq:bias-decomp}, we can establish the bias and MSE of $\widehat{\eta}_{kLR}$.
\begin{proposition}\label{prop:kLR-MSE}
	Suppose Assumption~\ref{assump:theta-df},~\ref{assump:n and k} and~\ref{assump:exp_family} hold. Then,
	$\E[\widehat{\eta}_{kLR}(\tboot)|\tboot,\hatt] -\eta(\tboot)
		= \cO\left(\frac{1}{rk}\right),
	$ and $\E[ ( \widehat{\eta}_{kLR}(\tboot) - \eta(\tboot) )^2| \tboot,\hatt]
		= \cO\left(\frac{1}{rk}\right)$.
\end{proposition}

Comparing Propositions~\ref{prop:kNN-MSE} and~\ref{prop:kLR-MSE}, we observe that incorporating the LR method effectively eliminates the dependence on $d$ in the MSE of the $k$NN estimator.
Meanwhile, the bias and MSE of the $k$LR ratio estimator are both of order $\cO(\frac{1}{rk})$, which matches the order of the standard ratio estimator obtained from $rk$ simulation runs at $\tboot$.
Although the constants in the leading terms for the bias and MSE of the $k$LR estimator may be larger, they remain bounded.
This highlights the advantage of combining $k$NN regression with the LR method---the performance degradation in a higher dimension for the $k$NN estimator is eliminated by the LR method, while the unbounded variance issue typically associated with the LR estimator is regulated by pooling only the $k$NNs of $\tboot$.

\subsection{Asymptotic coverage of the bootstrap confidence intervals}\label{subsec:CIs}
In this subsection, we first show that the empirical $\alpha$-quantile of the $k$NN and $k$LR ratio estimators converge to $q_\alpha(\eta(\tboot)|\hatt)$ with appropriately chosen values of $k, r, n$, and $\widetilde{n}$ for given $\hatt$.
Then, we proceed to show that for both methods, the percentile bootstrap CI constructed by Algorithm~\ref{algo: qt-CI for kNN & kLR} provides the $1-\alpha$ probability guarantee of covering $\eta(\ttrue)$ as the input data size, $m$, increases to infinity.

Starting with $\widehat{\eta}_{kNN}$, we first establish some notation.
Let $\Phi(x|\hatt) \triangleq \pr(\eta(\tboot)\leq x|\hatt)$ and $\phi(x|\hatt)$ be the conditional cdf and pdf of $\eta(\tboot)$, respectively, given $\hatt$.
The ecdf of $\Phi$ estimated from $\{\widehat{\eta}_{kNN} (\tboot_i)\}_{i\in\widetilde{n}}$ is $\Phi_{\widetilde{n},r}(x) = \frac{1}{\widetilde{n}}\sum_{i=1}^{\widetilde{n}} \mathbf{1}\{\widehat{\eta}_{kNN} (\tboot_i)\leq x\}$. Because $\widehat{\eta}_{kNN}(\tboot_1),\ldots,\widehat{\eta}_{kNN}(\tboot_{\widetilde{n}})$ are correlated,  $\Phi_{\widetilde{n},r}$ is constructed from non-i.i.d.\ observations.
Nevertheless, we show that $\Phi_{\widetilde{n},r}$ is a uniformly consistent estimator of $\Phi$ under the following assumptions.


\begin{assumption}\label{assump: ecdf}
	Define $\varepsilon_i$ to be the scaled error of the estimator of $\eta(\tboot_i)$.
	For the $k$NN and $k$LR estimators, let $\varepsilon_i = \sqrt{r k}(\widehat{\eta}_{kNN} (\tboot_i) - \eta(\tboot_i))$ and $\varepsilon_i = \sqrt{r k}(\widehat{\eta}_{kLR} (\tboot_i) - \eta(\tboot_i))$, respectively.
	For the standard estimator, let $\varepsilon_i = \sqrt{r}(\widehat{\eta}_{std} (\tboot_i) - \eta(\tboot_i))$.
	Moreover, let $h_i(\eta,\varepsilon|\hatt)$ and and $h_{i,j}(\eta_i,\eta_j,\varepsilon_i,\varepsilon_j|\hatt)$ be the conditional joint pdfs of $(\eta(\tboot_i), \varepsilon_i)$ and $(\eta(\tboot_i), \eta(\tboot_j),\varepsilon_i,\varepsilon_j)$, respectively.
	The following conditions hold:
	\begin{enumerate}[leftmargin=*]
		\item[(i)] $\phi(x|\hatt)$ is continuous; and both $\phi(x|\hatt)$ and $f(\cdot|\hatt)$ are bounded away from zero in $\Theta$.

		\item[(ii)] For any $\tboot\in\widetilde{\Theta}$ and any $i\in [\widetilde{n}]$, $h_i(\eta,\varepsilon|\hatt)$ is differentiable with respect to $\eta$.
		      There exist $p_{0,n,r}(\varepsilon)>0$ and $p_{1,n,r}(\varepsilon)>0$ such that
		      $h_i(\eta,\varepsilon|\hatt)\leq p_{0,n,r}(\varepsilon) \mbox{ and } \left\lvert\frac{\partial}{\partial \eta}h_i(\eta,\varepsilon| \hatt) \right\rvert \leq p_{1,n,r}(\varepsilon)$.
		      Moreover, $\sup_n\sup_{r}\int_{-\infty}^\infty \lvert \varepsilon\rvert^q p_{l,n,r}(\varepsilon)\rmd \varepsilon<\infty$ for $l=0,1$ and $0\leq q\leq 2$.

		\item[(iii)] For any $\tboot\in\widetilde{\Theta}$ and any $i,j\in [\widetilde{n}]$ with $i\neq j$, $h_{i,j}(\eta_i,\eta_j,\varepsilon_i,\varepsilon_j| \hatt)$ is differentiable with respect to both $\eta_i$ and $\eta_j$.
		      There exist $p_{0,n,r}(\varepsilon_i,\varepsilon_j)>0$ and $p_{1,n,r}(\varepsilon_i,\varepsilon_j)>0$ such that
		      $h_{i,j}(\eta_i, \eta_j,\varepsilon_i,\varepsilon_j|\hatt)\leq p_{0,n,r}(\varepsilon_i,\varepsilon_j)$ and $\max\left\{\left\lvert\frac{\partial}{\partial \eta_i}h_{i,j}(\eta_i,\eta_j,\varepsilon_i,\varepsilon_j|\hatt) \right\rvert, \left\lvert\frac{\partial}{\partial \eta_j}h_{i,j}(\eta_i,\eta_j,\varepsilon_i,\varepsilon_j|\hatt) \right\rvert \right\} \leq p_{1,n,r}(\varepsilon_i,\varepsilon_j)$.
		      Moreover, $\sup_n\sup_{r}\int_{-\infty}^\infty \int_{-\infty}^\infty \lvert \varepsilon_i\rvert^{q_i}\lvert \varepsilon_j\rvert^{q_j} p_{l,n,r}(\varepsilon)\rmd \varepsilon<\infty$ for $l=0,1$ and $0\leq q_i,q_j\leq 2$ with $q_i+q_j\leq 3$.
	\end{enumerate}
\end{assumption}
Assumption~\ref{assump: ecdf}(i) imposes smoothness and regularity conditions for $\phi$, as well as a regularity condition for $f(\cdot|\hatt)$.
Assuming $f(\cdot|\hatt)$ is bounded away from zero makes $\Theta$ a bounded set and Proposition~\ref{prop:kNN-MSE} holds uniformly for all $\tboot\in \widetilde{\Theta}$.
It is not a stringent condition because the user can choose $f(\cdot|\hatt)$ that satisfy it.
Parts~(ii) and (iii) state integrability conditions on the scaled estimation errors that are commonly adopted in the LR literature; they are similar to Assumption~1 in~\citet{gordy2010nested} and identical to Assumption~3 in~\cite{fengsong2024}.

Below, Lemma~\ref{lem: knn cdf convergence} derives the bias and variance of ecdf $\Phi_{{\widetilde{n}},r}(x)$ at each $x\in\R$ and establishes a Glivenko-Cantelli-type uniform weak consistency for $\Phi_{\widetilde{n},r}$.
Proposition~\ref{prop: knn qt conergence} states a weak consistency result for the $k$NN empirical quantile estimator, $q_{\alpha,\widetilde{n}}(\{\widehat{\eta}_{kNN}(\tboot_i)\}|\hatt)$, where the dependence on $\hatt$ is explicitly denoted.

\begin{lemma}\label{lem: knn cdf convergence}
	Suppose Assumptions~\ref{assump:theta-df},~\ref{assump:smoothness},~\ref{assump:n and k} and~\ref{assump: ecdf} hold.
	Then, $\E[\Phi_{{\widetilde{n}},r}(x)|\hatt] = \Phi(x) + \cO((\tfrac{k}{n})^{{2}/{d}}) + \cO(\tfrac{1}{k})$
	and
	$\var[\Phi_{\widetilde{n},r}(x)|\hatt] \leq \cO(\tfrac{1}{\widetilde{n}}) + \cO((\tfrac{k}{n})^{{4}/{d}} ) + \cO(\tfrac{1}{k})$.
	Furthermore, conditional on $\hatt$,
	$\sup\nolimits_{x\in \R}\lvert \Phi_{\widetilde{n},r}(x) - \Phi(x)\rvert
		= \cO_{\pr}(\tfrac{1}{\sqrt{\widetilde{n}}}) + \cO_{\pr}((\tfrac{k}{n})^{{2}/{d}} ) + \cO_{\pr}(\tfrac{1}{\sqrt{k}})$.
\end{lemma}

\begin{proposition}\label{prop: knn qt conergence}
	Suppose Assumptions~\ref{assump:theta-df},~\ref{assump:smoothness},~\ref{assump:n and k} and~\ref{assump: ecdf} hold.
	Then, conditional on $\hatt$, we have
	$\lvert q_{\alpha,\widetilde{n}}(\{\widehat{\eta}_{kNN}(\tboot_i)\}|\hatt) - q_\alpha(\eta(\tboot)|\hatt) \rvert
		= \cO_{\pr}(\tfrac{1}{\sqrt{\widetilde{n}}}) + \cO_{\pr}((\tfrac{k}{n})^{{2}/{d}} ) + \cO_{\pr}(\tfrac{1}{\sqrt{k}})$
	for $0<\alpha<1$.
\end{proposition}

Recall that the user may choose $n \neq \widetilde n$.
Since the total number of simulation runs is $nr$ for the $k$NN method, $\widetilde n$ can be chosen sufficiently large without inflating $nr$ so that the estimation error of $ q_{\alpha,\widetilde{n}}(\{\widehat{\eta}_{kNN}(\tboot_i)\}|\hatt)$ is not dominated by $\widetilde{n}$.
The convergence rate in Proposition~\ref{prop: knn qt conergence} can be maximized by choosing $n\sim k^{\frac{4}{d}+1}$.
Observe that the convergence rate degrades as $d$ increases.

As discussed in Section~\ref{sec:problem statement}, the percentile bootstrap CI~\eqref{eq:percentile.bootstrap.CI} has an empirical advantage over the basic bootstrap CI~\eqref{eq:basic.bootstrap.CI}, but requires an additional assumption to ensure its consistency:
\begin{assumption}\label{assump:mono-trans}
	There exists a strictly monotonic transformation $T$ such that
	$
		T( \eta(\hatt) - \eta(\ttrue))
	$
	and
	$T(\eta(\tboot) - \eta(\hatt))$ given $\hatt$ are symmetric around $0$ for sufficiently large $m$.
\end{assumption}
Note that Assumption~\ref{assump:mono-trans} only requires $T$ to exist, not known. Below is the first main result of this section.

\begin{theorem}\label{thm:qt-convergence-knn}
	Suppose Assumptions~\ref{assump:theta-df},~\ref{assump:smoothness},~\ref{assump:n and k},~\ref{assump: ecdf} and~\ref{assump:mono-trans} hold.
	Then, for $0< \alpha<1$, by choosing $k=\omega(m)$, $n=\omega(m^{d/4+1}), \widetilde{n}=\omega(m)$, and constant $r>0$, we have
	\begin{equation}
		\label{eq:boots-CI}
		\pr\left(\eta(\ttrue) \in
			[
				q_{\alpha/2,\widetilde{n}}(\{\widehat{\eta}_{kNN}(\tboot_i)\}|\hatt), q_{1-\alpha/2,\widetilde{n}}(\{\widehat{\eta}_{kNN}(\tboot_i)\} |\hatt)
			] \right)
		= 1-\alpha+o(1).
	\end{equation}
\end{theorem}
In addition to the conditions for $k$ and $n$ in Theorem~\ref{thm:qt-convergence-knn}, reall that Assumption~\ref{assump:n and k} prescribes ${k}/{n}\to 0$ and ${\log(n)}/{k} \to 0$.
In summary, $n$ and $\widetilde{n}$ are required to grow at slightly faster rates than input data size $m$, while $n$ and $k$ need to grow at suitable rates to ensure the effectiveness of the $k$NN regression.

Comparing Theorem~\ref{thm:qt-convergence-knn} to Proposition~\ref{prop:standard.est.convergence},
the $k$NN estimator-based CI is asymptotically more efficient than the standard ratio estimator-based CI when $d=1$, comparable for $d=2$, and less efficient for $d> 2$.
Recall that Algorithm~\ref{algo:parametric_bootstrapping} does not require a simulation parameter set, thus Proposition~\ref{prop:standard.est.convergence} only controls $\tilde n$ and $r$ for the standard ratio estimator, which prescribes the total simulation budget to grow in $\omega(m^{3/2})$ regardless of $d$.
On the other hand, the total simulation budget of Algorithm~\ref{algo: qt-CI for kNN & kLR} is $nr$, which grows in $\omega(m^{d/4 +1})$.

Next, we show that the $k$LR estimator-based CI is more efficient than the standard ratio estimator-based CI for regardless of dimension $d$. We start by establishing the weak uniform consistency of the ecdf $\widetilde \Phi_{\widetilde{n},r}(x) \triangleq \frac{1}{\widetilde{n}}\sum_{i=1}^{\widetilde{n}} \mathbf{1}\{\widehat{\eta}_{kLR} (\tboot_i)\leq x\}$ and the weak consistency of the empirical quantile estimator constructed from $\{\widehat{\eta}_{kLR} (\tboot_i)\}_{i\in [\widetilde{n}]}$ in Lemma~\ref{lem: klr cdf convergence} and Proposition~\ref{prop: klr qt conergence}, respectively.

\begin{lemma}\label{lem: klr cdf convergence}
	Suppose Assumptions~\ref{assump:theta-df},~\ref{assump:smoothness},~\ref{assump:n and k},~\ref{assump:exp_family} and~\ref{assump: ecdf} hold.
	Then, $\E[\widetilde \Phi_{{\widetilde{n}},r}(x)|\hatt] = \Phi(x) + \cO(\tfrac{1}{r k})$
	and
	$\var[\widetilde \Phi_{\widetilde{n},r}(x)|\hatt] \leq \cO\left(\frac{1}{\widetilde{n}}\right) +  \cO\left(\tfrac{1}{r k}\right)$.
	Further, conditional on $\hatt$,
	$ \sup\nolimits_{x\in \R}\lvert \widetilde \Phi_{\widetilde{n},r}(x) - \Phi(x)\rvert
		= \cO_{\pr}(\tfrac{1}{\sqrt{\widetilde{n}}}) + \cO_{\pr}(\tfrac{1}{\sqrt{r k}})$.
\end{lemma}

\begin{proposition}\label{prop: klr qt conergence}
	Suppose Assumptions~\ref{assump:theta-df},~\ref{assump:smoothness},~\ref{assump:n and k},~\ref{assump:exp_family} and~\ref{assump: ecdf} hold.
	Then, conditional on $\hatt$, we have
	$\lvert q_{\alpha,\widetilde{n}}(\{\widehat{\eta}_{kLR}(\tboot_i)\}|\hatt) - q_\alpha(\eta(\tboot)|\hatt) \rvert
		= \cO_{\pr}(\tfrac{1}{\sqrt{\widetilde{n}}}) + \cO_{\pr}(\tfrac{1}{\sqrt{r k}})$
	for $0<\alpha<1$.
\end{proposition}

The following theorem establishes the consistency of the  CI constructed from the $k$LR estimators.
\begin{theorem}\label{thm:qt-convergence-kLR}
	Suppose Assumptions~\ref{assump:theta-df},~\ref{assump:smoothness},~\ref{assump:n and k},~\ref{assump:exp_family},~\ref{assump: ecdf} and~\ref{assump:mono-trans} hold.
	Then, by choosing $k = \omega(m)$ and $\widetilde{n} = \omega(m)$, and constant $r>0$, we have
	$
		\pr\left(\eta(\ttrue) \in
			[
				q_{\alpha/2,\widetilde{n}}(\{\widehat{\eta}_{kLR}(\tboot_i)\}|\hatt), q_{1-\alpha/2,\widetilde{n}}(\{\widehat{\eta}_{kLR}(\tboot_i)\} |\hatt)
			]\right)
		= 1-\alpha+o(1)
	$  for $0< \alpha<1$.
\end{theorem}
In Theorem~\ref{thm:qt-convergence-kLR}, the required $\widetilde{n}$ does not grow in $d$.
Moreover, the total simulation budget of the $k$LR scheme is in $\omega(m)$, which is smaller than $\omega(m^{3/2})$ of the standard estimator regardless of $d$. This confirms that combining the $k$NN and LR methods indeed allow us to significantly improve the simulation efficiency of IUQ.


\section{Empirical Guidance on the Experiment Design}\label{sec:para selection}
Theorems~\ref{thm:qt-convergence-knn} and~\ref{thm:qt-convergence-kLR} in Section~\ref{sec:analysis} provide insight into how fast $n$ and $k$ should grow relative to $m$ to achieve the fastest asymptotic convergence rate as $m$ increases while $r$ remains constant.
However, for experiments with a finite simulation budget, further guidance is needed to determine $r$ and $k$ at each simulation parameter.
In this section, we describe heuristic procedures to select $r$ and $k$ via analysis of variance (ANOVA) and cross-validation (CV), respectively.
In addition, we propose two candidates for the simulation parameter sampling distribution, $f(\cdot|\hatt)$, whose performances are compared numerically in Section~\ref{sec:numerical}.

Although the asymptotic analysis prescribes a constant $r$, small $r$ may lead to large simulation error in the ratio estimator.
If the error is overwhelmingly large compared to IU, then the CI constructed from the ratio estimators may be too wide to be informative.
Meanwhile, since the total simulation budget of our experiment design is $nr$, we do not wish to choose $r$ larger than needed.

To address this, we  select $r$ by conducting the ANOVA experiment proposed by~\cite{ankenman2012quick} as a pilot experiment prior to running the main IUQ procedure.
To measure the IU variance relative to the stochastic error variance in $\bar{Y}(\tsim)$, they first decompose $\var[\bar{Y}(\tsim)|\hatt] = \E[\var[Y|\tsim]|\hatt]/r + \var[\E[Y|\tsim]|\hatt]$.
Under the homoscedasticity assumption such that $\var[Y|\tsim,\hatt]$ is constant for all $\tsim$, they propose the ratio, $\zeta_{Y}(r) = \nu_Y^2/(\sigma_Y^2/r)$, as a measure, where $\sigma_Y^2 \triangleq \var[Y|\tsim,\hatt]$ for all $\tsim\in\Theta$ and $\nu_{Y}^2 \triangleq\var[\E[Y|\tsim]|\hatt]$ given $\hatt$.
Small $\zeta_{Y}(r)$ implies that the simulation error in $\bar{Y}(\tsim)$ is too large for the IUQ to be meaningful, so $r$ should be increased.
The ANOVA adopted in~\cite{ankenman2012quick} samples $b$ bootstrap parameters,  $\{\tboot_i^\ast\}_{i\in [b]} \sim \widetilde{f}(\cdot|\hatt)$, and makes $s$ simulation runs at each $\tboot_i^\ast$ to generate $\{Y_j(\tboot_i^\ast)\}_{j\in [s], i\in[b]}$.
From these outputs, $\bar{Y}_s(\tboot_i^\ast) \triangleq \frac{1}{s}\sum_{j=1}^s Y_j(\tboot_i^\ast), i\in[b]$ and $\bar{\bar{Y}}_{bs} \triangleq \frac{1}{b}\sum_{i=1}^b\bar{Y}_s(\tboot_i^\ast)$ can be computed to estimate ${\zeta}_Y(r)$ with
\begin{equation} \label{eq:anova}
	\widehat{\zeta}_Y(r) = \frac{r}{s}\left(\frac{b(s-1)-2}{b(s-1)}\frac{\text{MSE}(Y)}{\text{MST}(Y)} - 1\right),
\end{equation}
where $\text{MSE}(Y) \triangleq \frac{\sum\nolimits_{i=1}^b\sum\nolimits_{j=1}^s (Y_j(\tboot_i^\ast) - \bar{Y}_s(\tboot_j^\ast))^2}{b(s-1)},\text{MST}(Y) \triangleq \frac{s\sum\nolimits_{i=1}^b (\bar{Y}_s(\tboot_i^\ast) - \bar{\bar{Y}}_{bs})^2}{b-1}$.

A similar heuristic can be applied to our problem by defining $\sigma_A^2, \nu_A^2$, and $\zeta_A(r)$  as well as MSE$(A)$ and MST$(A)$ for $A$.
Algorithm~\ref{algo:parameter selection} outlines the process of selecting $r$ by estimating $\zeta_Y(r)$ and $\zeta_A(r)$. 
Notice that we increment $s$ until the estimated $\widehat{\zeta}_Y(s)$ and $\widehat{\zeta}_A(s)$ are positive. 
Although $\zeta_Y(r)$  is strictly positive, its estimate can be negative if $\nu_Y^2/\sigma_Y^2$ is small. 
In this case, increasing $s$ helps, which makes the estimate  more precise. 
Once both $\widehat{\zeta}_Y(s)$ and $\widehat{\zeta}_A(s)$ are positive, then Algorithm~\ref{algo:parameter selection} returns $r$ that makes both $\widehat{\zeta}_Y(r)$ and $\widehat{\zeta}_A(r)$  sufficiently large, i.e., over threshold $c_\zeta>0$.
In our numerical experiments, we adopt $c_\zeta = 0.1$ implying that $r$ should be large enough to make IU variance  at least 10\% of
the simulation error variance.

Controlling the stochastic error variance of $\bar{Y}(\tboot_i)$ and $\bar{A}(\tboot_i)$ does not  directly translate to controlling the stochastic error variance in the ratio estimators. Nevertheless, the choice of $r$ returned by Algorithm~\ref{algo:parameter selection} proves to be effective in the numerical studies in Section~\ref{sec:numerical}.

\begin{algorithm}[tbp]
	\caption{Selection of $r$ via Analysis of Variance}
	\begin{algorithmic} \footnotesize
		\State \textbf{Inputs:} bootstrap sampling dist $\widetilde{f}(\cdot|\hatt)$, bootstrap size $b$, simulation run size $s$, incremental run size $\Delta s$ and threshold $c_\zeta$.
		\State Sample $\tboot^\ast_i \sim \widetilde{f}(\cdot|\hatt)$ and simulate $s$ runs at each $\tboot^\ast_i$ for all $i\in [b]$ to obtain $\{Y_j(\tboot_i^\ast), A_j(\tboot_i^\ast)\}_{j\in [s], i\in[b]}$
		\State Compute $\widehat{\zeta}_Y(s)$ and $\widehat{\zeta}_A(s)$ according to~\eqref{eq:anova}.
		\While{$\widehat{\zeta}_Y(s) <=0$ or $\widehat{\zeta}_A(s) <=0$}
		\State Make $\Delta s$ simulation runs at each $\tboot^\ast_i$ for all $i\in [b]$; let $s = s+\Delta s$; and recompute $\widehat{\zeta}_Y(s)$ and $\widehat{\zeta}_A(s)$.
		\EndWhile
		\State Output the largest integer $r$ such that $\min\{\widehat{\zeta}_Y(r),\widehat{\zeta}_A(r)\}\geq c_\zeta$.
	\end{algorithmic}
	\label{algo:parameter selection}
\end{algorithm}


In Algorithm~\ref{algo: qt-CI for kNN & kLR}, after $r$ runs are made at each of $n$ simulation parameters, we proceed to decide the value of $k$ via $K$-fold CV as outlined in Algorithm~\ref{algo: cross validation}.
Although the CV does not require additional simulation, testing a large number of values for $k$ can be demanding.
Algorithm~\ref{algo: cross validation} restricts the candidate values of $k$ within set $\mathcal K$ so that the user can control $|\mathcal K|$. Another user input is the number of folds, $K$.
For simplicity of exposition, suppose $K$ is chosen so that $n= cK$ for some integer $c$, so we can partition the index set $[n]$ into $K$ sets, i.e., $\Lambda_p \triangleq \{(p-1)c + 1, \ldots, pc\}$ for $p\in[K]$.
Let $\bar{\Lambda}_p = [n]\setminus \Lambda_p$ be the complement set of the $p$-th partition.
The inner for loop of Algorithm~\ref{algo: cross validation} pivots each $\Lambda_p$ as a test set and computes  $\bar{Y} (\tsim_{i_p}) = \frac{1}{r}\sum_{j=1}^r Y_j(\tsim_{i_p})$ for each $\tsim_{i_p} \in \Lambda_p$.
In the absence of $\E[Y|\tsim_{i_p}]$, its unbiased estimate, $\bar{Y} (\tsim_{i_p})$, serves as a benchmark to compute the MSE for the $k$NN estimator at each $\tsim_{i_p}$: $\widehat{Y}_{kNN}(\tsim_{i_p}) = \frac{1}{k}\sum_{j_p=1}^{k}\bar{Y} \left(\tsim_{(j_p)}\right)$, where
$\tsim_{(j_p)}$ is the $j_p$th NN of $\tsim_{i_p}$ among the parameters in $\bar{\Lambda}_p$. Then, the average of the approximate MSE for all $\tsim_{i_p}\in\Lambda_p$, $D_{Y,p}(k) = \frac{1}{|\Lambda_p|}\sum_{i_p \in \Lambda_p} (\bar{Y} (\tsim_{i_p}) - \widehat{Y}_{kNN}(\tsim_{i_p}))^2$ is adopted as a loss function of the $p$th testing set.
Algorithm~\ref{algo: cross validation} returns $k\in\mathcal K$ that minimizes the average loss over all $K$ folds.


Since we perform $k$NN regressions for both $Y$ and $A$ for the $k$NN ratio estimators,
we can run a CV for $A$ to find the loss-minimizing $k_A$ that may differ from $k_Y$.
In the numerical experiments, we adopt $k_Y$ and $k_A$ for $Y$ and $A$ separately---this contrasts with the theoretical analyses in Section~\ref{sec:analysis} where we apply the same $k$ for both for simplicity.
However, the same conclusions hold asymptotically even if $k_Y\neq k_A$.
A similar CV procedure can be applied to find $k$ for the $k$LR estimator~\eqref{def:kLR} by replacing $\widehat{Y}_{kNN}(\tsim_{i_p})$ with $\widehat{Y}_{kLR}(\tsim_{i_p})$.

\begin{algorithm}[tb]
	\caption{Selection of $k$ for the $k$NN regression via cross validation}
	\begin{algorithmic} \footnotesize
		\State \textbf{Inputs:} $n$ simulation parameters with $r$ runs at each; set $\mathcal{K}$ of candidate values for $k$; number of partitions $K$.
		\For{$k\in\mathcal{K}$}
		\For{$p=1,\ldots, K$}
		\State Calculate $\bar{Y} (\tsim_{i_p}) = \frac{1}{r}\sum_{j=1}^r Y_j(\tsim_{i_p})$ and $\widehat{Y}_{kNN}(\tsim_{i_p})=\frac{1}{k}\sum_{j_p=1}^{k}\bar{Y} \left(\tsim_{(j_p)}\right)$ for all $i_p \in \Lambda_p$.
		\State Calculate $D_{Y,p}(k) = \frac{1}{|\Lambda_p|}\sum_{i_p \in \Lambda_p} (\bar{Y} (\tsim_{i_p}) - \widehat{Y}_{kNN}(\tsim_{i_p}))^2$.

		\EndFor
		\State Calculate $D_Y(k) = \frac{1}{K} \sum_{p=1}^K D_{Y,p}(k)$.
		\EndFor
		\State Return $k_Y = \arg\min\{D_Y(k): k \in \mathcal{K}\}$.
	\end{algorithmic}
	\label{algo: cross validation}
\end{algorithm}

Lastly, we consider two choices of simulation parameter sampling distribution $f(\cdot|\hatt)$.
The first is to simply adopt the bootstrap parameter sampling distribution, i.e., $f(\cdot|\hatt) = \widetilde{f}(\cdot|\hatt)$.
Even with this choice, the bootstrap and simulation parameter sets are generated independently in the empirical study in Section~\ref{sec:numerical}.
The second choice is inspired by Lemma~\ref{lem: knn bias and var}. Since $f(\tboot|\hatt)$ appears in the denominators of~\eqref{eq:lemma1eq1} and~\eqref{eq:lemma1eq2}, the bias and the variance bounds for the $k$NN-based estimator $\widehat{Y}_{kNN}^s (\tboot)$ are large when $f(\tboot|\hatt)$ is small, i.e., when some bootstrap parameters lie in a rare-event region of simulation parameter sampling distribution.
Motivated by this observation, we construct a uniform distribution $f(\cdot|\hatt)$ over the smallest ellipsoid containing the bootstrap parameter set $\{\tboot_i\}_{i\in [\widetilde{n}]}$.
Under this construction, $f(\cdot|\hatt) \neq \widetilde{f}(\cdot|\hatt)$ and the former depends on $\{\tboot_i\}_{i\in [\widetilde{n}]}$, but our asymptotic analysis remains valid.
The numerical results in Section~\ref{sec:numerical} show that the latter choice indeed results in better coverage probabilities for $\eta(\bt^c)$.

\section{Numerical Studies}\label{sec:numerical}
In this section, we compare the performances of the standard, $k$NN, and $k$LR ratio estimators with  three numerical examples: (1) 13-dimensional stochastic activity network (SAN); (2) $M/M/1/10$ queueing model; and (3) enterprise risk management (ERM) example.
The SAN example demonstrates the effectiveness of our ratio estimators in higher dimensions.
The queueing example shows how our proposed design can be applied for estimating the steady-state performance measures in regenerative simulation models. 
The ERM example illustrates their applicability in a more realistic simulation problem.

Two primary performance measures are the empirical coverage probability of the bootstrap CI and its width. The former  targets the nominal probability, $1-\alpha$.
With equal coverage, a shorter CI is better. We test $m\in\{50, 100, 200, 500, 1000\}$ and $1-\alpha\in\{ 0.95,0.90\}$  for each example.
All estimators are computed from the same number of simulation runs; we first set the budget for the $k$NN and $k$LR methods and allocate the same for the standard method.
To differentiate the notation from the parameters of Algorithm~\ref{algo: qt-CI for kNN & kLR}, we refer to $\tilde{n}$ and $r$ in Algorithm~\ref{algo:parametric_bootstrapping} for the standard estimator by $\tilde{n}_s$ and $r_s$, respectively, in this section.

According to the asymptotic sampling guidance for the $k$LR estimator discussed in Theorem~\ref{thm:qt-convergence-kLR}, we generate $n = \lfloor m^{6/5} \rfloor$ simulation parameters and $\widetilde{n} = \max\{n,1000\}$ bootstrap parameters.
Although these choices are not asymptotically optimal for $k$NN, we adopt the same $n$ and $\tilde{n}$ so that the total simulation budgets for the $k$NN and $k$LR methods are equal.
In each example, we run Algorithm~\ref{algo:parameter selection}  $1000$ times for $m=50$ and average the returned values to determine $r$, then adopt the same $r$ for other values of $m$.
For each macro run, Algorithm~\ref{algo: cross validation} is applied to determine $k$ for the $k$NN and $k$LR estimators, respectively.
We also compare the two different choices of $f(\cdot|\hatt)$ discussed in Section~\ref{sec:para selection}:  bootstrap vs.\ ellipsoid sampling.

\begin{figure}[tbp]
	\centering
	\includegraphics[width=0.5\textwidth]{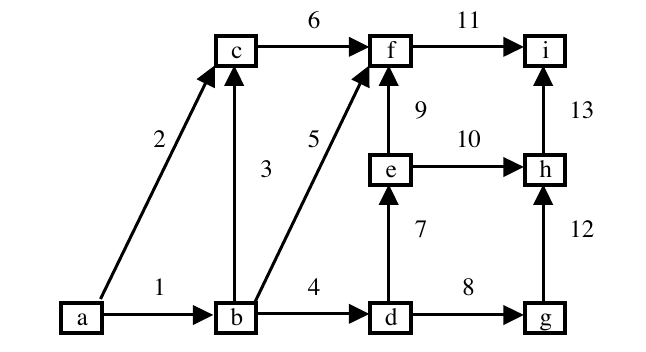}
	\caption{Network representation of a 13-dimensional stochastic activity network (SAN) problem.}
	\label{fig:SAN}
\end{figure}

\subsection{Stochastic activity network (SAN)}\label{sec:numerical-SAN}
We first consider a SAN problem in Figure~\ref{fig:SAN}~\citep{avramidis1996integrated,simopt}.
The network is a directed graph with $13$ arcs and $9$ nodes, where each arc represents a job with an exponentially distributed completion time and each node indicates the prerequisites of each job.
That is, the directions of the arcs denote the order of jobs to be finished and a job cannot start until its preceding jobs are complete.
If the length of an arc corresponds to the job completion time, then the completion time between any two nodes is length of the longest path between them.
We assume that the true mean time to completion equals $1$ for all $13$ jobs, i.e., $\bt^c=(1,\ldots,1)$ and $m$ i.i.d.\ observations of all $13$ activity times are available.

Let $V$ be the completion time from Node $a$ to Node $i$ and $T$ be the time to complete all activities up to Nodes $d$ and $f$.
We are interested in the conditional expectation, $\eta = \mathbb{E}[V|T<2.4] = \E[Y]/\E[A]$, where $Y \triangleq V\mathbf{1}\{T<2.4\}$ and $A\triangleq \mathbf{1}\{T<2.4\}$.

In each macro run, a new set of $m$ i.i.d.\ input data are generated using $\bt^c$.
The data are then fed to Algorithms~\ref{algo:parametric_bootstrapping} and~\ref{algo: qt-CI for kNN & kLR} to generate the percentile bootstrap CIs from the standard ratio estimators  and the $k$NN and $k$LR estimators, respectively.
Because $\E[A|\bt^c] = \pr(T<2.4)\approx 0.091$, estimating $\eta_{std}(\tboot)$ for some $\tboot$ can be challenging as the estimate of $\E[A|\tboot]$  in the denominator may be $0$ for smaller $r$. In this case, $\eta_{std}(\tboot)$ is replaced with the $k$LR estimator with $k=1$ by finding the nearest neighbor of $\tboot$ among the bootstrap parameter set at which the average of $A$ is nonzero.
For $\eta_{kNN}(\tboot)$ and $\eta_{kLR}(\tboot)$, we only pool from the simulation parameters at which the average $A$ is nonzero.
We set $r=99$ by running Algorithm~3 for $m=50$.

\begin{table}[tb]\centering
	\caption{Comparison of empirical coverage probabilities and CI widths with $1\!,\!000$ macro-runs on the SAN problem. The numbers in parentheses are the respective standard errors.}
	\label{tab:SAN}
	\scalebox{0.85}{
		\begin{tabular}{@{}ccccccccc@{}}
			\toprule
			$m$    & $\widetilde{n}$ & $n$    & $\widehat{\eta}_{kNN}$  $(f\sim\widetilde{f})$ & $\widehat{\eta}_{kNN}$ ($f\sim$ Ellip) & $\widehat{\eta}_{kLR}$ $(f\sim\widetilde{f})$ & $\widehat{\eta}_{kLR}$ ($f\sim$ Ellip) & $\widehat{\eta}_{std}$ (Opt) & $\widehat{\eta}_{std}$ (Even) \\
			\midrule
			$50$   & $1000$          & $109$  & $86\% \ (1\%)$                                 & $93\% \ (1\%)$                         & $97\% \ (1\%)$                                & $97\% \ (1\%)$                         & $100\% \ (0\%)$              & $100\% \ (0\%)$               \\
			$100$  & $1000$          & $251$  & $85\% \ (1\%)$                                 & $93\% \ (1\%)$                         & $96\% \ (1\%)$                                & $97\% \ (1\%)$                         & $100\% \ (0\%)$              & $100\% \ (0\%)$               \\
			$200$  & $1000$          & $577$  & $87\% \ (1\%)$                                 & $94\% \ (1\%)$                         & $96\% \ (1\%)$                                & $97\% \ (1\%)$                         & $100\% \ (0\%)$              & $100\% \ (0\%)$               \\
			$500$  & $1732$          & $1732$ & $88\% \ (1\%)$                                 & $95\% \ (1\%)$                         & $96\% \ (1\%)$                                & $96\% \ (1\%)$                         & $100\% \ (0\%)$              & $100\% \ (0\%)$               \\
			$1000$ & $3981$          & $3981$ & $86\% \ (1\%)$                                 & $95\% \ (1\%)$                         & $96\% \ (1\%)$                                & $97\% \ (1\%)$                         & $100\% \ (0\%)$              & $100\% \ (0\%)$               \\
			\bottomrule
		\end{tabular}
	}
	\subcaption{Empirical coverage probabilities with target coverage $95\%$.}
	\label{tab:SAN coverage 95}

	\scalebox{0.85}{
		\begin{tabular}{@{}ccccccccc@{}}
			\toprule
			$m$    & $\widetilde{n}$ & $n$    & $\widehat{\eta}_{kNN}$  $(f\sim\widetilde{f})$ & $\widehat{\eta}_{kNN}$ ($f\sim$ Ellip) & $\widehat{\eta}_{kLR}$ $(f\sim\widetilde{f})$ & $\widehat{\eta}_{kLR}$ ($f\sim$ Ellip) & $\widehat{\eta}_{std}$ (Opt) & $\widehat{\eta}_{std}$ (Even) \\
			\midrule
			$50$   & $1000$          & $109$  & $0.91 \ (0.02)$                                & $1.19 \ (0.02)$                        & $1.08 \ (0.01)$                               & $1.11 \ (0.01)$                        & $4.96 \ (0.02)$              & $2.48 \ (0.01)$               \\
			$100$  & $1000$          & $251$  & $0.61 \ (0.01)$                                & $0.79 \ (0.01)$                        & $0.70 \ (0.01)$                               & $0.73 \ (0.01)$                        & $4.47 \ (0.01)$              & $1.94 \ (0.01)$               \\
			$200$  & $1000$          & $577$  & $0.44 \ (0.01)$                                & $0.54 \ (0.01)$                        & $0.48 \ (0.00)$                               & $0.51 \ (0.01)$                        & $3.93 \ (0.01)$              & $1.50 \ (0.00)$               \\
			$500$  & $1732$          & $1732$ & $0.27 \ (0.00)$                                & $0.34 \ (0.00)$                        & $0.30 \ (0.00)$                               & $0.30 \ (0.00)$                        & $3.22 \ (0.00)$              & $1.09 \ (0.00)$               \\
			$1000$ & $3981$          & $3981$ & $0.18 \ (0.00)$                                & $0.24 \ (0.00)$                        & $0.21 \ (0.00)$                               & $0.22 \ (0.00)$                        & $2.72 \ (0.00)$              & $0.88 \ (0.00)$               \\
			\bottomrule
		\end{tabular}
	}
	\subcaption{Average CI widths with target coverage $95\%$.}
	\label{tab:SAN CI length 95}

	\scalebox{0.85}{
		\begin{tabular}{@{}ccccccccc@{}}
			\toprule
			$m$    & $\widetilde{n}$ & $n$    & $\widehat{\eta}_{kNN}$  $(f\sim\widetilde{f})$ & $\widehat{\eta}_{kNN}$ ($f\sim$ Ellip) & $\widehat{\eta}_{kLR}$ $(f\sim\widetilde{f})$ & $\widehat{\eta}_{kLR}$ ($f\sim$ Ellip) & $\widehat{\eta}_{std}$ (Opt) & $\widehat{\eta}_{std}$ (Even) \\
			\midrule
			$50$   & $1000$          & $109$  & $81\% \ (1\%)$                                 & $89\% \ (1\%)$                         & $94\% \ (1\%)$                                & $94\% \ (1\%)$                         & $100\% \ (0\%)$              & $100\% \ (0\%)$               \\
			$100$  & $1000$          & $251$  & $80\% \ (1\%)$                                 & $88\% \ (1\%)$                         & $91\% \ (1\%)$                                & $92\% \ (1\%)$                         & $100\% \ (0\%)$              & $100\% \ (0\%)$               \\
			$200$  & $1000$          & $577$  & $82\% \ (1\%)$                                 & $90\% \ (1\%)$                         & $92\% \ (1\%)$                                & $93\% \ (1\%)$                         & $100\% \ (0\%)$              & $100\% \ (0\%)$               \\
			$500$  & $1732$          & $1732$ & $83\% \ (1\%)$                                 & $92\% \ (1\%)$                         & $92\% \ (1\%)$                                & $93\% \ (1\%)$                         & $100\% \ (0\%)$              & $100\% \ (0\%)$               \\
			$1000$ & $3981$          & $3981$ & $81\% \ (1\%)$                                 & $92\% \ (1\%)$                         & $92\% \ (1\%)$                                & $93\% \ (1\%)$                         & $100\% \ (0\%)$              & $100\% \ (0\%)$               \\
			\bottomrule
		\end{tabular}
	}
	\subcaption{Empirical coverage probabilities with target coverage $90\%$}
	\label{tab:SAN coverage 90}

	\scalebox{0.85}{
		\begin{tabular}{@{}ccccccccc@{}}
			\toprule
			$m$    & $\widetilde{n}$ & $n$    & $\widehat{\eta}_{kNN}$  $(f\sim\widetilde{f})$ & $\widehat{\eta}_{kNN}$ ($f\sim$ Ellip) & $\widehat{\eta}_{kLR}$ $(f\sim\widetilde{f})$ & $\widehat{\eta}_{kLR}$ ($f\sim$ Ellip) & $\widehat{\eta}_{std}$ (Opt) & $\widehat{\eta}_{std}$ (Even) \\
			\midrule
			$50$   & $1000$          & $109$  & $0.77 \ (0.01)$                                & $1.00 \ (0.02)$                        & $0.90 \ (0.01)$                               & $0.92 \ (0.01)$                        & $3.99 \ (0.01)$              & $1.98 \ (0.01)$               \\
			$100$  & $1000$          & $251$  & $0.52 \ (0.01)$                                & $0.66 \ (0.01)$                        & $0.59 \ (0.01)$                               & $0.61 \ (0.01)$                        & $3.59 \ (0.01)$              & $1.59 \ (0.01)$               \\
			$200$  & $1000$          & $577$  & $0.37 \ (0.01)$                                & $0.45 \ (0.01)$                        & $0.41 \ (0.00)$                               & $0.43 \ (0.00)$                        & $3.16 \ (0.01)$              & $1.24 \ (0.00)$               \\
			$500$  & $1732$          & $1732$ & $0.23 \ (0.00)$                                & $0.29 \ (0.00)$                        & $0.26 \ (0.00)$                               & $0.25 \ (0.00)$                        & $2.60 \ (0.00)$              & $0.92 \ (0.00)$               \\
			$1000$ & $3981$          & $3981$ & $0.15 \ (0.00)$                                & $0.20 \ (0.00)$                        & $0.17 \ (0.00)$                               & $0.18 \ (0.00)$                        & $2.22 \ (0.00)$              & $0.73 \ (0.00)$               \\
			\bottomrule
		\end{tabular}
	}
	\subcaption{Average CI widths with target coverage $90\%$}
	\label{tab:SAN CI length 90}
\end{table}

Table~\ref{tab:SAN} summarizes the empirical coverage probabilities and the widths of CIs averaged from $1000$ macro runs.
For each row, all methods share the same simulation budget.
The columns labeled $f\sim\widetilde{f}$ and $f\sim\text{Ellip}$, for $\widehat{\eta}_{kNN}$ and $\widehat{\eta}_{kLR}$ respectively indicate the bootstrap sampling and the ellipsoid uniform sampling for simulation parameters.
For $\widehat{\eta}_{std}$(Opt), we sample $\tilde{n}_s = (nr)^{2/3}$ bootstrap parameters and make $r_s = (nr)^{1/3}$ simulation runs at each parameter, which is the asymptotically optimal allocation stipulated by Proposition~\ref{prop:standard.est.convergence}.
For $\widehat{\eta}_{std}$(Even), we set $\tilde{n}_s = r_s = \sqrt{nr}$.
We find that $\widehat{\eta}_{std}$(Even) shows better numerical performances than  $\widehat{\eta}_{std}$(Opt) not only in this, but also across all examples.

From Tables~\ref{tab:SAN coverage 95} and~\ref{tab:SAN coverage 90} observe that the CIs from $\widehat{\eta}_{std}$ consistently exhibit overcoverage. Meanwhile, the CIs from $\widehat{\eta}_{kNN}$ with $f\sim\tilde f$ show undercoverage, which is then significantly improved by $f\sim\text{Ellip}$. This affirms our motivation for choosing $f\sim\text{Ellip}$ discussed in Section~\ref{sec:para selection}.
The CIs based on $\widehat{\eta}_{kLR}$ show the most robust coverage probabilities; the choice for $f$ has little impact on the coverages, which reflects that the LR method effectively reduces error in the ratio estimator even if the bootstrap parameter corresponds to the tail quantile and its NNs are farther apart.
The advantage of $\widehat{\eta}_{kNN}$ and $\widehat{\eta}_{kLR}$ over $\widehat{\eta}_{std}$ is more pronounced when comparing CI lengths in Tables~\ref{tab:SAN CI length 95} and~\ref{tab:SAN CI length 90}.
The CIs built upon $\widehat{\eta}_{std}$ are twice as wide as those upon $\widehat{\eta}_{kLR}$  when IU is high ($m=50$), and four times when IU is low  ($m=1000$).



\begin{figure} [tbp]
	\centering
	\begin{subfigure}[b]{0.45\textwidth}
		\centering
		\includegraphics[scale=0.17]{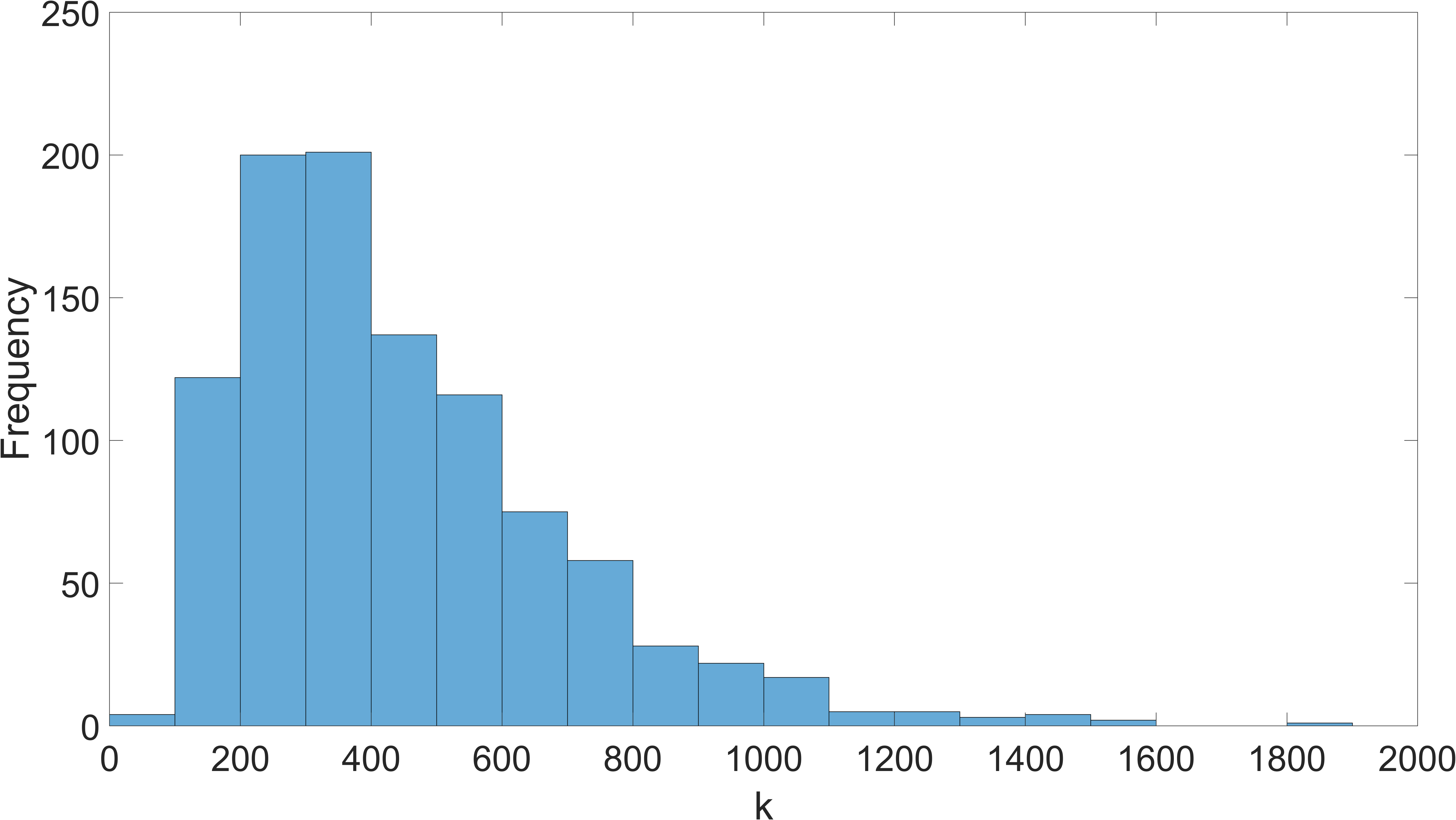}
		\caption{$k$NN with $f=\widetilde{f}$}
		\label{subfig: k_distro_boots_knn}
	\end{subfigure}
	\hfill
	\begin{subfigure}[b]{0.45\textwidth}
		\centering
		\includegraphics[scale=0.17]{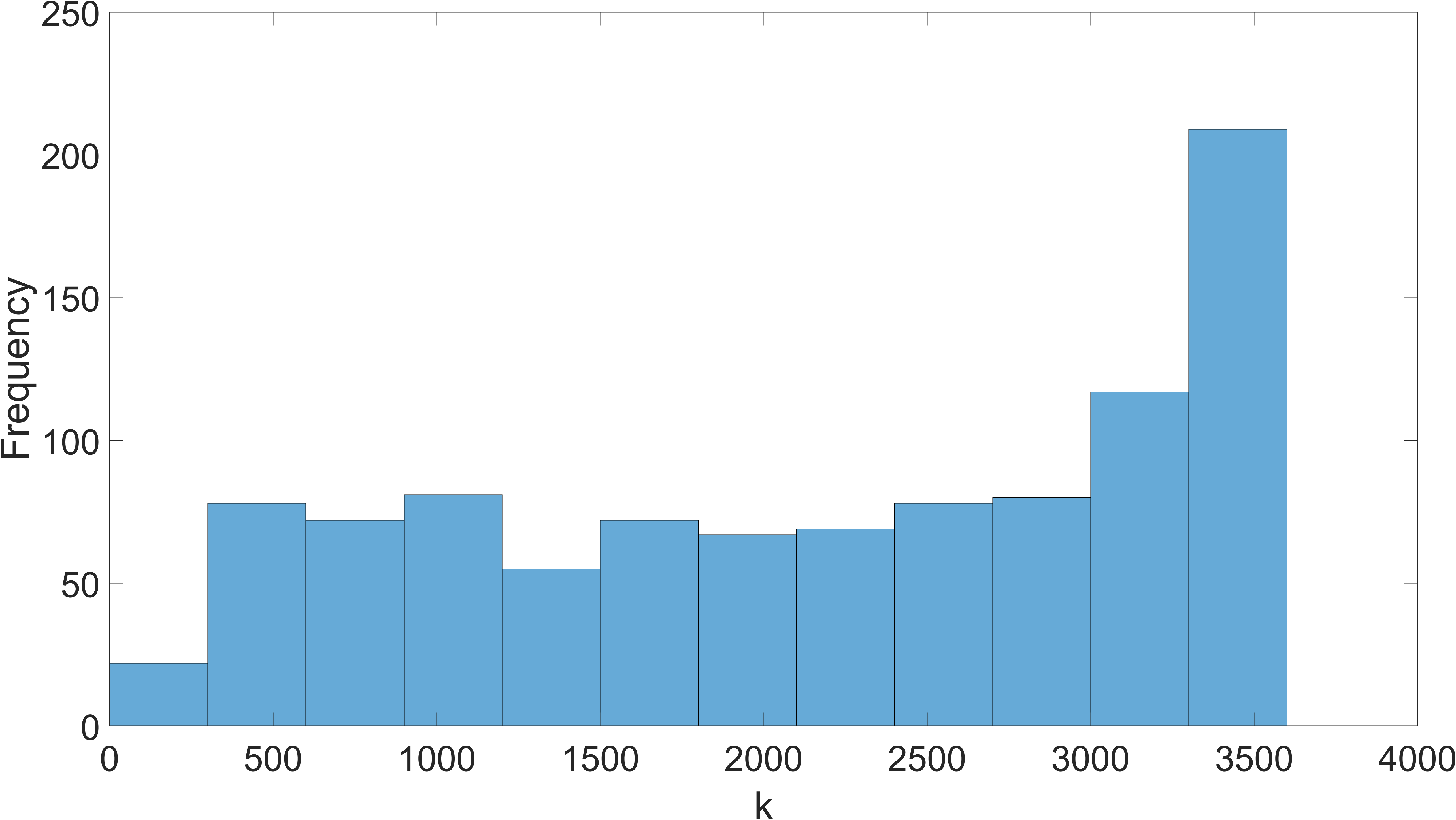}
		\caption{$k$LR with $f=\widetilde{f}$}
		\label{subfig: k_distro_boots_klr}
	\end{subfigure}
	\\
	\begin{subfigure}[b]{0.45\textwidth}
		\centering
		\includegraphics[scale=0.17]{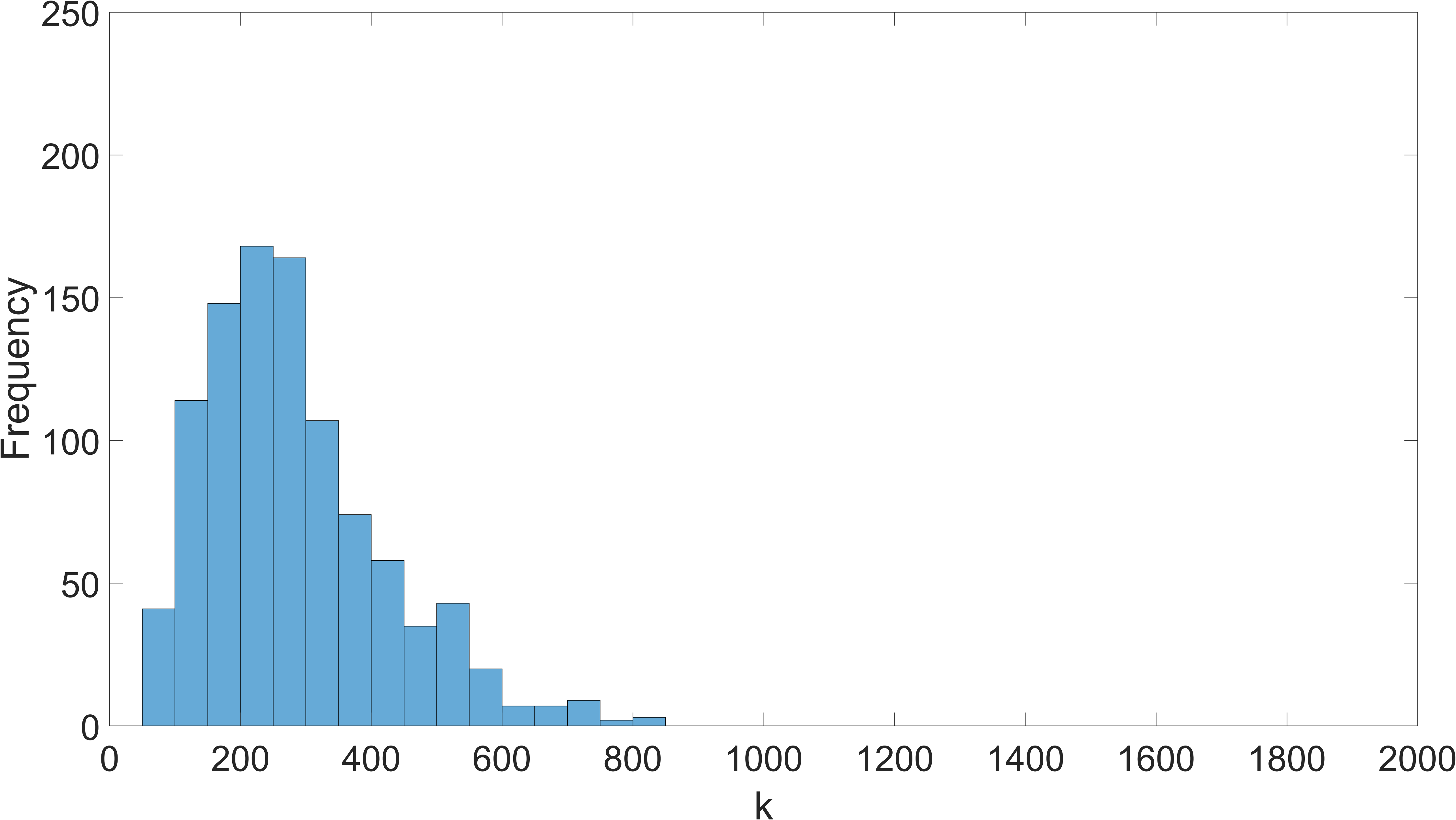}
		\caption{$k$NN with $f\sim \text{Ellip}$}
		\label{subfig: k_distro_ell_knn}
	\end{subfigure}
	\hfill
	\begin{subfigure}[b]{0.45\textwidth}
		\centering
		\includegraphics[scale=0.17]{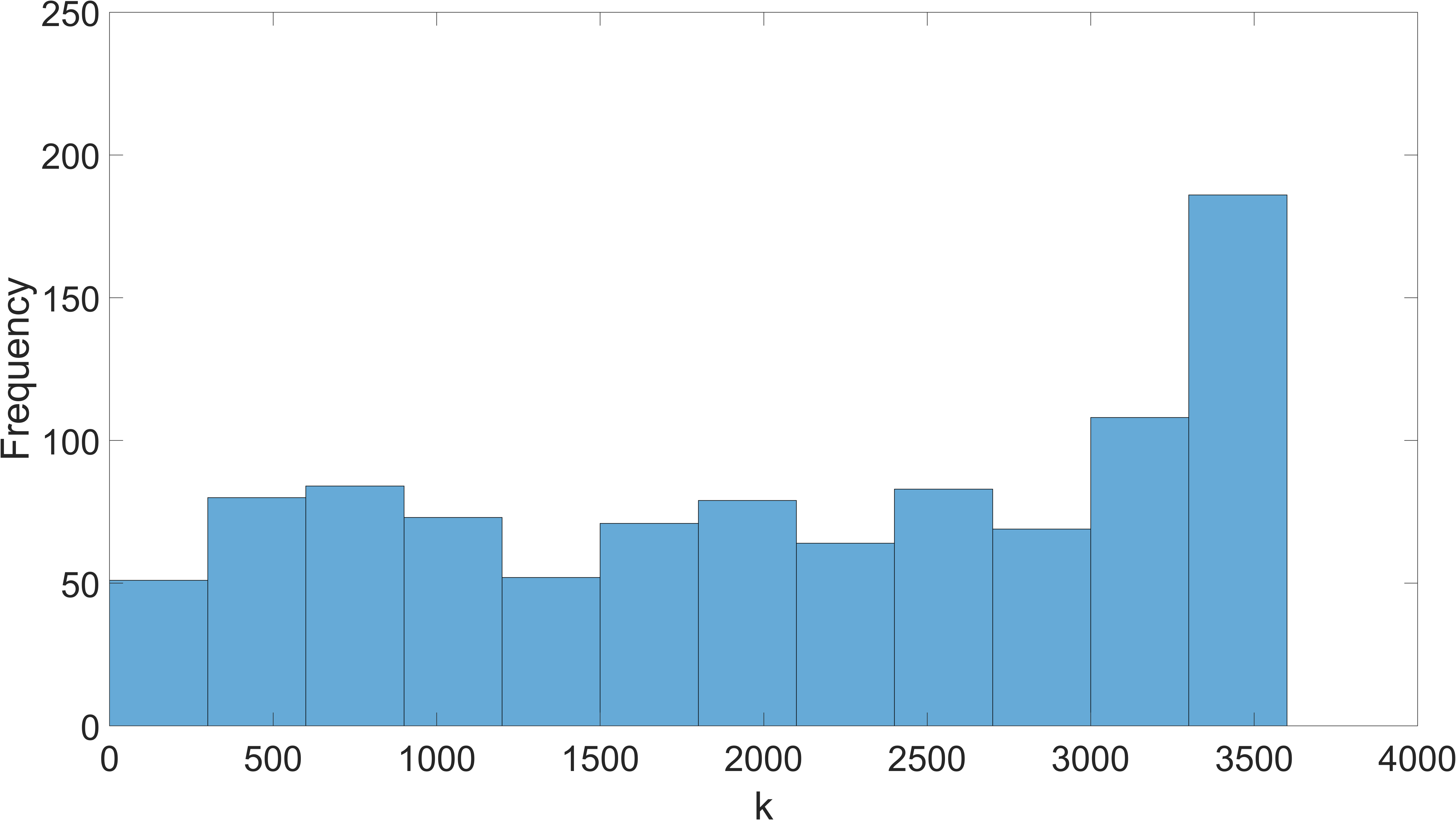}
		\caption{$k$LR with $f\sim \text{Ellip}$}
		\label{subfig: k_distro_ell_klr}
	\end{subfigure}
	\caption{Histogram of $k_Y$  selected by Algorithm~\ref{algo: cross validation} for $m=1\!,\!000$ from $1\!,\!000$ macro runs.}
	\label{fig:k_histogram}
\end{figure}

To further contrast the $k$NN and $k$LR methods, Figure~\ref{fig:k_histogram} displays the histograms of $k_Y$ determined by Algorithm~\ref{algo: cross validation} for $m=1000$.
Comparing (a) and (c), notice that the ellipsoid sampling produces $k_Y$ within a narrower range for $k$NN, thereby reducing the variability in the $k$NN estimators.
Comparing (a) and (b), notice that the $k$LR estimator tends to require larger $k_Y$, thus pools more simulation outputs than the $k$NN estimator.
This is because the LR method removes the bias in $k$NN estimators thus can afford to pool the simulation parameters farther away from the bootstrap parameter. No significant difference between (b) and (d) can be observed, which matches the results in Table~\ref{tab:SAN} that $k$LR is robust to the choice of $f$.

\subsection{An $M/M/1/10$ queue}\label{sec:numerical-MM1c}
An $M/M/1/10$ queue can be represented as a regenerative stochastic process. Let $\Xi(\bt,t),t\geq 0$ be the number in system at time $t$.
Then, $\Xi(\bt,t)$ hits the regenerative state whenever $\Xi(\bt,t)$ becomes $0$.
We define $A$ as the regenerative cycle length and $Y$ as the cumulative number in system during the cycle.
Denoting the starting and end moment of a regenerative cycle as $t_0$ and $t_1$ respectively, $Y = \frac{1}{A}\int_{t_0}^{t_1} \Xi(\bt,s)\rmd s$.
The performance measure of interest is the steady-state expected number in system,
$
	\eta(\tsim)
	= \E[Y|\bt]/\E[A|\bt].
$

Although we assume the distribution families (exponential) of the interarrival and service times are known, their rate parameters $\ttrue = (\lambda^c,\mu^c) = (0.5,1.5)$ need to be estimated from input data.
In each macro replication, we generate $m$ i.i.d.\ observations from each  distribution, then treat them as input data.

The value of $r$ determined by Algorithm~\ref{algo:parameter selection} for this example is $r=7$, which is smaller than in the SAN example because the simulation error variance  is smaller relative to IU in this example.
The number of inputs generated within each simulation run varies in this example depending on the length of the regenerative cycle.
Thus, the number of density products computed for the $k$LR method varies in each simulation run.

Table~\ref{tab:light MM1c} presents the empirical coverage probabilities and CI widths averaged over $1000$ macro runs.
We make similar observations as in Section~\ref{sec:numerical-SAN}:
The CIs from $\widehat{\eta}_{std}$ exhibit overcoverage and are wider.
The CIs from $\widehat{\eta}_{kNN} (f \sim \widetilde{f})$ show undercoverage and are narrower, which are improved by $f\sim\text{Ellip}$.
The CIs from the $k$LR estimator consistently show robust coverage across all $m$ and $\alpha$ values regardless of $f$.



\begin{table}[h]\centering
	\caption{Comparison of empirical coverage probabilities and CI width with $1\!,\!000$ macro-runs on an $M/M/1/10$ system with $\lambda^c = 0.5$ and $\mu^c = 1.5$. Numbers in parenthesis are the respective standard errors.}

	\scalebox{0.8}{
		\begin{tabular}{@{}ccccccccc@{}}
			\toprule
			$m$    & $\widetilde{n}$ & $n$    & $\widehat{\eta}_{kNN}$  $(f\sim\widetilde{f})$ & $\widehat{\eta}_{kNN}$ ($f\sim$ Ellip) & $\widehat{\eta}_{kLR}$  $(f\sim\widetilde{f})$ & $\widehat{\eta}_{kLR}$ ($f\sim$ Ellip) & $\widehat{\eta}_{std}$ (Opt) & $\widehat{\eta}_{std}$ (Even) \\
			\midrule
			$50$   & $1000$          & $109$  & $78\% \ (1\%)$                                 & $94\% \ (1\%)$                         & $94\% \ (1\%)$                                 & $95\% \ (1\%)$                         & $100\% \ (0\%)$              & $99\% \ (0\%)$                \\
			$100$  & $1000$          & $251$  & $80\% \ (1\%)$                                 & $95\% \ (1\%)$                         & $94\% \ (1\%)$                                 & $95\% \ (1\%)$                         & $100\% \ (0\%)$              & $99\% \ (0\%)$                \\
			$200$  & $1000$          & $577$  & $81\% \ (1\%)$                                 & $96\% \ (1\%)$                         & $95\% \ (1\%)$                                 & $96\% \ (1\%)$                         & $100\% \ (0\%)$              & $100\% \ (0\%)$               \\
			$500$  & $1732$          & $1732$ & $84\% \ (1\%)$                                 & $94\% \ (1\%)$                         & $94\% \ (1\%)$                                 & $94\% \ (1\%)$                         & $100\% \ (0\%)$              & $100\% \ (0\%)$               \\
			$1000$ & $3981$          & $3981$ & $84\% \ (1\%)$                                 & $96\% \ (1\%)$                         & $95\% \ (1\%)$                                 & $94\% \ (1\%)$                         & $100\% \ (0\%)$              & $100\% \ (0\%)$               \\
			\bottomrule
		\end{tabular}
		\label{tab:light MM1c}
	}
	\subcaption{Empirical coverage probabilities with target coverage $95\%$.}

	\scalebox{0.8}{
		\begin{tabular}{@{}ccccccccc@{}}
			\toprule
			$m$    & $\widetilde{n}$ & $n$    & $\widehat{\eta}_{kNN}$  $(f\sim\widetilde{f})$ & $\widehat{\eta}_{kNN}$ ($f\sim$ Ellip) & $\widehat{\eta}_{kLR}$ $(f\sim\widetilde{f})$ & $\widehat{\eta}_{kLR}$ ($f\sim$ Ellip) & $\widehat{\eta}_{std}$ (Opt) & $\widehat{\eta}_{std}$ (Even) \\
			\midrule
			$50$   & $1000$          & $109$  & $0.66 \ (0.02)$                                & $0.98 \ (0.02)$                        & $0.82 \ (0.02)$                               & $0.87 \ (0.02)$                        & $1.30 \ (0.02)$              & $1.18 \ (0.02)$               \\
			$100$  & $1000$          & $251$  & $0.38 \ (0.01)$                                & $0.56 \ (0.01)$                        & $0.50 \ (0.01)$                               & $0.54 \ (0.01)$                        & $1.10 \ (0.01)$              & $0.69 \ (0.01)$               \\
			$200$  & $1000$          & $577$  & $0.25 \ (0.00)$                                & $0.36 \ (0.00)$                        & $0.33 \ (0.00)$                               & $0.34 \ (0.00)$                        & $0.94 \ (0.01)$              & $0.59 \ (0.00)$               \\
			$500$  & $1732$          & $1732$ & $0.15 \ (0.00)$                                & $0.22 \ (0.00)$                        & $0.20 \ (0.00)$                               & $0.20 \ (0.00)$                        & $0.77 \ (0.00)$              & $0.42 \ (0.00)$               \\
			$1000$ & $3981$          & $3981$ & $0.11 \ (0.00)$                                & $0.15 \ (0.00)$                        & $0.14 \ (0.00)$                               & $0.14 \ (0.00)$                        & $0.68 \ (0.00)$              & $0.32 \ (0.00)$               \\
			\bottomrule
		\end{tabular}
	}
	\subcaption{Average CI widths with target coverage $95\%$.}

	\scalebox{0.85}{
		\begin{tabular}{@{}ccccccccc@{}}
			\toprule
			$m$    & $\widetilde{n}$ & $n$    & $\widehat{\eta}_{kNN}$  $(f\sim\widetilde{f})$ & $\widehat{\eta}_{kNN}$ ($f\sim$ Ellip) & $\widehat{\eta}_{kLR}$  $(f\sim\widetilde{f})$ & $\widehat{\eta}_{kLR}$ ($f\sim$ Ellip) & $\widehat{\eta}_{std}$ (Opt) & $\widehat{\eta}_{std}$ (Even) \\
			\midrule
			$50$   & $1000$          & $109$  & $74\% \ (1\%)$                                 & $91\% \ (1\%)$                         & $89\% \ (1\%)$                                 & $90\% \ (1\%)$                         & $100\% \ (0\%)$              & $95\% \ (1\%)$                \\
			$100$  & $1000$          & $251$  & $76\% \ (1\%)$                                 & $90\% \ (1\%)$                         & $90\% \ (1\%)$                                 & $91\% \ (1\%)$                         & $100\% \ (0\%)$              & $98\% \ (0\%)$                \\
			$200$  & $1000$          & $577$  & $76\% \ (1\%)$                                 & $92\% \ (1\%)$                         & $90\% \ (1\%)$                                 & $89\% \ (1\%)$                         & $100\% \ (0\%)$              & $100\% \ (0\%)$               \\
			$500$  & $1732$          & $1732$ & $80\% \ (1\%)$                                 & $91\% \ (1\%)$                         & $90\% \ (1\%)$                                 & $90\% \ (1\%)$                         & $100\% \ (0\%)$              & $100\% \ (0\%)$               \\
			$1000$ & $3981$          & $3981$ & $79\% \ (1\%)$                                 & $92\% \ (1\%)$                         & $90\% \ (1\%)$                                 & $90\% \ (1\%)$                         & $100\% \ (0\%)$              & $100\% \ (0\%)$               \\
			\bottomrule
		\end{tabular}
	}
	\subcaption{Empirical coverage probabilities with target coverage $90\%$.}

	\scalebox{0.85}{
		\begin{tabular}{@{}ccccccccc@{}}
			\toprule
			$m$    & $\widetilde{n}$ & $n$    & $\widehat{\eta}_{kNN}$  $(f\sim\widetilde{f})$ & $\widehat{\eta}_{kNN}$ ($f\sim$ Ellip) & $\widehat{\eta}_{kLR}$ $(f\sim\widetilde{f})$ & $\widehat{\eta}_{kLR}$ ($f\sim$ Ellip) & $\widehat{\eta}_{std}$ (Opt) & $\widehat{\eta}_{std}$ (Even) \\
			\midrule
			$50$   & $1000$          & $109$  & $0.55 \ (0.01)$                                & $0.80 \ (0.02)$                        & $0.65 \ (0.01)$                               & $0.69 \ (0.01)$                        & $1.04 \ (0.01)$              & $0.84 \ (0.01)$               \\
			$100$  & $1000$          & $251$  & $0.33 \ (0.01)$                                & $0.47 \ (0.01)$                        & $0.41 \ (0.01)$                               & $0.44 \ (0.01)$                        & $0.86 \ (0.01)$              & $0.58 \ (0.01)$               \\
			$200$  & $1000$          & $577$  & $0.22 \ (0.00)$                                & $0.31 \ (0.00)$                        & $0.28 \ (0.00)$                               & $0.28 \ (0.00)$                        & $0.76 \ (0.00)$              & $0.46 \ (0.00)$               \\
			$500$  & $1732$          & $1732$ & $0.14 \ (0.00)$                                & $0.18 \ (0.00)$                        & $0.17 \ (0.00)$                               & $0.17 \ (0.00)$                        & $0.63 \ (0.00)$              & $0.34 \ (0.00)$               \\
			$1000$ & $3981$          & $3981$ & $0.10 \ (0.00)$                                & $0.13 \ (0.00)$                        & $0.12 \ (0.00)$                               & $0.12 \ (0.00)$                        & $0.55 \ (0.00)$              & $0.27 \ (0.00)$               \\
			\bottomrule
		\end{tabular}
	}
	\subcaption{Average CI width with target coverage $90\%$.}
\end{table}

\subsection{Enterprise Risk Management (ERM) Example}
In this section, we consider an ERM example that estimates the conditional expected value of an option portfolio given that the total value of the underlying assets is below a threshold.
This example measures the option portfolio's risk when the underlying stocks perform poorly, i.e., when the stocks are at risk.

Consider a portfolio of $10L$ options written on $L$ underlying stocks with 5 European calls and 5 European puts (with different strikes) written on each stock.
The portfolio's value is the sum of all option values.
Let time-$0$ be the current time and $T >0$ be the common expiry date of all the options.
The goal is to estimate the option portfolio's conditional expected value at future time $0< \tau < T$ given that the total value of the stocks is below $K^*$.
We assume that the $L$ underlying stocks' price dynamics follow the multi-variate Black-Scholes model, i.e., the time-$t$ price of the stocks $\bm{S}_t = \{S_{t}^1, \ldots, S_{t}^L\}$ evolve as
$
	dS_t^{\ell} = \mu'_\ell S_t^\ell dt + \sigma_\ell S_t^\ell d W_t^\ell,
$
for $\ell\in [L]$, where $\mu'_\ell$ equals the expected return, $\mu$, or the risk-free rate, $ \chi$, under the real-world or risk-neutral measures, respectively.
Also, $\bm{W}_t = \{W_{t}^1, \ldots, W_{t}^L\}$ is a $d$-dimensional Brownian motion with correlation structure: $dW_t^i dW_t^j = \rho_{ij} dt$ for $i,j\in[L]$.
The covariance between two stocks is then $dS_t^i dS_t^j = \sigma_i\sigma_j\rho_{ij} S_t^iS_t^jdt$ for $i,j\in[L]$.
From the model, at any time $0 \leq t_1 < t_2$, $\bm{S}_{t_2}$ given $\bm{S}_{t_1}$ follows a multi-variate log-normal distribution.
Namely, $\bm{S}_{t_2}|\bm{S}_{t_1}$ can be simulated as $S_{t_2}^\ell = S_{t_1}^\ell \exp\left(Z^\ell \right)$ for $\ell\in [L]$ by drawing $\bm{Z}=(Z^1,\ldots,Z^\ell)$ from the multivariate normal distribution with mean $\left(\bm{\mu}'_\ell - \frac{1}{2}\bm{\sigma}^2_\ell\right)(t_2-t_1)$ and covariance matrix $\sqrt{t_2-t_1}\bm{\Sigma}$, where $\bm{\Sigma}=[\sigma_i\sigma_j\rho_{ij}]$.


For simplicity, we test on $L=2$ stocks and assume that $\boldsymbol{\mu} = \ttrue = (0.05, 0.1)$ is estimated from data thus is subject to IU.
We set the expiration as $T=2$ years and $\tau$ as $4$ weeks, or $\frac{4}{52}$ years.
We fix $\chi=0.02$, $\sigma=(0.15,0.35)$ and 
$\rho_{ij}=0.5$ for $i\neq j$.
In practice, the expected returns of stocks are estimated by the average of historical returns within some time window.
In each macro replication, we collect $m$ observations of i.i.d. random vectors generated from multivariate normal distribution with mean vector $\ttrue = (\mu_1^c,\ldots,\mu_L^c)$ and covariance vector $\Sigma$ as treat them as historical observations of the stock values.

Given each simulation parameter $\tsim_i = (\mu_1,\ldots,\mu_L)$, the simulation model first generates $\bm{S}_\tau|\bm{S}_0$ as $\bm{S}_\tau = \bm{S}_{0}^\ell \exp\left(\bm{Z}_\tau \right)$, where $\bm{Z}_\tau \sim MVN\left(\left(\tsim_i - \frac{1}{2}\bm{\sigma}^2\right)\tau, \sqrt{\tau}\bm{\Sigma}\right)$.
Then, given  $\bm{S}_\tau$, we calculate the option portfolio's value at time-$\tau$, $V(\bm{S}_\tau)$, via the Black-Scholes pricing formulas for European calls and puts.
Define $Y \triangleq V(\bm{S}_\tau)\mathbf{1}\{\sum_{\ell=1}^{L} S_\tau^{\ell}<K^*\}$ and $A\triangleq \mathbf{1}\{\sum_{\ell=1}^{L} S_\tau^{\ell}<K^*\}$.
Then, the performance measure is $\mathbb{E}[V(\bm{S}_\tau)|\sum_{\ell=1}^{L} S_\tau^{\ell}<K^*] = \E[Y]/\E[A]$, where
$K^*$ is chosen as the sum of the $5\%$-tiles of the stocks.

\begin{table}[tbp]\centering
	\caption{Comparison of empirical coverage probabilities and CI width with $1\!,\!000$ macro-runs on an $ERM$ system. Numbers in parenthesis are the respective standard errors.}
	\label{tab:ERM}

	\scalebox{0.80}{
		\begin{tabular}{@{}ccccccccc@{}}
			\toprule
			$m$    & $\widetilde{n}$ & $n$    & $\widehat{\eta}_{kNN}$  $(f\sim\widetilde{f})$ & $\widehat{\eta}_{kNN}$ ($f\sim$ Ellip) & $\widehat{\eta}_{kLR}$  $(f\sim\widetilde{f})$ & $\widehat{\eta}_{kLR}$ ($f\sim$ Ellip) & $\widehat{\eta}_{std}$ (Opt) & $\widehat{\eta}_{std}$ (Even) \\
			\midrule
			$50$   & $1000$          & $109$  & $80\% \ (1\%)$                                 & $89\% \ (1\%)$                         & $82\% \ (1\%)$                                 & $88\% \ (1\%)$                         & $100\% \ (0\%)$              & $100\% \ (0\%)$               \\
			$100$  & $1000$          & $251$  & $82\% \ (1\%)$                                 & $91\% \ (1\%)$                         & $84\% \ (1\%)$                                 & $92\% \ (1\%)$                         & $100\% \ (0\%)$              & $100\% \ (0\%)$               \\
			$200$  & $1000$          & $577$  & $84\% \ (1\%)$                                 & $94\% \ (1\%)$                         & $88\% \ (1\%)$                                 & $90\% \ (1\%)$                         & $100\% \ (0\%)$              & $100\% \ (0\%)$               \\
			$500$  & $1732$          & $1732$ & $87\% \ (1\%)$                                 & $94\% \ (1\%)$                         & $89\% \ (1\%)$                                 & $94\% \ (1\%)$                         & $100\% \ (0\%)$              & $100\% \ (0\%)$               \\
			$1000$ & $3981$          & $3981$ & $88\% \ (1\%)$                                 & $96\% \ (1\%)$                         & $91\% \ (1\%)$                                 & $93\% \ (1\%)$                         & $100\% \ (0\%)$              & $100\% \ (0\%)$               \\
			\bottomrule
		\end{tabular}
	}
	\subcaption{Empirical coverage probabilities with target coverage $95\%$.}

	\scalebox{0.80}{
		\begin{tabular}{@{}ccccccccc@{}}
			\toprule
			$m$    & $\widetilde{n}$ & $n$    & $\widehat{\eta}_{kNN}$  $(f\sim\widetilde{f})$ & $\widehat{\eta}_{kNN}$ ($f\sim$ Ellip) & $\widehat{\eta}_{kLR}$ $(f\sim\widetilde{f})$ & $\widehat{\eta}_{kLR}$ ($f\sim$ Ellip) & $\widehat{\eta}_{std}$ (Opt) & $\widehat{\eta}_{std}$ (Even) \\
			\midrule
			$50$   & $1000$          & $109$  & $1.77 \ (0.21)$                                & $2.57 \ (0.31)$                        & $1.66 \ (0.23)$                               & $1.70 \ (0.18)$                        & $11.74 \ (0.04)$             & $6.03 \ (0.02)$               \\
			$100$  & $1000$          & $251$  & $1.32 \ (0.17)$                                & $1.51 \ (0.18)$                        & $1.27 \ (0.16)$                               & $1.14 \ (0.13)$                        & $10.96 \ (0.02)$             & $4.87 \ (0.01)$               \\
			$200$  & $1000$          & $577$  & $1.06 \ (0.13)$                                & $0.92 \ (0.09)$                        & $0.69 \ (0.07)$                               & $1.03 \ (0.13)$                        & $9.99 \ (0.01)$              & $3.86 \ (0.01)$               \\
			$500$  & $1732$          & $1732$ & $0.51 \ (0.05)$                                & $0.56 \ (0.06)$                        & $0.45 \ (0.04)$                               & $0.51 \ (0.05)$                        & $8.47 \ (0.01)$              & $2.83 \ (0.00)$               \\
			$1000$ & $3981$          & $3981$ & $0.29 \ (0.03)$                                & $0.33 \ (0.03)$                        & $0.38 \ (0.05)$                               & $0.38 \ (0.04)$                        & $7.30 \ (0.00)$              & $2.31 \ (0.00)$               \\
			\bottomrule
		\end{tabular}
	}
	\subcaption{Average CI widths with target coverage $95\%$.}

	\scalebox{0.85}{
		\begin{tabular}{@{}ccccccccc@{}}
			\toprule
			$m$    & $\widetilde{n}$ & $n$    & $\widehat{\eta}_{kNN}$  $(f\sim\widetilde{f})$ & $\widehat{\eta}_{kNN}$ ($f\sim$ Ellip) & $\widehat{\eta}_{kLR}$  $(f\sim\widetilde{f})$ & $\widehat{\eta}_{kLR}$ ($f\sim$ Ellip) & $\widehat{\eta}_{std}$ (Opt) & $\widehat{\eta}_{std}$ (Even) \\
			\midrule
			$50$   & $1000$          & $109$  & $76\% \ (1\%)$                                 & $86\% \ (1\%)$                         & $79\% \ (1\%)$                                 & $85\% \ (1\%)$                         & $100\% \ (0\%)$              & $100\% \ (0\%)$               \\
			$100$  & $1000$          & $251$  & $79\% \ (1\%)$                                 & $88\% \ (1\%)$                         & $81\% \ (1\%)$                                 & $88\% \ (1\%)$                         & $100\% \ (0\%)$              & $100\% \ (0\%)$               \\
			$200$  & $1000$          & $577$  & $82\% \ (1\%)$                                 & $91\% \ (1\%)$                         & $84\% \ (1\%)$                                 & $87\% \ (1\%)$                         & $100\% \ (0\%)$              & $100\% \ (0\%)$               \\
			$500$  & $1732$          & $1732$ & $85\% \ (1\%)$                                 & $92\% \ (1\%)$                         & $84\% \ (1\%)$                                 & $90\% \ (1\%)$                         & $100\% \ (0\%)$              & $100\% \ (0\%)$               \\
			$1000$ & $3981$          & $3981$ & $86\% \ (1\%)$                                 & $92\% \ (1\%)$                         & $88\% \ (1\%)$                                 & $89\% \ (1\%)$                         & $100\% \ (0\%)$              & $100\% \ (0\%)$               \\
			\bottomrule
		\end{tabular}
	}
	\subcaption{Empirical coverage probabilities with target coverage $90\%$.}

	\scalebox{0.85}{
		\begin{tabular}{@{}ccccccccc@{}}
			\toprule
			$m$    & $\widetilde{n}$ & $n$    & $\widehat{\eta}_{kNN}$  $(f\sim\widetilde{f})$ & $\widehat{\eta}_{kNN}$ ($f\sim$ Ellip) & $\widehat{\eta}_{kLR}$ $(f\sim\widetilde{f})$ & $\widehat{\eta}_{kLR}$ ($f\sim$ Ellip) & $\widehat{\eta}_{std}$ (Opt) & $\widehat{\eta}_{std}$ (Even) \\
			\midrule
			$50$   & $1000$          & $109$  & $1.53 \ (0.18)$                                & $2.21 \ (0.26)$                        & $1.43 \ (0.20)$                               & $1.44 \ (0.15)$                        & $9.34 \ (0.02)$              & $4.77 \ (0.02)$               \\
			$100$  & $1000$          & $251$  & $1.14 \ (0.14)$                                & $1.28 \ (0.15)$                        & $1.08 \ (0.13)$                               & $0.96 \ (0.11)$                        & $8.71 \ (0.01)$              & $3.97 \ (0.01)$               \\
			$200$  & $1000$          & $577$  & $0.92 \ (0.12)$                                & $0.79 \ (0.08)$                        & $0.59 \ (0.06)$                               & $0.88 \ (0.11)$                        & $7.93 \ (0.01)$              & $3.18 \ (0.01)$               \\
			$500$  & $1732$          & $1732$ & $0.44 \ (0.05)$                                & $0.48 \ (0.05)$                        & $0.39 \ (0.04)$                               & $0.44 \ (0.04)$                        & $6.78 \ (0.00)$              & $2.38 \ (0.00)$               \\
			$1000$ & $3981$          & $3981$ & $0.25 \ (0.02)$                                & $0.28 \ (0.03)$                        & $0.32 \ (0.04)$                               & $0.32 \ (0.03)$                        & $5.88 \ (0.00)$              & $1.92 \ (0.00)$               \\
			\bottomrule
		\end{tabular}
	}
	\subcaption{Average CI width with target coverage $90\%$.}
\end{table}

Table~\ref{tab:ERM} presents the results from $1000$ macro runs. The $\widehat{\eta}_{std}$-based CIs show severe overcoverage and are much wider in this example reflecting the large simulation error. Similar to the previous two examples, the CIs from $\widehat{\eta}_{kNN}(f\sim\widetilde{f})$ show undercoverage, but are improved by $f\sim\text{Ellip}$ when $m$ is larger.
On the other, the $k$LR-based CIs show undercoverage regardless of $f$, although mitigated  for larger $m$.

\section{Conclusion}\label{sec:conclusion}
In this paper, we propose two new ratio estimators, the $k$NN and $k$LR estimators, in the context of IUQ.
Our asymptotic analyses establish convergence rates of the biases, variances, and MSEs of the estimators as well as a CLT for the $k$NN estimators.
We further prove convergence of the quantile estimators of the ratio and coverage probabilities of the corresponding percentile bootstrap CIs.

Unlike the standard estimator that requires  the number of simulation runs, $r$, at each bootstrap parameter to grow with the input data size to ensure consistency of the bootstrap CI,  we show that the consistency for the proposed estimators can be achieved with constant $r$. While the curse of dimensionality affects the sample size requirements for the $k$NN estimator, we prove that the $k$LR estimator's asymptotic performance is not affected by the dimension. Furthermore, the $k$LR is shown to be more efficient than the standard ratio estimator regardless of the problem dimension. 
The finite-sample performance of our estimators can be further improved by carefully selecting a simulation parameter sampling distribution. 

Empirical studies strongly indicate that the $k$NN and $k$LR estimators produce CIs that are both narrower and closer to the target coverage than the standard estimator when the same simulation budget is adopted. In particular, the $k$LR estimator shows robust performances across all examples and different input data sizes.



\bibliographystyle{informs2014} 
\bibliography{refbib} 









\newpage
\begin{APPENDICES}
\noindent This document is intended for the Online Supplement of ``Efficient input uncertainty quantification for ratio estimator.''
\section{Proofs in Section~\ref{subsec:kNN}}\label{appendix: knn fundamental}
We first present three lemmas that lay the groundwork for Lemma~\ref{lem: knn bias and var} and Proposition~\ref{prop:kNN-MSE}.
The first lemma can be found in~\cite{mack1979multivariate}.
Fix $\tboot\in\widetilde{\Theta}$, define $\gpr{t}\triangleq \pr(\lVert \tsim - \tboot\rVert \leq t|\tboot,\hatt)$.
\begin{lemma}\label{lem:G(r) approx}
Suppose Assumption~\ref{assump:theta-df}(ii) holds and $\tboot\in \Theta$. Then as $t\to 0$,
\[
\gpr{t} = V_d f(\tboot|\hatt) t^d + \cO(t^{d+1}) 
=V_d f(\tboot|\hatt) t^d + o(t^d). 
\]
Further, suppose $f(\tboot|\hatt)>0$.
Then for any $\lambda \neq 0$,
$
t^\lambda
= \left(\tfrac{ \gpr{t} }{V_d f(\tboot|\hatt)}\right)^{\tfrac{\lambda}{d}} + o(G^{\frac{\lambda}{d}} (t |\tboot,\hatt) )
$
as $t\to 0.$
\end{lemma}
Essentially, Lemma~\ref{lem:G(r) approx} states that $\gpr{t}$ can be effectively approximated by the product of $f(\tboot|\hatt)$ and the volume of the $d$-dimensional ball.
From Lemma~\ref{lem:G(r) approx}, we proceed to show the first and second moments of $Y(\tsim_{(i)})$ conditional on $R_{n,k+1}$ in Lemma~\ref{lem:ball_integration} and moments of $R_{n,k+1}$ and $\gpr{R_{n,k+1}}$ in Lemma~\ref{lem:R_moments}.
The proof of Lemma~\ref{lem:ball_integration} can be derived from the proof of Proposition $3$ in~\citet{mack1981local}.
\begin{lemma}\label{lem:ball_integration}
Suppose Assumption~\ref{assump:theta-df}(i)(ii) and Assumption~\ref{assump:smoothness} hold.
Then for $q\in\{1,2\}$ and $i\in [k]$,
\footnotesize
\begin{align*}
\E[Y^q \left(\tsim_{(i)}\right) |R_{n,k+1}, \tboot, \hatt ]
= & \E[Y^q|\tboot] f(\tboot|\hatt) V_d \frac{R_{n,k+1}^d}{\gpr{R_{n,k+1}}}   \\
& +\frac{V_d}{2(d+2)} \Delta_{\tsim}(\E[Y^q|\tboot]f(\tboot|\hatt)) \frac{R_{n,k+1}^{d+2}}{\gpr{R_{n,k+1}}}
+ o \left( \frac{R_{n,k+1}^{d+2}}{\gpr{R_{n,k+1}}}\right).
\end{align*}\small
\end{lemma}
\begin{lemma}\label{lem:R_moments}
Suppose Assumption~\ref{assump:theta-df}  and Assumption~\ref{assump:smoothness} hold.
Then, for $p,q\in\mathbb{Z}$, $q\geq -k$ and $\frac{p}{d}+q>0$,
\[
\E[R_{n,k+1}^p G^q(R_{n,k+1}|\tboot,\hatt) |\tboot, \hatt]
= ({V_d f(\tboot|\hatt)})^{-\frac{p}{d}} (\tfrac{k}{n})^{ q+\frac{p}{d} } + o((\tfrac{k}{n})^{ q+\frac{p}{d} }).
\]
Moreover,
$\E[\frac{k!(n+q)!}{(k+q)!n!}G^q(R_{n,k+1}|\tboot,\hatt)|\tboot, \hatt]= 1.$
\end{lemma}
\proof{Proof.}
Let $G^\prime(t|\tboot,\hatt)$ be the derivative of $\gpr{t}$ with respect to $t$.
\cite{mack1979multivariate} show that the density of $R_{n, k+1}$ is given by
$
\tfrac{n!}{k!(n-k-1)!} G^k(t|\tboot,\hatt) (1-\gpr{t} )^{n-k-1}G^\prime(t|\tboot,\hatt)
$
for $t>0.$
As $R_{n,k+1}\to 0$ almost surely when $n,k\to\infty$ and $\frac{k}{n}\to 0$~\citep{biau2015lectures}, we can invoke Lemma~\ref{lem:G(r) approx} and have
\begin{align*}
& \E[R_{n,k+1}^p  G^q (R_{n,k+1} |\tboot,\hatt) |\tboot, \hatt] \\
=&  \int_0^\infty t^p G^q(t|\tboot,\hatt) \tfrac{n!}{k!(n-k-1)!} G^k(t|\tboot, \hatt)(1-\gpr{t} )^{n-k-1}  G^\prime(t|\tboot,\hatt) \rmd t \\
= & \int_0^\infty \tfrac{n!}{k!(n-k-1)!} \left(\left(\tfrac{\gpr{t} }{V_d f(\tboot|\hatt)}\right)^{\tfrac{p}{d}} + o\left( G^{\tfrac{p}{d}} (R_{n,k+1} |\tboot,\hatt) \right)\right) G^{k+q}(t|\tboot, \hatt)(1-\gpr{t} )^{n-k-1}G^\prime(t|\tboot,\hatt) \rmd t \\
= & ({V_d f(\tboot|\hatt)})^{-\tfrac{p}{d}} \int_0^\infty \tfrac{n!}{k!(n-k-1)!}G^{k+q+\tfrac{p}{d}}(t|\tboot, \hatt)(1-\gpr{t} )^{n-k-1}G^\prime(t|\tboot, \hatt) \rmd t (1+o(1)).
\end{align*}
The last inequality holds by the dominated convergence theorem.
By substituting $\gpr{t} = y$, the integral becomes
\[
\frac{n!}{k!(n-k-1)!} \int_0^1 y^{k+q+\frac{p}{d}}(1-y)^{n-k-1} \rmd y
=  \frac{n!\Gamma(k+q+1+p/d)\Gamma(n-k)}{k!(n-k-1)!\Gamma(n+q+1+p/d)}
=  \frac{n!}{k!} \frac{\Gamma(k+1+q+p/d)}{\Gamma(n+1+q+p/d)}.
\]
If $q+p/d \in \mathbb{Z}$, the Gamma functions can be expanded as
\[
\frac{n!}{k!} \frac{\Gamma(k+1+q+p/d)}{\Gamma(n+1+q+p/d)}
= \frac{n!}{k!} \frac{(k+q+p/d)!}{(n+q+p/d)!} = \frac{(k+1)\cdots(k+q+p/d)}{(n+1)\cdots(n+q+p/d)}\to\left(\frac{k}{n}\right)^{q+\frac{p}{d}} \ \text{as}\ k,n\to\infty.
\]
If $q+p/d$ is a fraction, Gautschi's inequality applies:
for $x\in \R_+$ and $y\in(0,1)$,
$
x^{1-y}<\frac{\Gamma(x+1)}{\Gamma(x+y)}<(x+1)^{1-y}.
$
Then, if $q+p/d>0$,
\begin{align*}
& \frac{n!}{k!} \frac{\Gamma(k+1+q+p/d)}{\Gamma(n+1+q+p/d)}
=\frac{n!}{k!} \frac{\Gamma(k+1+\lfloor q+p/d \rfloor + ( q+p/d -\lfloor q+p/d \rfloor))}{\Gamma(n+1+\lfloor q+p/d \rfloor + ( q+p/d -\lfloor q+p/d \rfloor))}\\
\leq & \frac{n!}{k!} \frac{\Gamma(k+\lfloor q+p/d \rfloor+2)}{\left(k+1+\lfloor q+p/d \rfloor\right)^{1- ( q+p/d -\lfloor q+p/d \rfloor)}}\frac{\left(n+\lfloor q+p/d \rfloor+2\right)^{1- ( q+p/d -\lfloor q+p/d \rfloor)}}{\Gamma(n+\lfloor q+p/d \rfloor+2)} \\
=    & \frac{n!}{k!} \frac{(k+\lfloor q+p/d \rfloor+1)!}{(n+\lfloor q+p/d \rfloor+1)!}\left(\frac{k+\lfloor q+p/d \rfloor+1}{n+\lfloor q+p/d \rfloor+2}\right)^{ q+p/d -\lfloor q+p/d \rfloor-1} \\
=    & \frac{(k+1)\cdots(k+\lfloor q+p/d \rfloor)}{(n+1)\cdots(n+\lfloor q+p/d \rfloor)} \frac{n+\lfloor q+p/d \rfloor+2}{n+\lfloor q+p/d \rfloor+1}\left(\frac{k+\lfloor q+p/d \rfloor+1}{n+\lfloor q+p/d \rfloor+2}\right)^{q+p/d -\lfloor q+p/d \rfloor}     \\
\to  & \left(\frac{k}{n}\right)^{\lfloor q+p/d \rfloor} \left(\frac{k}{n}\right)^{q+p/d -\lfloor q+p/d \rfloor}
= \left(\frac{k}{n}\right)^{ q+p/d } \ \text{as} \ n,k\to \infty.
\end{align*}
When $q+p/d<0$, we can apply similar calculation and the same conclusion follows.
\hfill \Halmos

\subsection{Proof of Lemma~\ref{lem: knn bias and var}}
The proof of the bias is given by Proposition $3$ in~\cite{mack1981local}.
We rewrite the proof with our notations and accommodate for the simulation error that is not considered in~\cite{mack1981local}.
We first derive the conditional expectation of $\widehat{Y}_{kNN}^s(\tboot)$.
Because $\tsim_{(i)},$ $i\in [k]$ are i.i.d.\  given $R_{n,k+1}$, and $Y_j(\tsim_{(i)})$ and $Y_l(\tsim_{(i)})$ are i.i.d.\  given $\tsim_{(i)}$ for $j\neq l$, it follows that
\begin{align}
&\E[\widehat{Y}_{kNN}^s(\tboot)| R_{n,k+1}, \tboot, \hatt]
\overset{\eqref{eq:Y_kNN-scaled}}{=}\frac{\sum_{i=1}^k \E[\frac{1}{r}\sum_{j=1}^r \E[{Y}_j|\tsim_{(i)}]|R_{n,k+1}, \tboot, \hatt] }{nV_d R_{n,k+1}^d f(\tboot|\hatt)}
=  \frac{k \E[ Y (\tsim_{(i)}) |R_{n,k+1}, \tboot, \hatt]}{n V_d R_{n,k+1}^d f(\tboot|\hatt)} \nonumber \\
= &\frac{k}{n } \frac{\E[Y|\tboot] }{\gpr{R_{n,k+1}}} + \frac{\Delta_{\tsim}(\E[Y|\tboot]f(\tboot|\hatt)) }{2(d+2)f(\tboot|\hatt)}\frac{k}{n} \frac{R_{n,k+1}^2}{\gpr{R_{n,k+1}}} + o \left(\frac{k}{n} \frac{R_{n,k+1}^2}{\gpr{R_{n,k+1}}}\right). \label{eq: knn bias result1}  
\end{align}
The last equality holds by Lemma~\ref{lem:ball_integration}.
Then,~\eqref{eq:lemma1eq1} is obtained by taking expectation of~\eqref{eq: knn bias result1} and applying Lemma~\ref{lem:R_moments}.

For the variance~\eqref{eq:lemma1eq2}, we begin with the law of total variance,
\begin{equation}\label{eq: knn law of total variance}
\var[\widehat{Y}_{kNN}^s(\tboot) |\tboot, \hatt]
= \var[	\E[ \widehat{Y}_{kNN}^s(\tboot) |R_{n,k+1}] |\tboot,\hatt] 
+ \E[\var[ \widehat{Y}_{kNN}^s(\tboot)|R_{n,k+1}] |\tboot,\hatt].
\end{equation}
Let $\circled{1} \triangleq \E[ \widehat{Y}_{kNN}^s(\tboot) |R_{n,k+1},\tboot,\hatt]$ and $\circled{2}\triangleq \var[ \widehat{Y}_{kNN}^s(\tboot)|R_{n,k+1}],\tboot,\hatt]$.
Inserting~\eqref{eq: knn bias result1} into the expression of $\circled{1}$,
\footnotesize
\begin{align*}
& \var[\circled{1}|\tboot,\hatt]
=  \var\left[\left.\frac{k}{n } \frac{\E[Y|\tboot] }{\gpr{R_{n,k+1}}} + \frac{k\Delta_{\tsim} (\E[Y|\tboot]f(\tboot|\hatt))}{2n(d+2)f(\tboot|\hatt)}\frac{R_{n,k+1}^2}{\gpr{R_{n,k+1}}} + o \left(\frac{k}{n} \frac{R_{n,k+1}^2}{\gpr{R_{n,k+1}}}\right) \right| \tboot, \hatt\right]\\
\leq & \frac{2k^2}{n^2 }(\E[Y|\tboot])^2 \var[G^{-1}(R_{n,k+1}|\tboot,\hatt)|\tboot, \hatt] 
+ \frac{2k^2}{4(d+2)^2n^2}\left(\frac{\Delta_{\tsim} (\E[Y|\tboot]f(\tboot|\hatt))}{f(\tboot|\hatt)} \right)^2 \var\left[\left. \frac{R_{n,k+1}^2 + o (R_{n,k+1}^2)}{\gpr{R_{n,k+1}}} \right|\tboot, \hatt\right].
\end{align*} \small
The inequality follows from that $\var[A+B]\leq 2\var[A]+2\var[B]$ for arbitrary random variables $A$ and $B$.
From Lemma~\ref{lem:R_moments},
$
\var[G^{-1}(R_{n,k+1}|\tboot,\hatt)|\tboot, \hatt] 
= \E[G^{-2}(R_{n,k+1}|\tboot,\hatt)|\tboot, \hatt] - (\E[G^{-1}(R_{n,k+1}|\tboot,\hatt)|\tboot, \hatt])^2
= \frac{n(n-1)}{k(k-1)} - \frac{n^2}{k^2}
= \frac{n(n-k)}{k^2(k-1)},
$
and
$$
\var\left[\left. \frac{R_{n,k+1}^2 + o (R_{n,k+1}^2)}{\gpr{R_{n,k+1}}} \right|\tboot, \hatt \right]
\leq \E\left[\left. \frac{R^4_{n,k+1} + o(R^4_{n,k+1})}{G^2(R_{n,k+1}|\tboot,\hatt)} \right|\tboot,\hatt\right]
= ({V_d f(\tboot|\hatt)})^{-\frac{4}{d}} \left(\frac{k}{n}\right)^{ -2+\frac{4}{d} } + o\left(\left(\frac{k}{n}\right)^{ -2+\frac{4}{d} }\right).
$$
Therefore, we have
\footnotesize
\[
\var[\circled{1} |\tboot,\hatt]
= \frac{2(\E[Y|\tboot])^2 }{k}
+ o\left(\frac{1}{k}\right) 
+ \frac{(\Delta_{\tsim} (\E[Y|\tboot]f(\tboot|\hatt)))^2}{2(d+2)^2V_d^{\frac{4}{d}}(f(\tboot|\hatt))^{2+\tfrac{4}{d}} } \left(\frac{k}{n}\right)^{\tfrac{4}{d}}
+ o\left(\left(\frac{k}{n}\right)^{\tfrac{4}{d}}\right).
\] \small
For the second term of~\eqref{eq: knn law of total variance}, we rewrite 
\[
\circled{2}
=\var\left[\left.\frac{ \sum_{i=1}^k \bar{Y} (\tsim_{(i)})}{n V_d R_{n,k+1}^d f(\tboot|\hatt)} \right|R_{n,k+1},\tboot,\hatt \right]
=\frac{\var[ \sum_{i=1}^k \bar{Y}(\tsim_{(i)})|R_{n,k+1},\tboot,\hatt]}{( n V_d R_{n,k+1}^d f(\tboot|\hatt))^2}. 
\]
Because $\tsim_{(i)}$ are i.i.d.\  given $R_{n,k+1}$, and the simulation outputs are i.i.d.\ conditional on specific parameter $\tsim_{(i)}$,
$
\var\left[ \sum_{i=1}^k \bar{Y} (\tsim_{(i)}) | R_{n,k+1}, \tboot, \hatt \right]
= k \var\left[\frac{1}{r}\sum_{j=1}^r Y_j(\tsim_{(i)})|R_{n,k+1}, \tboot, \hatt\right]
=\frac{k}{r} \var[Y_j(\tsim_{(i)}) |R_{n,k+1}, \tboot, \hatt].
$
Again, from Lemmas~\ref{lem:ball_integration} and~\ref{lem:R_moments}, we obtain
$
\E[({n V_d R_{n,k+1}^d f(\tboot|\hatt)})^{-2} | \tboot,\hatt] = \frac{1}{k^2} + o(\frac{1}{k^2}),
$
and therefore
$
\E[\circled{2} |\tboot ]
= \frac{\var[Y|\tboot]}{rk} + o(\frac{1}{rk}).
$
Combining all pieces,~\eqref{eq:lemma1eq2} follows.
\hfill \Halmos

\subsection{Proof of Proposition~\ref{prop:kNN-MSE}}
For the bias, we let $\widehat{\eta}$ be $\widehat{\eta}_{kNN}(\tboot) = \widehat{Y}_{kNN}^s (\tboot)/ \widehat{A}_{kNN}^s (\tboot)$ in~\eqref{eq:bias-decomp} and take the expectation.
The first order terms gives
$
    \tfrac{\E[d(Y)|\tboot,\hatt] }{\E[A|\tboot]} -\tfrac{\eta(\tboot)\E[d(A)|\tboot,\hatt]}{\E[A|\tboot]} = \cO ((\tfrac{k}{n})^{\frac{2}{d}}),
$
the cross product term is bounded via the Cauchy–Schwarz inequality,
$
    {\E[d(A) d(Y)|\tboot,\hatt]}
    \leq (\E[d(A)^2|\tboot,\hatt] \E[d(Y)^2|\tboot,\hatt])^{\frac{1}{2}}
    = \cO ((\tfrac{k}{n})^{\frac{4}{d}}) +\cO(\frac{1}{k}),
$
and
$
    \frac{\E[Y|\tboot]}{(\E[A|\tboot])^3}\E[(d(A))^2|\tboot,\hatt]
    = \cO ((\tfrac{k}{n})^{\frac{4}{d}}) +\cO(\frac{1}{k})
$
by Lemma~\ref{lem: knn bias and var}.
Consequently,
$
    \E[\widehat{\eta}_{kNN}(\tboot)|\tboot,\hatt] -\eta(\tboot)
    =\cO((\tfrac{k}{n})^{\frac{2}{d}}) + \cO(\tfrac{1}{k}).
$

For the MSE, we square both sides of~\eqref{eq:bias-decomp}, then take the expectation.
Then, the dominant terms are $\E[d(Y)^2|\tboot,\hatt]$ and $\E[d(A)^2|\tboot,\hatt] $, which are in orders of
$\cO((\tfrac{k}{n})^{\frac{4}{d}}) +\cO(\frac{1}{k})$ by Lemma~\ref{lem: knn bias and var}.
\hfill \Halmos

\subsection{Proof of Theorem~\ref{thm:CLT}}
We establish this theorem by demonstrating a CLT for the first order terms of~\eqref{eq:bias-decomp} and showing that the higher order terms are of $o_{\pr}(1)$.
Given $\tsim$, let $X_j(\tsim) \triangleq Y_j(\tsim) - \eta(\tboot)A_j(\tsim)$, where $j$ labels the $j$th simulation run, and $\bar{X}(\tsim) = \bar{Y}(\tsim) - \eta(\tboot) \bar{A}(\tsim)$ be the $r$-average.
Then, the first order terms can be rewritten as
\[
 \frac{d(Y)}{\E[A|\tboot]} - \frac{\E[Y|\tboot]d(A)}{(\E[A|\tboot])^2}
= \frac{\widehat{Y}_{kNN}(\tboot) - \eta(\tboot) \widehat{A}_{kNN}(\tboot)}{\E[A|\tboot]} 
= \frac{\frac{1}{k}\sum_{i=1}^{k}(\bar{Y}(\tsim_{(i)}) - \eta(\tboot) \bar{A}(\tsim_{(i)}))}{\E[A|\tboot]} 
=\frac{\frac{1}{k} \sum_{i=1}^{k} \bar{X}(\tsim_{(i)}) }{\E[A|\tboot]}, 
\]
and thus can also be interpreted as a $k$NN estimator.

We follow to invoke Theorem 3 in~\cite{mack1981local}, which states the CLT for a general $k$NN regression estimator.
To verify the smoothness conditions, let $g_X(\tsim,x|\hatt)$ be the joint density of $\tsim$ and $ \bar{X}(\tsim)$, conditional on $\hatt$.
For $\beta\in\{ 0,1,2\}$, consider the integral
$
\int x^\beta g_X(\tsim, x|\hatt)\rmd x = \int x^\beta g_X( x|\tsim, \hatt)\rmd x f(\tsim|\hatt) = \E[\bar{X}^\beta|\tsim]f(\tsim|\hatt).
$
Since $\{X_j(\tsim)\}_{j\in [r]}$ are i.i.d. conditional on $\tsim$, $\E[\bar{X}^\beta|\tsim]$ can be expressed by the sum of $\E[Y^\beta|\tsim]$, $\E[A^\beta|\tsim]$ and their products.
Therefore, from Assumption~\ref{assump:smoothness}, we can shown that $\int x^\beta g_X(\tsim, x|\hatt)\rmd x $ is bounded and thrice differentiable.
The other requirements of~\cite{mack1981local} Theorem 3 are satisfied by Assumption~\ref{assump:smoothness} and~\ref{assump:n and k}.
It follows that
\[
\sqrt{k}\left( \frac{1}{k} \sum_{i=1}^{k} \bar{X}(\tsim_{(i)}) - \E\left[\left. \frac{1}{k} \sum_{i=1}^{k}\bar{X}(\tsim_{(i)})\right|\tboot, \hatt \right]\right)
\overset{\mathcal{D}}{\to}\mathcal{N} \left( \mathbf{0}, \frac{\var[ \bar{X}(\tboot)|\tboot, \hatt] }{V_d} \right)
=\mathcal{N} \left( \mathbf{0}, \frac{\var[X(\tboot)|\tboot, \hatt] }{r V_d} \right).
\]
Notice that $r$ shows up in the denominator of the asymptotic variance.
Then, by scaling~\eqref{eq:bias-decomp} with $\sqrt{kr}$, we have
\begin{align}
  & \sqrt{kr}(\widehat{\eta}_{kNN}(\tboot) - \eta(\tboot))
= \sqrt{kr} \left(\frac{d(Y) -\eta(\tboot)d(A)}{\E[A|\tboot]} + \frac{\eta(\tboot)(d(A))^2-d(A) d(Y)}{(\E[A|\tboot])^2} + o((d(A))^2)\right)  \nonumber     \\
= & \frac{\sqrt{kr}}{\E[A|\tboot]} \left( \frac{1}{k} \sum_{i=1}^{k} \bar{X}(\tsim_{(i)}) - \E\left[\left. \frac{1}{k} \sum_{i=1}^{k}\bar{X}(\tsim_{(i)})\right|\tboot, \hatt \right]\right)
+ \frac{\sqrt{kr}}{\E[A|\tboot]}\E\left[\left. \frac{1}{k} \sum_{i=1}^{k}\bar{X}(\tsim_{(i)})\right|\tboot, \hatt \right]  \nonumber \\
  & +\sqrt{kr} \left( \frac{\eta(\tboot)(d(A))^2-d(A) d(Y)}{(\E[A|\tboot])^2} + o((d(A))^2)\right). \label{eq:CLT-result0}
\end{align}
For the second term of~\eqref{eq:CLT-result0},
we rewrite
$
\E[\frac{1}{k} \sum_{j=1}^{k}\bar{X}(\tsim_{(j)})|\tboot, \hatt ]
=\E[\widehat{Y}_{kNN}(\tboot)|\tboot,\hatt] - \eta(\tboot)\E[ \widehat{A}_{kNN}(\tboot)|\tboot,\hatt]
$
and proceed to derive the bias of the $\widehat{Y}_{kNN}$ in a similar fashion as in the proof of Lemma~\ref{lem: knn bias and var}, 
$
\E[\widehat{Y}_{kNN}(\tboot)|\tboot,\hatt] = \E[Y|\tboot] + \cO((\frac{k}{n})^{\frac{1}{d}}).
$
The same rate holds for the bias of $\widehat{A}_{kNN}$.
Therefore, the second term of~\eqref{eq:CLT-result0} is of the order $\cO(\sqrt{kr}(\frac{k}{n})^{1/d})$.

The last term of~\eqref{eq:CLT-result0} is of order $\cO_{\pr}(\sqrt{kr}((\tfrac{k}{n})^{\frac{2}{d}}+\frac{1}{k}))$ by following a calculation similar to what was done in the proof of Lemma~\ref{lem: knn bias and var}.
By taking $k=o(n^{\frac{2}{2+d}})$, the second term of~\eqref{eq:CLT-result0} becomes $o(1)$ and the third term tends to $o_{\pr}(1)$.
Therefore, by Slutsky's theorem, the CLT follows.
\hfill \Halmos

\section{Proofs in Section~\ref{subsec:kLR}}
\subsection{Proof of Lemma~\ref{lem:Y-kLR-MSE}}
In this proof, we derive the variance of $\widehat{Y}_{kLR}(\tboot)$ conditional on $\tboot$ and $\hatt$, with similar results applicable to $\widehat{A}_{kLR}(\tboot)$.
Given $R_{n,k+1}$, $\{\tsim_{(i)}\}_{i\in [k]}$, are i.i.d. and the simulation runs at $\tsim_{(i)}$ are independent of those run at $\tsim_{(l)}$ for any $i\neq l$, thus the estimators $\bar{Y}_{LR} (\tsim_{(i)})$ and $\bar{Y}_{LR} (\tsim_{(l)}) $ are independent.
Because each $\bar{Y}_{LR} (\tsim_{(i)})$ is an unbiased estimator of of $\E[Y|\tboot]$, the following holds.
\begin{align}
      & \E \left[ \left. ( \widehat{Y}_{kLR}(\tboot) - \E[Y|\tboot])^2 \right| R_{n,k+1}, \tboot, \hatt\right]
    = \E\left[\left. \left( \frac{1}{k} \sum_{i=1}^k ( \bar{Y}_{LR} (\tsim_{(i)}) - \E[Y|\tboot] ) \right)^2 \right| R_{n,k+1}, \tboot, \hatt \right] \nonumber \\
    = & \frac{1}{k^2}\E\left[\left. \sum_{i=1}^k ( \bar{Y}_{LR} \left(\tsim_{(i)}\right) - \E[Y|\tboot] )^2 \right|R_{n,k+1}, \tboot, \hatt\right]
    =\frac{1}{k}\E\left[\left. ( \bar{Y}_{LR} \left(\tsim_{(i)}\right) - \E[Y|\tboot] )^2 \right| R_{n,k+1}, \tboot, \hatt\right]. \label{eq:Y_kLR-MSE-result1}
\end{align}
Conditional on specific $\tsim_{(i)}$, the simulation outputs in each run multiplied by the corresponding LR are also i.i.d and unbiased with respect to $\E[Y|\tboot]$.
Thus,~\eqref{eq:Y_kLR-MSE-result1} can be rewritten as
\begin{align*}
    \eqref{eq:Y_kLR-MSE-result1}
    = & \frac{1}{k} \E\left[\left. \left( \frac{1}{r}  \sum_{j=1}^r Y_j(\tsim_{(i)}) W_j(\tsim_{(i)};\tboot )- \E[Y|\tboot] \right)^2 \right|R_{n,k+1}, \tboot, \hatt\right] \\
    = & \frac{1}{r^2k} \E\left[\left. \E\left[\left. \left( \sum_{j=1}^r (Y_j(\tsim_{(i)}) W_j(\tsim_{(i)};\tboot)- \E[Y|\tboot] )\right)^2 \right|\tsim_{(i)} \right] \right| R_{n,k+1}, \tboot, \hatt\right] \\
    = & \frac{1}{rk} \E[ \E[( Y_j(\tsim_{(i)}) W_j(\tsim_{(i)};\tboot)- \E[Y|\tboot] )^2|\tsim_{(i)} ]| R_{n,k+1}, \tboot, \hatt]               \\
    = & \frac{1}{rk} \E[ ( Y_j(\tsim_{(i)}) W_j(\tsim_{(i)};\tboot) )^2| R_{n,k+1}, \tboot, \hatt] - \frac{1}{rk} (\E[Y|\tboot])^2.
\end{align*}
Conditional on any $\tsim$, the randomness in the $j$th simulation run is determined by the simulation inputs $\Z_j(\tsim)$.
We can therefore express $Y_j(\tsim_{(i)})$ as a function of $\Z_j(\tsim_{(i)})$, $Y_j(\tsim_{(i)}) \triangleq y(\Z_j(\tsim_{(i)}))$.
From Assumption~\ref{assump:exp_family} that the density of $\Z_j(\tsim_{(i)})$ belong to an exponential family in the canonical form, we have
\footnotesize
\begin{align}
  & \E[ ( Y_j(\tsim_{(i)}) W_j(\tsim_{(i)};\tboot) )^2| \tsim_{(i)}, \tboot, \hatt,  R_{n,k+1}] \\
=& \int y^2(\Z) p_b(\Z) \exp((2\tboot - \tsim_{(i)})^\top U(\Z)) \exp(L( \tsim_{(i)}) - 2 L(\tboot))\rmd \Z \nonumber \\
= & \exp(L( \tsim_{(i)}) - 2 L(\tboot) + L(2\tboot - \tsim_{(i)}))\int y^2(\Z) p_b(\Z) \exp((2\tboot - \tsim_{(i)})^\top U(\Z) - L(2\tboot - \tsim_{(i)}))\rmd \Z \nonumber \\
= & \frac{\exp(L( \tsim_{(i)}))}{\exp(L( \tboot))} \frac{\exp(L(2\tboot - \tsim_{(i)}))}{\exp(L( \tboot))} \E[Y^2|2\tboot- \tsim_{(i)}]. \label{eq:Y_kLR-MSE-result2}
\end{align} \small
Suppose $n$ and $k$ are sufficiently large such that $\tsim_{(i)}$ and $2\tboot - \tsim_{(i)}$ lie in the neighborhood $N(\tboot)$ stated in Assumption~\ref{assump:exp_family}.
By invoking the mean value theorem, we have
\begin{subequations}
\label{eqs: Y_kLR-MSE-result3}
\begin{align}
    \exp(L( \tsim_{(i)})) =  & \exp(L( \tboot)) + \nabla_{\tsim} (\exp(L( \tsim^\prime)))^\top (\tsim_{(i)} - \tboot),       \\
    \exp(L(2\tboot - \tsim_{(i)})) = & \exp(L( \tboot)) + \nabla_{\tsim} (\exp(L( \tsim^{\prime\prime})))^\top (\tboot-\tsim_{(i)}), \\
    \E[Y^2|2\tboot- \tsim_{(i)}] =   & \E[Y^2|\tboot] + \nabla_{\tsim}\E[Y^2|\tsim^{\prime\prime\prime}]^\top (\tboot-\tsim_{(i)}),
\end{align}
\end{subequations}
where $\tsim^\prime$ lies between $\tboot$ and $\tsim_{(i)}$, and $\tsim^{\prime\prime}$ and $\tsim^{\prime\prime\prime}$ lie between $\tboot$ and $2\tboot - \tsim_{(i)}$, i.e., $\tsim^\prime, \tsim^{\prime\prime}$ and $\tsim^{\prime\prime\prime}$ lie within $N(\tboot)$.
Interchanging the integral and differential operator, for any $\tsim \in N(\tboot)$,
\begin{align}
    \nabla_{\tsim} \exp(L(\tsim))
     & =\nabla_{\tsim} \int p_b(\Z_j)\exp(\tsim^\top U(\Z_j))\rmd \Z
    = \int p_b(\Z_j) \nabla_{\tsim}\exp(\tsim^\top U(\Z_j))\rmd \Z \nonumber   \\
     & = \int U(\Z) p_b(\Z_j) \exp(\tsim^\top U(\Z_j))\rmd \Z
    =\E[U(\Z)|\tsim] <\infty. \label{eq: Y_kLR-MSE-result4}
\end{align}
Combining~\eqref{eqs: Y_kLR-MSE-result3} and~\eqref{eq: Y_kLR-MSE-result4} and plugging into~\eqref{eq:Y_kLR-MSE-result2}, we obtain
\footnotesize
\begin{align*}
    \eqref{eq:Y_kLR-MSE-result2}
    = & \left(1 + \frac{\E[U(\Z)|\tsim^\prime]^\top (\tsim_{(i)} - \tboot)}{\exp(L( \tboot))}\right) \left(1+ \frac{\E[U(\Z)|\tsim^{\prime\prime}]^\top (\tboot-\tsim_{(i)} )}{\exp(L( \tboot))}\right) \left(\E[Y^2|\tboot] + \nabla_{\tsim}\E[Y^2|\tsim^{\prime\prime\prime}]^\top (\tboot-\tsim_{(i)})\right) \\
    = & (1 + \cO\lVert \tboot-\tsim_{(i)} \rVert)^2 (\E[Y^2|\tboot] + \cO\lVert \tboot-\tsim_{(i)} \rVert),
\end{align*}
\small
where the last equality follows from Assumption~\ref{assump:exp_family}.
Combining all pieces,
\begin{align*}
      & \E\left[\left. ( \widehat{Y}_{kLR}(\tboot) - \E[Y|\tboot] )^2 \right| \tboot,\hatt\right]
    =\E \left[ \left.\E \left[ \left. ( \widehat{Y}_{kLR}(\tboot) - \E[Y|\tboot])^2 \right| R_{n,k+1}\right]\right| \tboot \hatt\right]  \\
    = & \frac{1}{rk}\E\left[\left. \E\left[\left. (1 + \cO\lVert \tboot-\tsim_{(i)} \rVert)^2 (\E[Y^2|\tboot] + \cO\lVert \tboot-\tsim_{(i)} \rVert) \right|R_{n,k+1}\right] -(\E[Y|\tboot])^2 \right|\tboot,\hatt\right] \\
    = & \frac{1}{rk}\E\left[\left. \E[Y^2|\tboot] -(\E[Y|\tboot])^2\right|\tboot, \hatt \right] + o\left(\frac{1}{rk}\right)
    = \frac{1}{rk}\var[Y|\tboot] + o\left(\frac{1}{rk}\right).
\end{align*}
\hfill \Halmos

\subsection{Proof of Proposition~\ref{prop:kLR-MSE}}
We start with the bias expression.
Because $\widehat{Y}_{kLR}$ and $\widehat{A}_{kLR}$ are unbiased, $\E[d(A)|\tboot, \hatt] = \E[d(Y)|\tboot, \hatt] = 0$.
If we take expectation of both sides of~\eqref{eq:bias-decomp}, we have
\[
    \E[\widehat{\eta}_{kLR}(\tboot) - \eta(\tboot)|\tboot, \hatt]
    = {\E[\eta(\tboot)d(A)^2-d(A) d(Y)|\tboot, \hatt]}/{(\E[A|\tboot])^2} + \E[o((d(A))^2)|\tboot, \hatt].
\]
The numerator of the dominant (first) term can be bounded by the Cauchy–Schwarz inequality:
$
\E[\eta(\tboot)d(A)^2-d(A) d(Y)|\tboot, \hatt]
\leq  \eta(\tboot)\E[d(A)^2|\tboot, \hatt] + (\E[d(Y)^2|\tboot, \hatt] \E[d(A)^2|\tboot, \hatt])^{\frac{1}{2}}                   
= \frac{\eta(\tboot)\var[A|\tboot]}{rk} + \frac{(\var[A|\tboot]\var[Y|\tboot])^{\frac{1}{2}}}{rk}+ o\left(\frac{1}{rk}\right)
= \cO(\frac{1}{rk}).
$

For the MSE, by squaring and taking expectation of both sides of~\eqref{eq:bias-decomp},
\[
    \E[(\widehat{\eta}_{kLR}(\tboot) - \eta(\tboot))^2| \tboot,\hatt]
    = \E[(d(Y)- \eta(\tboot)d(A))^2 | \tboot,\hatt] /(\E[A|\tboot])^2 
    + \E[o(d(A)^2)+ o(d(A)d(Y))| \tboot,\hatt].
\]
Applying the Cauchy–Schwarz inequality to the dominant term's numerator, we have
$
\E[ (d(Y)- \eta(\tboot)d(A))^2|\tboot, \hatt]
\leq 2\E[d(Y)^2|\tboot, \hatt] + 2\eta(\tboot)^2\E[ d(A)^2|\tboot, \hatt] 
= \frac{2}{rk}(\var[Y|\tboot] + \eta(\tboot)^2\var[A|\tboot]) + o\left(\frac{1}{rk}\right)
= \cO(\frac{1}{rk}).
$
Consequently,
$
    \E[(\widehat{\eta}_{kLR}(\tboot) - \eta(\tboot))^2|\tboot, \hatt] = \cO(\frac{1}{rk})
$
follows.
\hfill \Halmos

\section{Proofs in Section~\ref{subsec:CIs}}
We first define some auxiliary quantities that are used in the proofs in this section.
Given the bootstrap parameters $\{\tboot_{i}\}_{i\in [\widetilde{n}]}$, let
$\Phi_{\widetilde{n}}(x) \triangleq \frac{1}{\widetilde{n}}\sum_{i=1}^{\widetilde{n}} \mathbf{1}\{\eta (\tboot_i)\leq x\}$
denote the ecdf constructed from the true ratios, $\eta$s, without simulation error. As an intermediate step towards showing that the ecdf constructed from the ratio estimators, we show that the ecdf converges to $\Phi_{\widetilde{n}}(x)$. 
We additionally define $\widetilde{\varepsilon}_j = \mathbf{1}\{\widehat{\eta} (\tboot_j)\leq x\} - \mathbf{1}\{\eta (\tboot_j)\leq x\}$, where $\widehat{\eta}$ may be replaced with $\widehat{\eta}_{std}, \widehat{\eta}_{kNN}$, or $\widehat{\eta}_{kLR}$ depending on the proof.

\subsection{Proof of Lemma~\ref{lem: knn cdf convergence}}
Since $\Phi_{\widetilde{n}}$ is an unbiased estimator for $\Phi$, at any $x\in\mathbb{R}$,
$
    \E[\Phi_{\widetilde{n},r}(x)|\hatt] - \Phi(x|\hatt)
    =\E[\Phi_{\widetilde{n},r}(x) - \Phi_{\widetilde{n}}(x)|\hatt]
    = \frac{1}{\widetilde{n}}\sum_{j=1}^{\widetilde{n}} \E[\mathbf{1}\{\widehat{\eta}_{kNN} (\tboot_j)\leq x\} - \mathbf{1}\{\eta(\tboot_j)\leq x\}|\hatt]
    = \frac{1}{\widetilde{n}}\sum_{j=1}^{\widetilde{n}} \E[\widetilde{\varepsilon}_j|\hatt].
$
We can express the expectation as an integral,
\begin{equation}
    \E[\widetilde{\varepsilon}_j|\hatt]
    =  \int_{-\infty}^0 \int_{x}^{ x-\frac{\varepsilon}{\sqrt{r k}}} h_j(\eta,\varepsilon|\hatt)\rmd \eta \rmd \varepsilon - \int_{0}^\infty \int_{ x-\frac{\varepsilon}{\sqrt{r k}} }^x h_j(\eta,\varepsilon|\hatt)\rmd \eta \rmd \varepsilon
    = \int_{-\infty}^{\infty} \int_x^{x-\frac{\varepsilon}{\sqrt{r k}}} h_j(\eta,\varepsilon|\hatt)\rmd \eta \rmd \varepsilon. \label{eq:Phi-bias-MVT}
\end{equation}
By applying the mean value theorem on $h_j(\eta,\varepsilon|\hatt)$, there exists $\widetilde{x}$ lying between $x$ and $\eta$ such that
$
    h_j(\eta,\varepsilon|\hatt)
    = h_j(x,\varepsilon|\hatt) + \frac{\partial}{\partial \eta} h_j(\widetilde{x},\varepsilon|\hatt) (\eta-x).
$
Then, from Assumption~\ref{assump: ecdf}(ii),
\begin{align*}
    \E[\widetilde{\varepsilon}_j|\hatt]
    =    & \int_{-\infty}^{\infty} \int_x^{x-\frac{\varepsilon}{\sqrt{r k}}} h_j(x,\varepsilon|\hatt) \rmd \eta \rmd \varepsilon  +  \int_{-\infty}^{\infty} \int_x^{x-\frac{\varepsilon}{\sqrt{r k}}} \frac{\partial}{\partial \eta} h_j(\widetilde{x},\varepsilon|\hatt) (\eta-x)\rmd \eta \rmd \varepsilon \\
    \leq & -\frac{1}{\sqrt{r k}}\int_{-\infty}^\infty h_j(x,\varepsilon|\hatt) \varepsilon\rmd \varepsilon  + \frac{1}{2 r k} \int_{-\infty}^{\infty} p_{1,n,r}(\varepsilon) \varepsilon^2 \rmd \varepsilon.
\end{align*}
Because the conditional pdf of $\varepsilon_j$ given $\eta(\tboot_j)$ and $\hatt$ is
$
h_j(\varepsilon|\eta,\hatt)=h_j(\eta,\varepsilon|\hatt)/\phi(\eta|\hatt),
$
the first term can be written as
\begin{equation}
    -\frac{1}{\sqrt{r k}}\int_{-\infty}^\infty h_j(x,\varepsilon|\hatt) \varepsilon\rmd \varepsilon
    =-\frac{1}{\sqrt{r k}}\int_{-\infty}^\infty h_j(\varepsilon|x,\hatt) \phi(x|\hatt) \varepsilon\rmd \varepsilon
    = -\phi(x|\hatt) \E[ \tfrac{\varepsilon_j}{\sqrt{r k}} | \eta(\tboot_j) = x,\hatt]. \label{eq: Phi-bias-cond pdf}
\end{equation}
From the definition, $\tfrac{\varepsilon_j}{\sqrt{r k}} = \widehat{\eta}_{kNN} (\tboot_j) - \eta(\tboot_j)$ represents the estimation error, whose point-wise bias is given in Proposition~\ref{prop:kNN-MSE}. 
Recall that the proof of Proposition~\ref{prop:kNN-MSE} utilizes the bias and variance expressions given by Lemma~\ref{lem: knn bias and var} to show the point-wise MSE of the $k$NN ratio estimator. 
As Assumption~\ref{assump: ecdf}(i) assumes that $f(\tilde{\bt}|\hatt)$ is bounded away from zero, the upper bounds on the bias and variance in Lemma~\ref{lem: knn bias and var} hold for all $\tilde{\bt} \in \widetilde{\Theta}$ and thus Proposition~\ref{prop:kNN-MSE} holds uniformly.
Then, define $T_x\triangleq \{\tboot: \eta(\tboot) = x\} \subseteq \widetilde{\Theta}$, we have
$
    \E[  \tfrac{\varepsilon_j}{\sqrt{r k}}| \eta(\tboot_j) = x,\hatt]
    = \E[ \tfrac{\varepsilon_j}{\sqrt{r k}}| \tboot_j\in T_x,\hatt]
    = \cO((\tfrac{k}{n})^{\frac{2}{d}}) + \cO(\tfrac{1}{k}).
$ 
Consequently, we have
$
\E[\Phi_{\widetilde{n},r}(x)|\hatt] - \Phi(x|\hatt)
= \cO((\tfrac{k}{n})^{\frac{2}{d}}) + \cO(\tfrac{1}{k}).
$

For the variance,
$
    \var[\Phi_{\widetilde{n},r}(x)|\hatt]
    = \var[\Phi_{\widetilde{n}}(x) + \Phi_{\widetilde{n},r}(x) - \Phi_{\widetilde{n}}(x)|\hatt]
    \leq 2\var[ \Phi_{\widetilde{n}}(x)|\hatt] + 2\var[\Phi_{\widetilde{n},r}(x) - \Phi_{\widetilde{n}}(x)|\hatt].
$
The first term $\var[ \Phi_{\widetilde{n}}(x)|\hatt] = \cO(\tfrac{1}{\widetilde{n}})$
because $\Phi_{\widetilde{n}}(x)$ is the ecdf of i.i.d. samples $\{\eta(\tboot_i)\}_{i\in [\widetilde{n}]}$.
We decompose the second term into
$
    \var[\Phi_{\widetilde{n},r}(x) - \Phi_{\widetilde{n}}(x)|\hatt]
    = \frac{1}{\widetilde{n}^2}\sum_{j=1}^{\widetilde{n}}\var[\widetilde{\varepsilon}_j |\hatt]
    + \frac{1}{{\widetilde{n}}^2}\sum_{i=1}^{\widetilde{n}}\sum_{j=1, i\neq j}^{\widetilde{n}} \cov[\widetilde{\varepsilon}_i, \widetilde{\varepsilon}_j|\hatt].
$
Since $\widetilde{\varepsilon}_j$ is bounded by definition, $\var[\widetilde{\varepsilon}_j|\hatt]$ is bounded, and thus
$\frac{1}{\widetilde{n}^2}\sum_{j=1}^{\widetilde{n}}\var[\widetilde{\varepsilon}_j |\hatt] =\cO(\tfrac{1}{\widetilde{n}})$.
By rewriting the covariance into expectations,
\begin{equation}
    \frac{1}{{\widetilde{n}}^2}\sum_{i=1}^{\widetilde{n}}\sum_{j=1, i\neq j}^{\widetilde{n}} \cov[\widetilde{\varepsilon}_i, \widetilde{\varepsilon}_j|\hatt]
    = \frac{1}{{\widetilde{n}}^2}\sum_{i=1}^{\widetilde{n}}\sum_{j=1, i\neq j}^{\widetilde{n}} \E[ \widetilde{\varepsilon}_i\widetilde{\varepsilon}_j |\hatt]
    - \frac{1}{{\widetilde{n}}^2}\sum_{i=1}^{\widetilde{n}}\sum_{j=1, i\neq j}^{\widetilde{n}}  \E[ \widetilde{\varepsilon}_i |\hatt] \E[\widetilde{\varepsilon}_j |\hatt]. \label{eq: Phi-cov-expansion}
\end{equation}
The second term in~\eqref{eq: Phi-cov-expansion} can be bounded by
$
    \cO((\tfrac{k}{n})^{\frac{4}{d}}) + \cO(\tfrac{1}{k^2})
$
from the bias analysis;
and the first term can be rewritten as integrals in a similar fashion to~\eqref{eq:Phi-bias-MVT},
\[
    \E[ \widetilde{\varepsilon}_i\widetilde{\varepsilon}_j |\hatt ]
    = \int_{-\infty}^{\infty} \int_{-\infty}^{\infty} \int_x^{x-\frac{\varepsilon}{\sqrt{r k}}} \int_x^{x-\frac{\varepsilon}{\sqrt{k^2}}}
    h_{i,j}(\eta_i,\eta_j,\varepsilon_i,\varepsilon_j)
    \rmd \eta_i\rmd \eta_j \rmd \varepsilon_i\rmd \varepsilon_j,
\]
and use the same technique to obtain an upper bound.
Namely, by applying the first order Taylor expansion to $h_{i,j}(\eta_i,\eta_j,\varepsilon_i,\varepsilon_j)$, there exist $\widetilde{x}_i$ lying between $x$ and $\eta_i$ and $\widetilde{x}_j$ lying between $x$ and $\eta_j$ such that
$
    h_{i,j}(\eta_i,\eta_j,\varepsilon_i,\varepsilon_j) = h_{i,j}(x,x,\varepsilon_i,\varepsilon_j)
    + \frac{\partial}{\partial \eta_i} h_{i,j}(\widetilde{x}_i,\widetilde{x}_j,\varepsilon_i,\varepsilon_j)(\eta_i-x)
    + \frac{\partial}{\partial \eta_j} h_{i,j}(\widetilde{x}_i,\widetilde{x}_j,\varepsilon_i,\varepsilon_j)(\eta_j-x),
$
which is bounded from below/above by
$
    h_{i,j}(x,x,\varepsilon_i,\varepsilon_j)
    \mp p_{1,n,r}(\varepsilon_i,\varepsilon_j) (\lvert \eta_i-x \rvert + \lvert \eta_j-x \rvert)
$
from Assumption~\ref{assump: ecdf}(iii).

Let $\phi_{ij} (\cdot,\cdot|\hatt)$ be the joint pdf of $\eta(\tboot_i)$ and $\eta(\tboot_j)$.
Following the analysis similar to~\eqref{eq: Phi-bias-cond pdf} and invoking Proposition~\ref{prop:kNN-MSE}, we have
\begin{align}
         & \E[ \widetilde{\varepsilon}_i\widetilde{\varepsilon}_j|\hatt ]
    \leq  \int_{-\infty}^\infty\int_{-\infty}^\infty \frac{\varepsilon_i \varepsilon_j}{r k} h_{i,j}(x,x,\varepsilon_i,\varepsilon_j)\rmd \varepsilon_i \rmd\varepsilon_j
    + \int_{-\infty}^\infty\int_{-\infty}^\infty  \frac{\lvert \varepsilon_i^2\varepsilon_j+\varepsilon_i\varepsilon_j^2 \rvert}{2 (r k)^{\frac{3}{2}}}p_{1,n,r}(\varepsilon_i,\varepsilon_j) \rmd \varepsilon_i \rmd\varepsilon_j\nonumber   \\
    =    & \phi_{ij}(x,x) \E[\tfrac{\varepsilon_i \varepsilon_j}{r k}|\eta(\tboot_i) =x, \eta(\tboot_j)=x,\hatt] + \cO((rk)^{-\frac{3}{2}}) \nonumber    \\
    \leq & \phi_{ij}(x,x)(\E[ \tfrac{\varepsilon_i^2}{r k}|\eta(\tboot_i) =x, \eta(\tboot_j)=x,\hatt] \E [ \tfrac{\varepsilon_j^2}{r k}|\eta(\tboot_i) =x, \eta(\tboot_j)=x,\hatt])^{\frac{1}{2}} + \cO((rk)^{-\frac{3}{2}}) \label{eq: Phi-cov CS ieq} \\
    =    & \cO((\tfrac{k}{n})^{\frac{4}{d}} )+\cO(\tfrac{1}{k}) + \cO((rk)^{-\frac{3}{2}}). \nonumber
\end{align}
Combining all pieces, we arrive at
$
    \var[\Phi_{\widetilde{n},r}(x)|\hatt]
    = \cO(\tfrac{1}{\widetilde{n}}) + \cO((\tfrac{k}{n})^{\frac{4}{d}} ) +\cO(\tfrac{1}{k}).
$

Because $\Phi_{\widetilde{n},r}(x)$ converges in quadratic mean to $\Phi(x)$ for any $x\in \mathbb{R}$, the convergence in probability follows.
\hfill \Halmos

\subsection{Proof of Proposition~\ref{prop: knn qt conergence}}
Recall that $\Phi_{\widetilde{n},r}$ is the ecdf constructed from $\{\widehat{\eta}_{kNN}(\tboot_i)\}_{i\in [\widetilde{n}]}$ and $q_{\alpha,\widetilde{n}}(\{\widehat{\eta}_{kNN}(\tboot_i)\}|\hatt)$ is the $\alpha$-level quantile.
Then $\Phi_{\widetilde{n},r}( q_{\alpha,\widetilde{n}}(\{\widehat{\eta}_{kNN}(\tboot_i)\}|\hatt)) = {\lceil \widetilde{n}\alpha \rceil}/{\widetilde{n}}$, which differs from $\alpha$ by at most ${1}/{\widetilde{n}}$.
Recall that $\phi$ is the pdf of $\Phi$.
By applying the inverse function rule and invoking the mean value theorem (for some $\widetilde{q}$ between $ q_{\alpha,\widetilde{n}}(\{\widehat{\eta}_{kNN}(\tboot_i)\}|\hatt)$ and $q_\alpha(\eta(\tboot)|\hatt)$), we have
\begin{align*}
& \lvert  q_{\alpha,\widetilde{n}}(\{\widehat{\eta}_{kNN}(\tboot_i)\}|\hatt) - q_\alpha(\eta(\tboot)|\hatt)\rvert
= \lvert \Phi^{-1}( \Phi(  q_{\alpha,\widetilde{n}}( \{\widehat{\eta}_{kNN}(\tboot_i)\}|\hatt)) ) - \Phi^{-1}(\Phi(q_\alpha(\eta(\tboot)|\hatt))) \rvert \\
=    & \frac{1}{\phi(\widetilde{q})} \lvert \Phi(q_{\alpha,\widetilde{n}}(\{\widehat{\eta}_{kNN}(\tboot_i)\}|\hatt)) - \Phi(q_\alpha(\eta(\tboot)|\hatt))\rvert
=\frac{1}{\phi(\widetilde{q})} \lvert \Phi(q_{\alpha,\widetilde{n}}(\{\widehat{\eta}_{kNN}(\tboot_i)\}|\hatt)) - \alpha\rvert \\
\leq & \frac{1}{\phi(\widetilde{q})} (\lvert \Phi(q_{\alpha,\widetilde{n}}(\{\widehat{\eta}_{kNN}(\tboot_i)\}|\hatt)) - \Phi_{\widetilde{n},r}( q_{\alpha,\widetilde{n}}(\{\widehat{\eta}_{kNN}(\tboot_i)\}|\hatt)) \rvert + \lvert \Phi_{\widetilde{n},r}(q_{\alpha,\widetilde{n}}(\{\widehat{\eta}_{kNN}(\tboot_i)\}|\hatt)) - \alpha\rvert ) \\
=    & \cO_{\pr}(\tfrac{1}{\sqrt{\widetilde{n}}}) + \cO_{\pr}((\tfrac{k}{n})^{\frac{2}{d}} ) +\cO_{\pr}(\tfrac{1}{\sqrt{k}}).
\end{align*}
The inequality follows from the triangle inequality and the last equality holds from Lemma~\ref{lem: knn cdf convergence}.
\hfill \Halmos

\subsection{Proof of Theorem~\ref{thm:qt-convergence-knn}}
\proof{Proof.}
The classical bootstrap theory guarantees that the following holds under the stated assumptions:
$
\sqrt{m}(\eta(\hatt) - \eta(\ttrue)) \overset{D}{\to} \mathcal{N}(0,\sigma_I^2)
$
and
$
\sqrt{m} (\eta(\tboot) - \eta(\hatt))| \hatt \overset{D}{\to} \mathcal{N}(0,\sigma_I^2)
$
in probability.
Moreover, for $\alpha\in (0,1)$ let $Z_{\alpha}$ be the standard normal $\alpha$-quantile.
Then, we have
$
q_{\alpha}(\tfrac{\sqrt{m}}{\sigma_I} (\eta(\hatt) - \eta(\ttrue) ) ) = Z_{\alpha} + o(1)
$
and
$
q_{\alpha}( \tfrac{\sqrt{m}}{\sigma_I} (\eta(\tboot) - \eta(\hatt) )|\hatt ) = Z_{\alpha} + o_{\pr}(1).
$
To establish~\eqref{eq:boots-CI}, we first show that for arbitrary $\varepsilon>0$, we have
\begin{align}
         & \pr\left(\eta(\ttrue) \leq q_{\alpha,\widetilde{n}} ( \{\widehat{\eta}_{kNN}(\tboot_i) \}|\hatt) \right)
    =\pr\left(\eta(\ttrue) -q_{\alpha}({\eta}(\tboot)|\hatt) \leq q_{\alpha,\widetilde{n}} ( \{\widehat{\eta}_{kNN}(\tboot_i) \}|\hatt) - q_{\alpha}({\eta}(\tboot)|\hatt)\right) \nonumber \\
    \leq & \pr\left(\eta(\ttrue) - q_{\alpha}({\eta}(\tboot)|\hatt) \leq \varepsilon\right)
    + \pr\left( q_{\alpha,\widetilde{n}} ( \{\widehat{\eta}_{kNN}(\tboot_i) \}|\hatt) - q_{\alpha}({\eta}(\tboot)|\hatt) > \varepsilon\right). \label{ieq: boots-CI-result2}
\end{align}
To derive an upper bound for the first term in~\eqref{ieq: boots-CI-result2}, we consider the probability,
$
    \pr(\eta(\ttrue) \leq q_{\alpha}({\eta}(\tboot)|\hatt))
$
\footnotesize
\begin{align}
& \pr(\eta(\ttrue) \leq q_{\alpha}({\eta}(\tboot)|\hatt))
= \pr\left( \tfrac{\sqrt{m}}{\sigma_I} (\eta(\ttrue) - \eta(\hatt)) \leq q_{\alpha}( \tfrac{\sqrt{m}}{\sigma_I} (\eta(\tboot) - \eta(\hatt))|\hatt) \right) \nonumber \\
= & \pr\left( T(\tfrac{\sqrt{m}}{\sigma_I} (\eta(\ttrue) - \eta(\hatt))) \leq q_{\alpha}(  T(\tfrac{\sqrt{m}}{\sigma_I} (\eta(\tboot) - \eta(\hatt)) ) |\hatt) \right) \nonumber             \\
= & \pr\left( T(\tfrac{\sqrt{m}}{\sigma_I} (\eta(\ttrue) - \eta(\hatt))) \geq q_{1-\alpha}( T(\tfrac{\sqrt{m}}{\sigma_I} (\eta(\tboot) - \eta(\hatt)) ) |\hatt)\right) \nonumber             \\
= & \pr\left( T(\tfrac{\sqrt{m}}{\sigma_I} (\eta(\hatt) -\eta(\ttrue))) \leq -q_{1-\alpha}( T(\tfrac{\sqrt{m}}{\sigma_I} (\eta(\tboot) - \eta(\hatt)) ) |\hatt) \right) \nonumber  \\
= & \pr\left( T(\tfrac{\sqrt{m}}{\sigma_I} (\eta(\hatt) -\eta(\ttrue))) \leq q_{\alpha}( T(\tfrac{\sqrt{m}}{\sigma_I} (\eta(\tboot) - \eta(\hatt)) ) |\hatt) \right) 
=  \pr\left( \tfrac{\sqrt{m}}{\sigma_I} (\eta(\hatt) - \eta(\ttrue)) \leq q_{\alpha}( \tfrac{\sqrt{m}}{\sigma_I} (\eta(\tboot) - \eta(\hatt))  |\hatt) \right). \label{eq:boots-CI-result0}
\end{align} \small
The second and last equalities follow from the existence of strict monotonic transformation $T$ in Assumption~\ref{assump:mono-trans}.
The third to fifth equalities hold since the distributions of  
$T( \eta(\hatt) - \eta(\ttrue))$
and
$T(\eta(\tboot_i) - \eta(\hatt))$
given $\hatt$ are symmetric around $0$.

For any $\varepsilon^\prime>0$, let $\mathcal{B}_{\varepsilon^\prime}$ denote the event,
$
    \mathcal{B}_{\varepsilon^\prime} \triangleq
    \left\{
    q_{\alpha}( \tfrac{\sqrt{m}}{\sigma_I} (\eta(\tboot) - \eta(\hatt) )|\hatt )
    - q_{\alpha}(\tfrac{\sqrt{m}}{\sigma_I} (\eta(\hatt) - \eta(\ttrue) ) )
    \leq \varepsilon^\prime
    \right\},
$
and $\mathcal{C}(\mathcal{B}_{\varepsilon^\prime})$ be the complementary event.
Then,~\eqref{eq:boots-CI-result0} can be rewritten as 
\begin{align*}
    \eqref{eq:boots-CI-result0}
     & =  \pr\left( \tfrac{\sqrt{m}}{\sigma_I} (\eta(\hatt) - \eta(\ttrue)) \leq q_{\alpha}( \tfrac{\sqrt{m}}{\sigma_I} (\eta(\tboot) -\eta(\hatt) ) |\hatt ), \mathcal{B}_{\varepsilon^\prime} \right) \\
     & \; \; \; \; + \pr\left( \tfrac{\sqrt{m}}{\sigma_I} (\eta(\hatt) - \eta(\ttrue)) \leq q_{\alpha}( \tfrac{\sqrt{m}}{\sigma_I} (\eta(\tboot) -\eta(\hatt) ) |\hatt ), \mathcal{C}(\mathcal{B}_{\varepsilon^\prime}) \right)   \\
     & \leq \pr\left( \tfrac{\sqrt{m}}{\sigma_I} (\eta(\hatt) - \eta(\ttrue) ) \leq q_{\alpha}(\tfrac{\sqrt{m}}{\sigma_I}(\eta(\hatt) - \eta(\ttrue)) ) +\varepsilon^\prime \right) + \pr(\mathcal{C}(\mathcal{B}_{\varepsilon^\prime}) ),
\end{align*} 
which tends to $\alpha+o(1)$ if $\varepsilon^\prime\to 0$.
Therefore, by choosing $\varepsilon = o(\tfrac{1}{\sqrt{m}})$, the first term in~\eqref{ieq: boots-CI-result2} can be rewritten as
$
    \pr\left( \eta(\ttrue) \leq q_{\alpha} (  \eta(\tboot) |\hatt ) +\varepsilon \right)
    =  \pr\left( \tfrac{\sqrt{m}}{\sigma_I}  (\eta(\ttrue) - \eta(\hatt) ) \leq q_{\alpha} ( \tfrac{\sqrt{m}}{\sigma_I} (\eta(\tboot) - \eta(\hatt) )  |\hatt )  +\tfrac{\sqrt{m}}{\sigma_I}\varepsilon \right)
    = \alpha + o(1).
$ 
The $o(1)$ term follows from that the distributions of $\frac{\sqrt{m}}{\sigma_I}  (\eta(\ttrue) - \eta(\hatt))$ and $\tfrac{\sqrt{m}}{\sigma_I} (\eta(\tboot) - \eta(\hatt) )$ given $\hatt$ converge to the standard normal distribution, and by choosing $\varepsilon$ to shrink sufficiently fast in $m$,  
$\pr\left( \tfrac{\sqrt{m}}{\sigma_I}  (\eta(\ttrue) - \eta(\hatt)) \leq q_{\alpha} ( \tfrac{\sqrt{m}}{\sigma_I} (\eta(\tboot) - \eta(\hatt) )  |\hatt )  +\tfrac{\sqrt{m}}{\sigma_I}\varepsilon \right)$ can be made arbitrarily close to 
$\pr\left( \tfrac{\sqrt{m}}{\sigma_I}  (\eta(\ttrue) - \eta(\hatt)) \leq q_{\alpha} ( \tfrac{\sqrt{m}}{\sigma_I} (\eta(\tboot) - \eta(\hatt) )  |\hatt )\right)$.

For the second term in~\eqref{ieq: boots-CI-result2}, we rewrite
\begin{equation}
    \pr\left( q_{\alpha,\widetilde{n}} (\{\widehat{\eta}_{kNN}(\tboot_i)\} |\hatt) - q_{\alpha}({\eta}(\tboot) |\hatt ) > \varepsilon \right)
    = \pr\left( q_{\alpha,\widetilde{n}} (\{\tfrac{\sqrt{m}}{\sigma_I}\widehat{\eta}_{kNN}(\tboot_i)\} |\hatt) - q_{\alpha}( \tfrac{\sqrt{m}}{\sigma_I}{\eta}(\tboot) |\hatt) > \tfrac{\sqrt{m}}{\sigma_I}\varepsilon \right).    \label{eq: boots-CI-result3}
\end{equation}
From Proposition~\ref{prop: knn qt conergence},
$
    q_{\alpha,\widetilde{n}} (\{\tfrac{\sqrt{m}}{\sigma_I}\widehat{\eta}_{kNN}(\tboot_i)\} |\hatt) - q_{\alpha}( \tfrac{\sqrt{m}}{\sigma_I}{\eta}(\tboot) |\hatt)
    = \cO_{\pr}(\sqrt{m}(\tfrac{1}{\sqrt{\widetilde{n}}} + (\tfrac{k}{n})^{{2}/{d}}+ \tfrac{1}{\sqrt{k}})).
$
Therefore, by making $k=\omega(m)$, $n=\omega(m^{d/4+1}), \widetilde{n}=\omega(m)$,~\eqref{eq: boots-CI-result3} tends to $0$.

Combining all the pieces, we arrive at\footnotesize
\[
\limsup_{m\to\infty} \pr\left( \eta(\ttrue) \leq  q_{\alpha,\widetilde{n}} (\{\widehat{\eta}_{kNN}(\tboot_i)\} |\hatt) \right) \leq \alpha +g_1(\varepsilon)
\ \mbox{and} \
 \liminf_{m\to\infty}\pr\left( \eta(\ttrue) \leq  q_{\alpha,\widetilde{n}} (\{\widehat{\eta}_{kNN}(\tboot_i)\} |\hatt) \right) \geq \alpha -g_2(\varepsilon),
\] \small
for some functions $g_1$ and $g_2$ that can be made arbitrarily small by choosing small $\varepsilon$.
Finally, one can similarly show that
$
    \pr(\eta(\ttrue) \leq q_{\alpha/2,\widetilde{n}} (\widehat{\eta}_{kNN}(\tboot_i)| \hatt) ) = \alpha/2 + o(1)
$
and
$
    \pr(\eta(\ttrue) \geq q_{1-\alpha/2,\widetilde{n}}( \widehat{\eta}_{kNN}(\tboot_i) | \hatt)) = \alpha/2 + o(1)
$
to obtain~\eqref{eq:boots-CI}. \hfill \Halmos

\subsection{Proof of Lemma~\ref{lem: klr cdf convergence}}
The proof is mostly identical to the proof of Lemma~\ref{lem: knn cdf convergence}, by replacing $\widehat{\eta}_{kNN}$ with $\widehat{\eta}_{kLR}$ and $\Phi_{\widetilde{n},r}$ with $\widetilde \Phi_{{\widetilde{n}},r}$, respectively.
For the bias analysis, we follow the proof of Lemma~\ref{lem: knn cdf convergence} until~\eqref{eq: Phi-bias-cond pdf}.
From Proposition~\ref{prop:kLR-MSE},~\eqref{eq: Phi-bias-cond pdf} can be similarly bounded by $\cO(\tfrac{1}{{rk}})$, resulting in
$
    \E[\widetilde\Phi_{\widetilde{n},r}(x)|\hatt] - \Phi(x|\hatt)
    =  \cO(\tfrac{1}{{rk}}).
$

For the variance, we can similarly reach~\eqref{eq: Phi-cov-expansion}, whose second term is $\cO(\tfrac{1}{r^2 k^2})$ from the bias analysis.
We then use the same proof technique as in Lemma~\ref{lem: knn cdf convergence} to obtain~\eqref{eq: Phi-cov CS ieq} and bound it with $\cO(\tfrac{1}{rk}) + \cO((rk)^{-\frac{3}{2}})$, which gives
$
    \var[\widetilde\Phi_{\widetilde{n},r}(x)|\hatt]
    = \cO(\tfrac{1}{\widetilde{n}}) +\cO(\tfrac{1}{rk})  + \cO((rk)^{-\frac{3}{2}}) +\cO(\tfrac{1}{r^2 k^2})
    = \cO(\tfrac{1}{\widetilde{n}}) +\cO(\tfrac{1}{rk}).
$
\hfill \Halmos

\subsection{Proof of Proposition~\ref{prop: klr qt conergence}}
The proof proceeds in the same way as the proof of Proposition~\ref{prop: knn qt conergence} by replacing $\widehat{\eta}_{kNN}$ with $\widehat{\eta}_{kLR}$ and invoking results in Lemma~\ref{lem: klr cdf convergence} instead of Lemma~\ref{lem: knn cdf convergence}.

\subsection{Proof of Theorem~\ref{thm:qt-convergence-kLR}}
The proof proceeds identically as that of Theorem~\ref{thm:qt-convergence-knn} until~\eqref{eq: boots-CI-result3}, by replacing $\widehat{\eta}_{kNN}$ with $\widehat{\eta}_{kLR}$.
From Proposition~\ref{prop: klr qt conergence},
$
    q_{\alpha,\widetilde{n}} (\{\tfrac{\sqrt{m}}{\sigma_I}\widehat{\eta}_{kLR}(\tboot_i)\} |\hatt) - q_{\alpha}( \tfrac{\sqrt{m}}{\sigma_I}{\eta}(\tboot) |\hatt)
    = \cO_{\pr}(\sqrt{m}(\tfrac{1}{\sqrt{\widetilde{n}}} + \tfrac{1}{\sqrt{rk}})).
$
Since $r$ remains constant, if we take $k=\omega(m)$ and $\widetilde{n}=\omega(m)$,
\begin{equation*}
    \pr\left( q_{\alpha,\widetilde{n}} (\{\widehat{\eta}_{kLR}(\tboot_i)\} |\hatt) - q_{\alpha}({\eta}(\tboot) |\hatt ) > \varepsilon \right)
    = \pr\left( q_{\alpha,\widetilde{n}} (\{\tfrac{\sqrt{m}}{\sigma_I}\widehat{\eta}_{kLR}(\tboot_i)\} |\hatt) - q_{\alpha}( \tfrac{\sqrt{m}}{\sigma_I}{\eta}(\tboot) |\hatt) > \tfrac{\sqrt{m}}{\sigma_I}\varepsilon \right) = o(1).
\end{equation*}
The remainder of the proof is identical to that of Theorem~\ref{thm:qt-convergence-knn} and we arrive at the result as desired.
\hfill \Halmos

\section{Auxiliary Results and Proof of Proposition~\ref{prop:standard.est.convergence}}\label{appendix: std ratio estimator}
We first create several auxiliary lemmas and propositions prior to showing Proposition~\ref{prop:standard.est.convergence}.
We show the bias and MSE of $\widehat{\eta}_{std}$ in Proposition~\ref{prop:std-MSE};
and establish the the bias and variance of ecdf 
$\check{\Phi}_{\widetilde{n}, r} = \frac{1}{\widetilde{n}}\sum_{i=1}^{\widetilde{n}} \mathbf{1}\{\widehat{\eta}_{std} (\tboot_i)\leq x\}$
as well as its weak uniform consistency for $\Phi(x)$ in Lemma~\ref{lem: std cdf convergence}.
As a variant to Proposition~\ref{prop: knn qt conergence} and~\ref{prop: klr qt conergence}, we show the weak consistency of the empirical quantile estimator, $q_{\alpha,\widetilde{n}}(\{\widehat{\eta}_{std}(\tboot_i)\}|\hatt)$, in Proposition~\ref{prop: std qt conergence}.
Proposition~\ref{prop:standard.est.convergence} is finally proved in a similar fashion as in Theorem~\ref{thm:qt-convergence-knn} and~\ref{thm:qt-convergence-kLR}.

\begin{proposition}\label{prop:std-MSE}
    Conditional on $\hatt$, for any $\tboot \in \widetilde{\Theta}$,
    \[
        \E[\widehat{\eta}_{std}(\tboot) - \eta(\tboot)|\tboot,\hatt] = \cO(\tfrac{1}{r}) \mbox{, and }
        \E[(\widehat{\eta}_{std}(\tboot) - \eta(\tboot))^2 |\tboot,\hatt ] = \cO(\tfrac{1}{r}).
    \]
\end{proposition}
\proof{Proof.}
For any $\bt$,
the numerator and denominator of $\widehat{\eta}_{std}$ are unbiased estimators of $\E[Y|\bt]$ and $\E[A|\bt]$, respectively.
In addition, $\var[\bar{Y}(\bt)] = \var[\frac{1}{r}\sum_{j=1}^r Y_j(\tsim)] = \cO(\tfrac{1}{r})$ and the same order holds for $\var[\bar{A}(\bt)]$.
Then, by taking expectation on both sides of~\eqref{eq:bias-decomp} and invoke the Cauchy–Schwarz inequality,
$
    \E[\widehat{\eta}_{std}(\tboot) - \eta(\tboot)|\tboot,\hatt]  = \frac{\E[d(A) d(Y)]}{(\E[A|\tboot])^2} + \frac{\E[Y|\tboot]}{(\E[A|\tboot])^3}(\E[d(A)])^2 + o(\E[(d(A))^2]) = \cO(\tfrac{1}{r}).
$
For the MSE, we take the expectation after squaring both sides of~\eqref{eq:bias-decomp}.
Then, the dominant terms that remain are $\E[d(A)^2]$ and $\E[d(Y)^2]$ with order $\cO(\tfrac{1}{r})$.
\hfill \Halmos

We comment that~\citet{hendersonglynn2001} show the same order of bias and MSE for the regenerative ratio estimator under different assumptions.

\begin{lemma}\label{lem: std cdf convergence}
    Suppose Assumption~\ref{assump: ecdf} hold.
    Then, $ \E[\check\Phi_{{\widetilde{n}},r}(x)|\hatt] = \Phi(x) + \cO(\tfrac{1}{r})$
    and
    $\var[\check\Phi_{\widetilde{n},r}(x)|\hatt] = \cO(\tfrac{1}{\widetilde{n}})$.
    Furthermore, conditional on $\hatt$,
    $\sup\nolimits_{x\in \R}\lvert \check\Phi_{\widetilde{n},r}(x) - \Phi(x)\rvert
        = \cO_{\pr}(\tfrac{1}{\sqrt{\widetilde{n}}}) +\cO_{\pr}(\tfrac{1}{r})$.
\end{lemma}
\proof{Proof.}
We follow the same steps as in proof of Lemma~\ref{lem: knn cdf convergence}.
Note that the scaled error for the standard estimator is defined as $\varepsilon_i = \sqrt{r}(\widehat{\eta}_{std} (\tboot_i) - \eta(\tboot_i))$, while the scaling factor for $\widehat{\eta}_{kNN}$ and $\widehat{\eta}_{kLR}$ are both $rk$.
By replacing $\widehat{\eta}_{kNN}$ with $\widehat{\eta}_{std}$ and $\tfrac{\varepsilon}{rk}$ with $\tfrac{\varepsilon}{r}$,
we follow the same steps until~\eqref{eq: Phi-bias-cond pdf} and invoke Lemma~\ref{lem: std cdf convergence} to obtain
$
    \E[\widetilde{\varepsilon}_j|\hatt]
    = \cO(\tfrac{1}{r}),
$
which gives
$ \E[\check\Phi_{{\widetilde{n}},r}(x)|\hatt] = \Phi(x) + \cO(\tfrac{1}{r})$.

The variance proof is different from that of Lemma~\ref{lem: knn cdf convergence} and~\ref{lem: klr cdf convergence}.
Because $\widehat{\eta}_{std}$ does not involve pooling, the estimators $\{\widehat{\eta}_{std}(\tboot_i)\}_{i\in [\widetilde{n}]}$ are i.i.d. given $\hatt$.
Hence, the variance is
$
    \var[\check\Phi_{\widetilde{n},r}(x)|\hatt]
    = \frac{1}{\widetilde{n}^2}\var[ \sum_{i=1}^{\widetilde{n}} \mathbf{1}\{\widehat{\eta}_{std} (\tboot_j)\leq x\}|\hatt]
    =\frac{1}{\widetilde{n}}\var[ \mathbf{1}\{\widehat{\eta}_{std} (\tboot_j)\leq x\}|\hatt]
    = \cO(\tfrac{1}{\widetilde{n}}).
$

The MSE is then calculated as
$
    \var[\Phi_{\widetilde{n},r}(x)|\hatt] + (\E[\Phi_{\widetilde{n},r}(x)|\hatt] - \Phi(x|\hatt))^2
    = \cO(\tfrac{1}{\widetilde{n}}) + \cO(\tfrac{1}{r^2}),
$
which gives the uniform weak convergence
$
\sup\nolimits_{x\in \R}\lvert \Phi_{\widetilde{n},r}(x) - \Phi(x)\rvert
= \cO_{\pr}(\tfrac{1}{\sqrt{\widetilde{n}}}) + \cO_{\pr}(\tfrac{1}{r}).
$
\hfill \Halmos

\begin{proposition}\label{prop: std qt conergence}
    Suppose Assumption~\ref{assump: ecdf} hold.
    Then, for $0<\alpha<1$,
    $\lvert q_{\alpha,\widetilde{n}}(\{\widehat{\eta}_{std}(\tboot_i)\}|\hatt)- q_\alpha(\eta(\tboot)|\hatt) \rvert
        = \cO_{\pr}(r^{-1})+ \cO_{\pr}(\widetilde{n}^{-1/2})$.
\end{proposition}
\proof{Proof.}
The proof of Proposition~\ref{prop: std qt conergence} is identical to that of Proposition~\ref{prop: knn qt conergence}, 
where $\widehat{\eta}_{kNN}$ is replaced with $\widehat{\eta}_{std}$, followed by invoking results in Lemma~\ref{lem: std cdf convergence}.
\hfill \Halmos

\subsection{Proof of Proposition~\ref{prop:standard.est.convergence}}
We replace $\widehat{\eta}_{kNN}$ with $\widehat{\eta}_{std}$ in the proof of Theorem~\ref{thm:qt-convergence-knn} and follow the steps until~\eqref{eq: boots-CI-result3}.
By invoking Proposition~\ref{prop: std qt conergence},
$
    q_{\alpha,\widetilde{n}} (\{\tfrac{\sqrt{m}}{\sigma_I}\widehat{\eta}_{std}(\tboot_i)\} |\hatt) - q_{\alpha}( \tfrac{\sqrt{m}}{\sigma_I}{\eta}(\tboot) |\hatt)
    = \cO_{\pr}(\sqrt{m}(\tfrac{1}{\sqrt{\widetilde{n}}} + \tfrac{1}{r})).
$
If we take $k=\omega(m)$ and $r=\omega(\sqrt{m})$,
\begin{equation*}
    \pr\left( q_{\alpha,\widetilde{n}} (\{\widehat{\eta}_{std}(\tboot_i)\} |\hatt) - q_{\alpha}({\eta}(\tboot) |\hatt ) > \varepsilon \right)
    = \pr\left( q_{\alpha,\widetilde{n}} (\{\tfrac{\sqrt{m}}{\sigma_I}\widehat{\eta}_{std}(\tboot_i)\} |\hatt) - q_{\alpha}( \tfrac{\sqrt{m}}{\sigma_I}{\eta}(\tboot) |\hatt) > \tfrac{\sqrt{m}}{\sigma_I}\varepsilon \right) = o(1).
\end{equation*}
The remaining proof is identical to that of Theorem~\ref{thm:qt-convergence-knn} and we arrive at the result as desired.
\hfill \Halmos

\end{APPENDICES}

\end{document}